\documentclass[12pt]{article}
\usepackage{psfig}
\usepackage{verbatim}
\usepackage{color}
\usepackage{rotate}
\usepackage{graphics}
\usepackage{epsf}
\usepackage{epsfig}
\usepackage{graphicx}
\usepackage{latexsym}
\usepackage{amscd}
\usepackage{amssymb}
\newcommand{\beq}{\begin{equation}}
\newcommand{\eeq}{\end{equation}}
\newcommand{\beqn}{\begin{eqnarray}}
\newcommand{\eeqn}{\end{eqnarray}}
\newcommand\noi{\noindent} 
\newcommand\la{\langle}
\newcommand\ra{\rangle}
\newcommand\eps\varepsilon
\def\pom{{\bf I\!P}}
\def\reg{{\bf I\!R}}

\def\fm{\,\mbox{fm}}
\def\GeV{\,\mbox{GeV}}
\def\TeV{\,\mbox{TeV}}
\def\lsim{\mathrel{\rlap{\lower4pt\hbox{\hskip1pt$\sim$}}
    \raise1pt\hbox{$<$}}}         
\def\gsim{\mathrel{\rlap{\lower4pt\hbox{\hskip1pt$\sim$}}
    \raise1pt\hbox{$>$}}}         

\begin{document}

\hfill LA-UR-01-5770

\vspace*{2cm}

\begin{center}
{\Large
\bf
Nuclear effects in the Drell-Yan process\\ 

\smallskip
at very high energies}
\end{center}
\vspace{.5cm}

\begin{center}

 {\large B.Z.~Kopeliovich$^{a,b,c}$, J.~Raufeisen$^{d}$,
A.V.~Tarasov$^{a,b,c}$, and M.B.~Johnson$^d$}\\ 

\bigskip

{\sl $^a$Max-Planck Institut f\"ur Kernphysik, Postfach 103980,
69029 Heidelberg, Germany}\\

{\sl $^b$Institut f\"ur Theoretische Physik der Universit\"at, 93040
Regensburg, Germany}\\ %\vspace{0.3cm}

{\sl $^c$Joint Institute for Nuclear Research, Dubna, 141980 Moscow
Region, Russia}\\

{\sl $^d$Los Alamos National Laboratory, MS H846,
Los Alamos, NM 87545, USA}

\end{center}

\vspace{1cm}

\begin{abstract} \noi We study Drell-Yan (DY) dilepton production in
proton(deuterium)-nucleus and in
nucleus-nucleus collisions within the light-cone color
dipole formalism. This approach is especially suitable for predicting
nuclear effects in the DY cross section for heavy ion collisions, as it
provides the impact parameter dependence of nuclear shadowing
and transverse momentum broadening, quantities
that are not available from the standard parton model.  For $p(D)+A$
collisions we calculate nuclear shadowing and investigate nuclear
modification of the DY transverse momentum distribution at RHIC and LHC
for kinematics corresponding to coherence length much longer than the
nuclear size.  Calculations are performed separately for transversely and
longitudinally polarized DY photons, and predictions are presented for the
dilepton angular distribution. Furthermore, we calculate nuclear
broadening of the mean transverse momentum squared of DY dileptons as
function of the nuclear mass number and energy. We also predict nuclear
effects for the cross section of the DY process in heavy ion collisions.  
We found a substantial nuclear shadowing for valence quarks, stronger
than for the sea.
\medskip


\end{abstract}

\clearpage

\section{Introduction}

The cross section for the Drell-Yan (DY) process at the energies of the
SPS suggests rather weak nuclear effects, if any at all (although
measurements at small Feynman $x_F$ are the only data available).  
However, the fixed target experiment E772 at Fermilab at $800$ GeV
\cite{e772} shows a sizable nuclear suppression at large $x_F$. Although
this suppression results from a complicated interplay between energy loss
and shadowing \cite{eloss1,eloss2}, shadowing effects are expected to be
much stronger and span the entire range of $x_F$ at the energies of 
the Relativistic Heavy Ion Collider (RHIC)
and the Large Hadron Collider (LHC).

Relying on the standard parton model for proton-nucleus collisions, one
can predict the DY cross section integrated over transverse momentum
employing QCD factorization and data for the nuclear structure functions
measured in deep-inelastic scattering (DIS)\footnote{The analysis of data
\cite{ekr} based on the DGLAP evolution equations still neglects effects
of saturation \cite{glr} that should be important once shadowing sets in.
Additionally, no data for DIS on nuclei are available for small
Bjorken $x$ relevant for LHC. Existing data for DIS and DY are not 
sensitive to the amount of gluon shadowing, which had to be parameterized 
ad hoc in \cite{ekr}.}. However, no reliable means to
calculate nuclear effects in the transverse momentum distribution within
the parton model is available. Therefore,
one has to parameterize nuclear broadening 
according to Brownian motion in the plane of transverse momentum and fit 
to data (see \cite{report} for a review). Perhaps, this situation will
be improved by higher-twist factorization theorems \cite{luo} in the future.

Moreover, parton model predictions are doubtful even for the integrated DY 
cross section in nucleus-nucleus collisions.  Indeed, compared to $pA$
collisions, this case requires knowledge of the impact parameter dependence of
nuclear shadowing and, of course, the introduction of an additional 
integration over the
impact parameter [see Eq.~(\ref{ab-x-sect})]. Neither DIS nor the DY 
reaction on nuclei provides the information needed for this. %\footnote
{In principle one can access such information relying, for instance, on 
knowledge of the number of so-called gray tracks and using simple cascade 
models.  However, this possibility has never been realized either for the DIS 
or the DY process}. In view of this problem, it might be tempting to assume, 
as in a recent analysis \cite{esk-01}, that nuclear shadowing is independent of 
impact parameter. Clearly, this cannot be correct and can lead to 
unphysical results such as a DY cross section in heavy ion collisions 
independent of centrality. It is known, however, for many processes, that peripheral
collisions are similar to the free $NN$ interaction, while central
collisions should manifest the strongest nuclear effects.

\begin{figure}[ht]
  \centerline{\scalebox{0.8}{\includegraphics{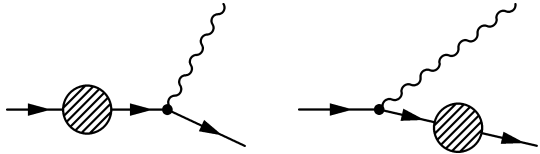}}}
    \center{\parbox[b]{13cm}{\caption{
      \label{fig:dy}\em
      In the target rest frame,
      DY dilepton production looks like bremsstrahlung. A quark 
      or an anti-quark from the
      projectile hadron scatters off the target color field 
	(denoted by the shaded circles)
	and radiates a
      massive photon, which subsequently decays into the lepton pair.
	The photon decay is not shown.
	The photon can be radiated before or after the quark scatters.}  
    }  }
\end{figure}

In this paper we calculate nuclear shadowing for the DY cross section
using the light-cone (LC) dipole approach suggested in \cite{boris}.   
This approach formulates the DY reaction in the rest frame of the target, 
which allows one to overcome these problems in a simple way by taking advantage
of the description of the DY process as bremsstrahlung of a heavy photon
from a beam quark that subsequently decays into the lepton pair as in 
fig.~\ref{fig:dy}.  Although this looks very different from the more familiar 
DY mechanism \cite{dy}, which interprets the DY reaction as quark-antiquark 
annihilation, it is known that the space-time interpretation of high-energy 
reactions is not Lorentz invariant and depends on the frame of reference.  
Indeed, it is only in a fast-moving frame where Feynman's picture of 
the colliding particles as bundles of non-interacting partons with no 
(or small) transverse momenta is valid and the DY reaction can be formulated 
consistently in terms of the parton density of the proton (or nucleus).

In the target rest frame, a quark of the incident hadron 
fluctuates into a state that contains a massive photon (dilepton) and a quark.  
Interaction with the target
breaks down the coherence of the fluctuation and the $\gamma^*$ is freed.
Correspondingly, the cross section of the process $qp\to \gamma^*\,X$ has
a factorized form \cite{boris,bhq,kst1,krt3,krt4},
 \beq\label{eq:dylctotal}
\frac{d\sigma(qp\to\gamma^*X)}{d\ln\alpha} =
\int d^2\rho\,\left|\Psi_{\gamma^*q}(\alpha,\rho)\right|^2\,
\sigma^N_{\bar qq}(\alpha\rho,x_2)\ .
\label{eq:10}
 \eeq 
where $\Psi_{\gamma^*q}(\alpha,\rho)$ is the LC distribution amplitude
in Eqs.~(\ref{eq:dylctT}) or (\ref{eq:dylctL}) for having a quark-photon
(transversely or longitudinally polarized) fluctuation with transverse
separation $\vec\rho$ and relative fractions $\alpha$ and $1-\alpha$ of
light-cone momenta carried by the photon and quark, respectively.  For
the DY reaction in $pp$-scattering, the dipole cross section needed in
(\ref{eq:10}) is the same $\sigma^N_{q\bar q}$ as in DIS off a proton
\footnote{This can be proven in leading-log order or otherwise
justified by referring to QCD factorization.}. Note, that in the two
graphs for bremsstrahlung, fig.\ \ref{fig:dy}, the quark scatters at
different impact parameters, depending on whether it scatters when in the
$|\gamma^*q\ra$-state (right) or not (left).  This leads to the
appearance of the dipole cross section $\sigma^N_{q\bar
q}(\alpha\rho,x_2)$ in (\ref{eq:10}), although there is actually no
physical dipole in this process \cite{boris,krt3}. The light-cone
Fock-state formulation of QCD is discussed in detail in the recent
review \cite{bpp97}.

We use standard kinematical variables,
 \beq\label{eq:x1x2}
x_1=\frac{2P_2\cdot q}{s}\quad,\quad x_2=\frac{2P_1\cdot q}{s},
 \eeq
with $x_1-x_2=x_F$ and $x_1x_2=(M^2+q_{T}^2)/s$, where $P_1$, $P_2$
and $q$ are the four-momenta of the beam, target and the virtual photon,
respectively; $M^2=q^2$ and $\vec q_T$ are the dilepton invariant mass 
squared and transverse momentum, respectively; $s=(P_1+P_2)^2$.  

The frame dependence of the space-time interpretation of the DY process
can be illustrated by the different meanings of $x_1$ in different
reference frames: it is well known that in the standard picture of DY
\cite{dy}, $x_1$ is the momentum fraction of the projectile quark
annihilating with the target antiquark. However, evaluating the scalar
product (\ref{eq:x1x2}) in the target rest frame shows that the
projectile quark carries momentum fraction $x=x_1/\alpha > x_1$ of its
parent hadron and, correspondingly, $x_1$ is the momentum fraction of the
proton taken away by the photon.  This is not a contradiction, since the
projectile quarks in the two reference frames are different particles.

In the case of a $pA$-scattering, one has to distinguish between two
limiting kinematical regimes for the DY reaction.  On the one hand, there
is the regime of short coherence time $t_c$, which can be interpreted as
the mean fluctuation lifetime. If $t_c$ is much shorter than the mean
internucleon separation no effect of coherence (shadowing) is expected.  
On the other hand, in the regime of long coherence time compared to the
nuclear radius, $t_c\gg R_A$, interference of the multiple interaction
amplitudes with bound nucleons will affect the probability of breaking
down the coherence of the fluctuation and releasing the dilepton on mass
shell, i.e. shadowing (sometimes antishadowing) occurs. These
interferences are controlled by the longitudinal momentum transfer
$q_c=1/t_c$ in the process of $\gamma^*$ radiation by a projectile quark
of energy $E_q$, $q\,N \to \gamma^*\,q\,X$,
 \beq
q_c=\frac{M^2_{\gamma^*q}-m_q^2}{2\,E_q}\ ,
\label{qc}
 \eeq
where we assume energy conservation, and the invariant mass squared of 
the $\gamma^*q$ pair is
 \beq
M^2_{\gamma^*q}=\frac{M^2}{\alpha} + 
\frac{m_q^2}{1-\alpha} +
\frac{q_T^2}{\alpha(1-\alpha)}\ .
\label{m-eff}
 \eeq
% Here $\alpha$ is the fraction of the LC momentum of the parent quark
%carried by the photon. 

One arrives at a similar estimate with help of the uncertainty principle.
Indeed, a quark can violate energy conservation by fluctuating into
$\gamma^*q$ for a time $\Delta t\sim 1/(M_{\gamma^*q}-m_q)$ in the quark's
rest frame. Applying the Lorentz gamma-factor $\gamma\sim
2E_q/(M_{\gamma^*q}+m_q)$ one reproduces the lifetime $t_c=1/q_c$ in the lab
frame as given by (\ref{qc}).

The intuitive space-time pattern related to the coherence time for DY
pair production off nuclei is rather obvious. In the limit of short
coherence time (relevant for the SPS energy) the initial state
interactions are predominantly soft, since the hard fluctuation
containing the heavy dilepton appears only deep inside the nucleus and is
immediately freed on mass shell via the interaction with a bound nucleon.

On the other hand, if the coherence length substantially exceeds the size
of the nucleus (as expected for the energies of RHIC for $x_F$ greater than
about 0.5 and the LHC for essentially all $x_F$), the
hard fluctuation is created long in advance of the interaction with the
nucleus, which acts as a whole in freeing the fluctuation. Since different
target nucleons compete with each other, the DY cross section is subject
to shadowing.  

In this paper, we study nuclear effects in the limit of long coherence
time. This is a most interesting regime, where interference effects are
maximal and all the nucleons having the same impact parameter participate
coherently in the DY process.  A special advantage of the color-dipole 
approach is that it allows one to incorporate nuclear shadowing via a
simple eikonalization of the dipole cross section $\sigma^N_{q\bar q}$
\cite{zkl,boris} (see next section).  This follows from the fact that in this 
limit the dipole size is ``frozen'' by Lorentz time dilation.

A projectile quark can develop more complicated fluctuations that involve
gluons in addition to the heavy photon.  These correspond to Fock
states $|q\gamma^*G\ra$, $|q\gamma^*2G\ra$, etc. Interaction of these
fluctuations with the nucleus is also affected by shadowing, which may be
even stronger than for $|q\gamma^*\ra$ provided that the
fluctuation lifetime is long compared with nuclear size. This additional
shadowing is, in the parton model, related to the shadowing of gluons
resulting from gluon fusion at small $x_2$ (see
Sect.~\ref{sec:glueshadow}).

One might think that in the case of $pA$ collisions shadowing for DY can
be easily predicted relying on QCD factorization and using data for
shadowing in DIS off nuclei. However, data at small $x$ are available
only at very low $Q^2$, where neither factorization nor DGLAP evolution
are expected to be valid.  Additionally, one should be cautious applying
factorization at large $x_1$ where, as we pointed out above, higher twist
corrections are rather large \cite{eloss2}.  In particular, the Bjorken $x$ of 
the target,
$x_2$, reaches its minimal value as $x_1\to 1$; therefore, factorization
predicts the maximal strength for shadowing. However, shadowing for the DY
process vanishes in this limit. Indeed, since $\alpha > x_1$, the
invariant mass Eq.(\ref{m-eff}) increases (for massive quarks and/or
nonzero $q_T$), leading to the disappearance of the coherence length.

We do not make a fit to the observed shadowing in DIS on nuclei.  We rather
follow the logic of the conventional Glauber approach, where one is
permitted to make fits to the data for nucleon-nucleon collisions but
then predict nuclear effects in a parameter-free way. Indeed,
we use the phenomenological dipole cross section on a nucleon target
\cite{Wuesthoff1} fitted to the data for $ep$ DIS from HERA that cover 
a range of
energies much higher than that available from fixed-target nuclear shadowing
data.  The DY cross section calculated with this phenomenological
cross section is supposed to include all higher-order corrections and
higher twist effects.

Another important advantage of our approach is the possibility of calculating
nuclear effects in the transverse momentum distribution of DY pairs.  This 
is a difficult problem within the parton model. The
phenomenon of nuclear broadening of the dilepton transverse momentum
looks very different at low (short $t_c$) and high (long $t_c$) energies.
If $t_c$ is short, a hard fluctuation containing the heavy dilepton is
created deep inside the nucleus just before the interaction that 
releases it. Meanwhile, the incident hadron may have soft initial state
interactions in the nucleus.  Although these do not generate DY dileptons, 
they do increase the mean transverse momentum of the fast partons of the
projectile. Indeed, a fast parton experiencing multiple interactions
performs something like Brownian motion in the plane of transverse momenta.
As a result, 
the parton arrives at the point of the DY pair creation with increased 
transverse momentum. Information about the enhanced transverse momentum of the
projectile quark in a nucleus compared to a proton target is carried away 
undisturbed by the dilepton pair.  Nuclear broadening in the limit of short 
coherence time was investigated previously in \cite{jkt}.

At first glance, the observed broadening of the dilepton transverse momentum 
in the parton model might be viewed as arising from an increased transverse 
momentum of quarks and antiquarks in the nucleus.  However, such a conclusion 
contradicts the usual picture of a nucleus boosted to the infinite momentum 
frame.  There, nucleons and their parton clouds are well separated and do not 
overlap at large $x_2$, just as in the nuclear rest frame.  We know that only 
at very small $x_2$ do the parton clouds overlap and fuse leading to nuclear
shadowing. Such a fusion process results not only in suppressing the parton
density, but it also increases the transverse momenta of the partons. Thus,
shadowing and $q_T$ broadening at small $x_2$ are closely related processes;
no broadening is possible without shadowing.  However, the regime of short 
$t_c$ corresponds to large $x_2$, where neither shadowing, nor $q_T$ 
broadening is expected for the nuclear parton distribution function. Thus, we 
face a puzzle:  nuclear broadening of the transverse momentum distribution of 
DY pairs calculated in the nuclear rest frame and observed experimentally has 
no analog within the parton model.  This puzzle was resolved a long time ago 
by Bodwin, Brodsky and Lepage \cite{bbl}, who found that in the regime of short
coherence length initial state interactions leading to $q_T$ broadening 
violate QCD factorization and should not be translated into a modification of 
the nuclear quark distribution function. Likewise, initial state energy
loss \cite{eloss1,eloss2} cannot be translated to a modification of
$x$-distribution of partons in nuclei. This is a general statement which
is applied to other hard reactions like high-$q_T$ hadron production,
etc.

In the regime of long coherence time $t_c\gg R_A$ relevant to RHIC and
LHC, a very different mechanism is responsible for broadening of the
transverse momentum distribution \cite{kst1}. A high-energy projectile
quark emits a dilepton fluctuation (via a virtual time-like photon) long
before its interaction with the nucleus. These components of the
fluctuation, the recoil quark and the dilepton, do not ``talk'' to each
other because of Lorentz time dilation.  Therefore, multiple interactions
of the quark in nuclear matter seem to have no further influence on the
produced dilepton and no broadening of the transverse momentum is
expected. However, this conclusion is not correct. While it is true that
different ingredients of the fluctuation cannot communicate, not all
fluctuations contribute to DY pair production in the same way: 
The harder the fluctuation, that is the larger the
intrinsic relative transverse momentum between the quark and dilepton,
the stronger the kick from the target required for loss of coherence,
i.e. for the fluctuation to be produced on mass shell.
Large $\gamma^*q$-fluctuations (which have a low intrinsic transverse 
momentum) have a high probability to lose coherence in the first scattering
on the surface of the nucleus. Therefore, 
the contribution to the DY cross section from these configurations scales
as $A^{2/3}$, resulting in a nuclear suppression of the DY cross section at low
$q_T$. 
Due to multiple interaction of the projectile quark a nucleus provides a 
larger transverse momentum than a nucleon target and hence is able to 
break up smaller fluctuations than a nucleon.
This enhances the DY cross section
at intermediate $q_T$.
At high $q_T$ multiple scatterings are not important any more, just a 
single high $q_T$ scattering dominates, eliminating the nuclear 
enhancement. This is a result of the power $q_T$ dependence of the cross 
section predicted by QCD.
Since small $q_T$ dileptons are suppressed while large $q_T$ dileptons are not,
the mean transverse momentum of DY pairs is increased. Thus, at long $t_c$,
the mechanism behind nuclear broadening is color filtering \cite{bbgg}.

This paper is organized as follows. We explain the main ideas of the
light-cone approach in the introduction. The key ingredient of this
method, the universal color dipole cross section for a $q\bar q$ pair
interacting with a nucleon, is known from phenomenology.
In Sect.~\ref{sec:sigmaa}, we explain how nuclear effects are treated in
the color dipole approach. In particular, we describe, how nuclear gluon
shadowing has to be included along with the $q\bar q$-nucleus cross section. 
The
results for gluon shadowing are presented in Sect.~\ref{sec:glueshadow}.
The results of our calculations for the DY cross section in $p(D)+A$ collisions,
and predictions for RHIC and LHC, can be found in Sect.~\ref{sec:shadow}. 
Nuclear modification of the DY pair transverse momentum 
is calculated for the energies of RHIC and LHC in
Sect.~\ref{sec:broad}. We find that the so called Cronin effect, nuclear
enhancement of the DY cross section at medium-large $q_T$, is nearly
compensated at RHIC energies but is expected to have a large magnitude
at LHC. We conclude that nuclear broadening of the DY transverse momentum
squared diverges logarithmically for transversely polarized photons if
nuclear shadowing occurs. Differences in nuclear effects for radiation of
longitudinally and transversely polarized photons lead to a specific nuclear
modification of the DY polarization.  Corresponding predictions are
presented in Sect.~\ref{sec:polarization}.  In addition, we find
substantial deviation from the so called Lam-Tung relation \cite{LT1}. 
Indeed, this
relation is not supported by data, which is difficult to explain within the
standard parton approach.  In Sect.~\ref{sec:aa} we address the more
difficult problem of nuclear effects in heavy ion collisions. We follow 
conventional wisdom and simplify the problem by employing QCD factorization. 
Nuclear shadowing for 
both sea and valence quarks is calculated within the LC dipole 
approach. 
In contrast to usual expectations,
shadowing for valence quarks turns out to be larger than for the sea.  
We summarize the
results and observations of this paper and present an outlook for further 
development and application of the LC dipole
approach in Sect.~\ref{sec:outlook}.

\section{The \boldmath$q\bar q$-nucleus cross section}\label{sec:sigmaa}

In order to calculate the DY cross section in $pA$ scattering, one has to
replace $\sigma^N_{q\bar q}$ in Eq.~(\ref{eq:10}) by the color dipole cross
section on a nucleus, $\sigma_{q\bar q}^A(\rho,x)$, which is easy to
calculate within the color dipole approach. 

In the limit of long coherence time, the projectile quark may be decomposed
into a series of Fock-states with frozen transverse separations.  Since
partonic configurations with fixed transverse separations in impact
parameter space are interaction eigenstates \cite{zkl}, $\sigma_{q\bar
q}^A(\rho,x)$ may be calculated using Glauber theory \cite{Glauber}, i.e.
via simple eikonalization of the $q\bar q$-nucleon cross section,
 \beq\label{eq:signuc}
\widetilde\sigma_{q\bar q}^A(\rho,x)=
2\int d^2b\left[1-\left(1-\frac{1}{2A}\,\sigma^N_{q\bar q}(\rho,x)
\,T_A(b)\right)^A\right]\,,
 \eeq
where $b$ is the impact parameter, $A$ is the nuclear mass number and
 \beq
T_A(b)=\int^\infty_{-\infty} dz \rho_A(b,z) 
 \eeq
is the nuclear thickness, i.e. the integral over the nuclear density. 
The possibility of eikonalization of a color dipole propagating through
a nucleus suggested in \cite{zkl} 
has been more rigorously justified recently in 
\cite{jkt,km98,Brodsky,Urs}.
We mark $\sigma_{q\bar q}^A$ in Eq.~(\ref{eq:signuc}) with a tilde to indicate
that it misses important contributions from higher Fock components, which 
motivates us to introduce an improved eikonal 
formula, Eq.~(\ref{eq:sigmanuc}), below.
Note that at larger $x_2\gsim 0.01$, when the coherence length becomes
shorter than the nuclear radius, the eikonal approximation breaks down
and one has to employ the Green function technique developed in \cite{kst1}.

The single-scattering term can be obtained by expanding (\ref{eq:signuc}) 
to first order in $\sigma^N_{q\bar q}(\rho,x)$. The dipole interacts with
the target by exchange of a gluonic colorless system, the so called
Pomeron. The unitarity cut of such an amplitude reveals multiple gluon
radiation that is related to higher Fock states within the LC approach in
the target rest frame.  Thus, for single scattering, $\sigma^N_{q\bar q}$
takes all Fock states of the projectile parton into account, not only
$|q\bar q\ra$. The energy dependence of the dipole cross section is
generated by the phase space of gluons from higher Fock states $|q\bar
qG\ra$, $|q\bar qGG\ra$, $\dots$.  Indeed, in the Born approximation, i.e. 
two gluon exchange, $\sigma^N_{q\bar q}$ would be independent of $x$. 

Calculation of $\sigma^N_{q\bar q}$ from first principles is still a
challenge.  We rely on phenomenology and employ the parameterization of
Golec-Biernat and W\"usthoff \cite{Wuesthoff1} motivated by the saturation
model,
 \beq\label{eq:wuestsigma}
\sigma^N_{q\bar q}(\rho,x)=
\sigma_0\left[1-\exp\left(-\frac{\rho^2Q_0^2}{4(x/x_0)^\lambda}\right)\right],
 \eeq
where $Q_0=1$ GeV and the three fitted parameters are $\sigma_0=23.03$ mb,
$x_0=0.0003$, and $\lambda=0.288$.  This dipole cross section vanishes
$\propto\rho^2$ at small distances, as implied by color transparency, and
levels off exponentially at large separations.
The authors of \cite{Wuesthoff1} are able to fit all
available HERA data with a quite low $\chi^2$ and can also describe
diffractive HERA data. 

\begin{figure}[t]
\centerline{
  \scalebox{0.6}{\includegraphics{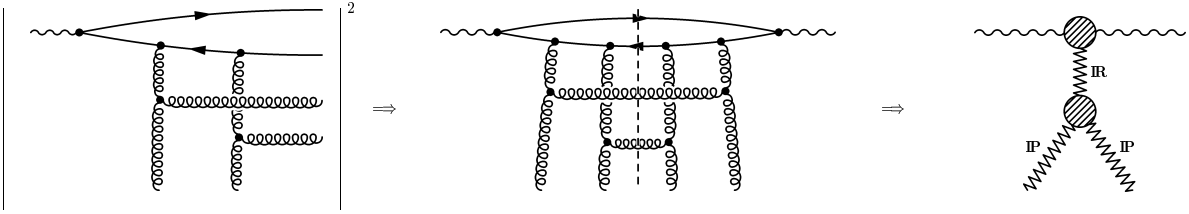}}
 }
\center{\parbox[b]{13cm}{
\caption{\label{fig:ppr}\em
The eikonal formula (\ref{eq:signuc}) takes only multiple rescatterings of
the $|q\bar q\ra$-Fock component into account. This figure illustrates
the amplitude for double scattering (left). When the amplitude is
squared (middle), the gluon rungs combine into gluon ladders (Pomerons), which
are enclosed into each other. In Regge theory, this contribution to the cross
section is expressed in terms of
the Pomeron-Pomeron-Reggeon vertex (right). }}}
\end{figure}

We now turn to the multiple scattering terms.  Describing shadowing for DY
is simplified if we make use of the fact that the dipole cross section 
entering the formula for dilepton production, Eq.~(\ref{eq:dylctotal}), is the 
same as that needed to calculate the DIS cross section.  
We may thus illustrate the physics of Eq.~(\ref{eq:signuc}) in
fig.\ \ref{fig:ppr}, where the double-scattering term for a $q\bar q$-dipole is
depicted. In terms of Regge phenomenology, the double scattering of
the $q\bar q$-pair corresponds to the Pomeron-Pomeron-Reggeon vertex.
Note that (\ref{eq:signuc}) accounts for not only the double-scattering
term, but also for all higher-order rescatterings of the $q\bar q$-pair.
The $n$-fold scattering graph has $n$ gluon ladders, which are enclosed
into each other. 

\begin{figure}[t]
\centerline{
  \scalebox{0.6}{\includegraphics{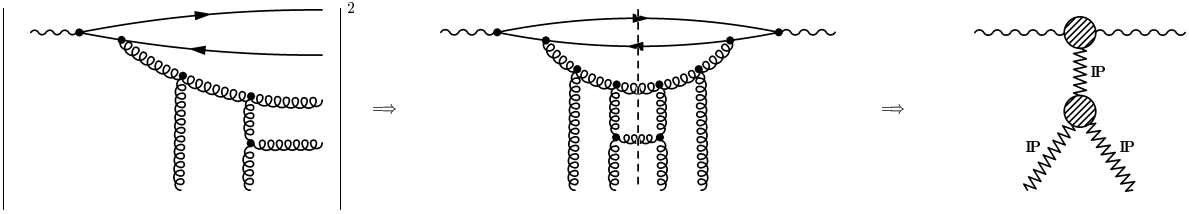}}
 }
\center{\parbox[b]{13cm}{
\caption{\label{fig:ppp}\em
At high energy, the lifetime of higher Fock states becomes long enough for
multiple scattering. Shown here is the double-scattering amplitude
for the $|q\bar qG\ra$-Fock state. In Regge theory, this process is expressed
in terms of the triple-Pomeron vertex (right). The eikonal formula 
(\ref{eq:signuc}) is improved by multiplying $\sigma_{q\bar q}$ 
by the gluon shadowing ratio $R_G$, Eq.~(\ref{eq:sigmanuc}), to include 
these contributions also. 
 }}}
\end{figure}

Rescatterings of higher Fock states, containing gluons are omitted in
(\ref{eq:signuc}). This equation corresponds to gluon radiation in the
Bethe-Heitler regime, where each interaction of the quarks leads to
independent radiation of the entire spectrum of gluons. However,
interferences known as the Landau-Pomeranchuk effect, lead to a suppression
of gluon radiation. This shadowing correction is taken into account
via multiple interactions of radiated gluons in the nucleus
\cite{mueller} and leads to an additional suppression of the DY cross
section. At low $x$ the lifetime of these higher Fock states becomes
significantly longer than the mean internucleon distance, and they can
scatter more than once inside the nucleus, as illustrated in fig.\
\ref{fig:ppp}.  In this case, which occurs at RHIC and LHC energies,
(\ref{eq:signuc}) needs to be modified to include also these
rescattering.

In order to include processes like the one illustrated in fig.\ \ref{fig:ppp},
it is useful to note that the rescattered gluon can be interpreted as the 
first rung 
of a single gluon ladder exchanged between the $q\bar q$-pair and the target.
In Regge phenomenology, rescattering of gluons leads to the triple-Pomeron
vertex
fig.\ \ref{fig:ppp}(right), which can be regarded as  a correction to
the single-scattering term. More precisely, it leads to a 
reduction of the nuclear
gluon density $G_A$ because the two Pomerons from the target in 
fig.\ \ref{fig:ppp}(right)
fuse into a single one before interacting with the pair.
Indeed, multiple 
scatterings of higher Fock states containing gluons are known as the effect of 
gluon shadowing \cite{mueller} and lead to an additional suppression of the DY 
cross section.  In the infinite momentum frame of the nucleus, gluon clouds of 
different nucleons overlap and fuse at small $x$ 
\cite{kancheli,glr}, fig.\ \ref{fig:ppp}(right), thereby
reducing  the gluon density at small $x$. Although the corrections 
for gluon rescatterings the nuclear rest frame and gluon fusion 
(in the nuclear infinite momentum frame) look very different, this is the 
same
phenomenon seen from different reference frames. Of course, observables
are Lorentz invariant, and both effects lead to a reduction of the DY
cross section.

Thus, the Pomeron-Pomeron fusion process in fig.~\ref{fig:ppp}(right) can
be taken into account by multiplying $\sigma^N_{q\bar q}(\rho,x)$ by the 
gluon shadowing ratio
\beq\label{eq:deltarg}
R_G(x,\widetilde Q^2)=\frac{G_A(x,\widetilde Q^2)}{AG_N(x,\widetilde Q^2)}
\equiv 1-\Delta R_G(x,\widetilde Q^2),
 \eeq
as calculated in Appendix A.  
The single-scattering term then reads 
 \beqn\label{eq:single}
\sigma^A_{q\bar q}(x,\rho)&=&\sigma^N_{q\bar q}(\rho,x)\int d^2b 
R_G(x,\widetilde Q^2,b)
T_A(b)+\dots,\nonumber\\
\label{eq:single2}
&=&A\sigma^N_{q\bar q}(\rho,x)
\left[1-{1\over A}\int d^2b 
\Delta R_G(x,\widetilde Q^2,b)
T_A(b)\right]+\dots,
 \eeqn
where gluon shadowing as a function of impact parameter $b$ is given in
Eqs.~(\ref{A.20}) and (\ref{A.21}).

The first term in the brackets of 
(\ref{eq:single2}) stands for the direct exchange of a
Pomeron, while the second (negative) term represents the correction due to
the Pomeron fusion process depicted in fig.~\ref{fig:ppp}.  This
recipe becomes even more clear from the relation, valid at $\rho \to 0$,
\cite{FS},
 \beq\label{eq:fs}
\sigma_{q\bar q}^N(\rho,x)\Bigr|_{\rho\to0}
=\frac{\pi^2}{3}\alpha_s\!
\left(\frac{\lambda}{\rho^2}\right)
\rho^2\,G_N\left(x,\frac{\lambda}{\rho^2}\right).
 \eeq
In (\ref{eq:single}), the gluon density of the proton in (\ref{eq:fs}) was
replaced by the average gluon density of the nucleus,
 \beq\label{eq:fs1}
\sigma^N_{q\bar q}(\rho,x)\Bigr|_{\rho\to0}
\Rightarrow\frac{\pi^2}{3}\alpha_s\!
\left(\frac{\lambda}{\rho^2}\right)
\rho^2\,{1\over A}\,
G_A\left(x,\frac{\lambda}{\rho^2}\right)\,.
 \eeq
It is clear from (\ref{eq:single}) that the effective dipole cross section
on a bound nucleon appears to be reduced due to gluon shadowing.  We also
see from (\ref{eq:fs})  that $R_G$ has to be evaluated at a scale
$\widetilde Q^2=\lambda/\rho^2$. 

The specific fusion processes that are included in (\ref{eq:single}) 
depends on the approximation in which
gluon shadowing $R_G$ is evaluated.  For our actual calculations (see
Sect.~\ref{sec:glueshadow}) nuclear shadowing for gluons is calculated
within the Green function formalism for a $|q\bar qG\ra$ fluctuation
propagating through nuclear medium as developed in \cite{kst2}. This means,
that the single-scattering term in (\ref{eq:single}) is corrected not only
for the $2\pom \to \pom$ Pomeron fusion term depicted in 
fig.~\ref{fig:ppp}(right), but also for all the $n\pom \to \pom$ 
fusion processes.
Moreover, the Green function approach properly describes the
finite lifetime of the $|q\bar qG\ra$-state. This is important, because
even when the $q\bar q$-fluctuation lives much longer than the nuclear
radius, the lifetime of the $|q\bar qG\ra$-state will be shorter.

For the rescattering terms, we can account for higher Fock states in the
same way as in (\ref{eq:single}), namely by the replacement \cite{agl,kth,knst}
 \beq\label{eq:recipe}
\sigma^N_{q\bar q}(\rho,x)\Rightarrow
\sigma^N_{q\bar q}(\rho,x)R_G(x,\widetilde 
Q^2,b)\ ,
 \eeq
i.e. the improved formula Eq.~(\ref{eq:signuc}) for the $q\bar q$-nucleus
section reads,
 \beq\label{eq:sigmanuc}
\sigma_{q\bar q}^A(\rho,x)=
2\int d^2b\left[1-\left(1-\frac{1}{2A}\,\sigma^N_{q\bar q}(\rho,x)
R_G(x,\lambda/\rho^2,b)
T_A(b)\right)^A\right]\,.
 \eeq
This expression includes also the contribution of higher Fock states
containing more than one gluon. The higher-order multiple interactions of the
$|q\bar qG\ra$ Fock state correspond, as was mentioned, to multi-Pomeron
fusion, $n\pom \to \pom$, while the Reggeon diagrams with $n\pom \to m\pom$
($m\geq2$) are missing.  Those diagrams should be incorporated via the Fock
components $|q\bar q\,mG\ra$ containing two or more gluons.  The modified
expression Eq.~(\ref{eq:sigmanuc}) sums multiple interactions of the
$q\bar q$ pair via $m\pom$ exchange (summed over $m$), each of which has the
form of a fan $n\pom \to \pom$ (summed over $n$).  We assume that each
gluon in the Fock state $|q\bar q\,mG\ra$ experiences multiple interactions
independently of other gluons. This assumption corresponds to the Gribov's
interpretation \cite{gribov} of Glauber eikonal shadowing; namely, the
unitarity cut of an $n$-fold scattering term must contain a simultaneous
cut of all $n$ Pomerons. It therefore corresponds to a nonplanar graph
describing the independent multiple interactions of $n$ projectile partons (see
discussion in \cite{kth}). 

Note that the GLR equation \cite{glr} can be obtained from 
Eq.~(\ref{eq:sigmanuc}) by expanding the r.h.s.\ up to quadratic order
in $\sigma^N_{q\bar q}$. Though the higher order terms in this
expansion may not be found in the large $N_c$, leading 
$\log(1/x)$ limit of QCD \cite{BK}, their numerical
influence on the DY cross section will be small
because of the large dilepton mass. 

\begin{figure}[t]
\centerline{
  \scalebox{0.43}{\includegraphics{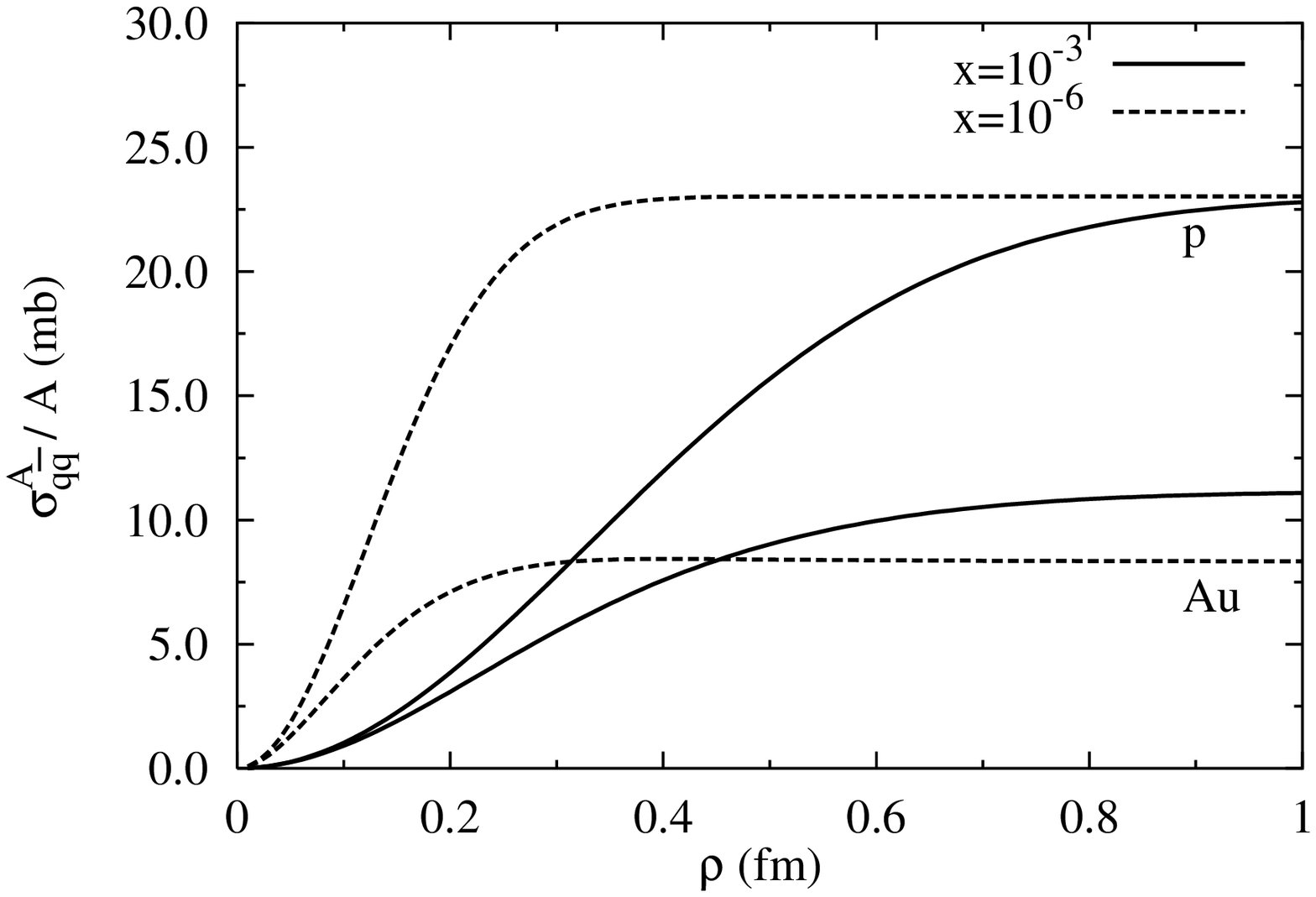}}
  \scalebox{0.43}{\includegraphics{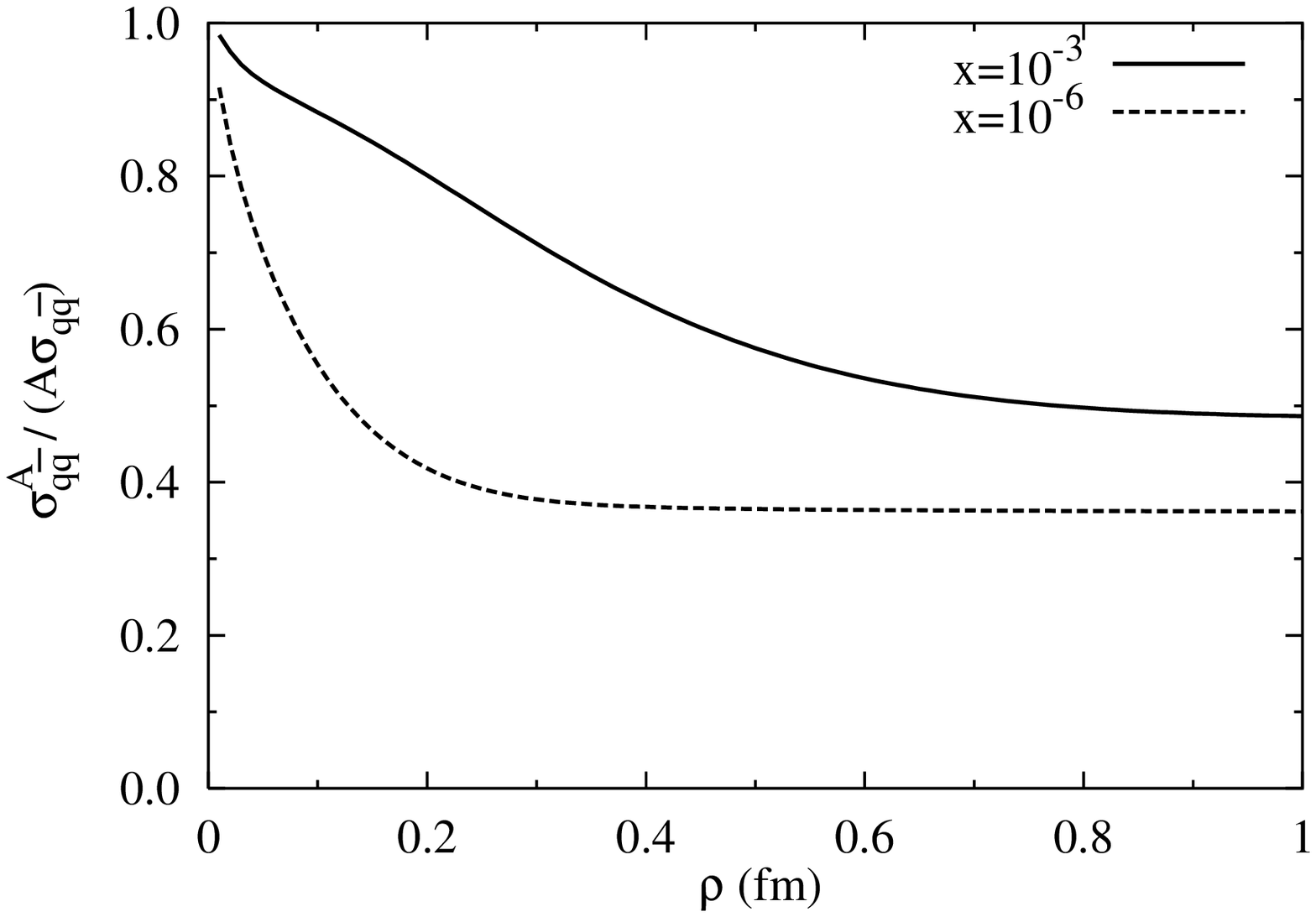}}
 }
\center{\parbox[b]{13cm}{
\caption{\label{fig:sigmaa}\em
  The figure on the left shows the $q\bar q$-nucleus cross section
(\ref{eq:sigmanuc}) divided by the nuclear mass number $A$ for two different
values of $x$. The two lower curves (solid and dashed) are calculated for
gold ($A=197$) and the two upper curves for a proton.  The figure on the
right shows the $q\bar q$-nucleus cross section divided by $A$ times the
dipole cross section (\ref{eq:wuestsigma}).  While large separations are
strongly suppressed, small size dipoles are much less affected by the
nucleus. Nuclear gluon shadowing is included in the calculation, as explained
in the text. It slowly vanishes at small $q\bar q$ separations, which correspond to
high $\widetilde Q^2$. 
 }}}
 \end{figure}

We can now proceed to calculate $\sigma^A_{q\bar q}(x,\rho)$ according to
(\ref{eq:sigmanuc}). We briefly summarize
our calculation of gluon shadowing ($R_G$) \cite{kst2}
in section \ref{sec:glueshadow}. 
The results for the nuclear dipole cross section
$\sigma^A_{q\bar q}(\rho,x)$ are depicted in fig.\ \ref{fig:sigmaa}. Since
(\ref{eq:sigmanuc}) is a high-energy approximation, valid when the lifetime of
the $q\bar q$-pair exceeds the nuclear radius, these results are relevant at
RHIC and LHC energies.  At lower energies however, one has to take
transitions between different eigenstates into account \cite{kst1}. 

The plot on the l.h.s.\ of fig.\ \ref{fig:sigmaa} shows the dipole cross
section itself. First, we discuss the $q\bar q$-proton cross section. The two
upper curves show this quantity for two different values of $x$ typical for
RHIC and LHC.  After a quadratic rise, $\sigma^N_{q\bar q}$ levels off and
assumes an energy-independent saturation value of $23.03$ mb. The onset of
saturation, i.e.\ the flattening of the dipole cross section as function of
$\rho$, is controlled by the saturation radius,
 \beq
R_s^2(x)=\frac{2}{Q_0^2}\left(\frac{x_0}{x}\right)^\lambda,
 \eeq
which decreases with energy. The energy dependence of (\ref{eq:wuestsigma})
correlates with $\rho$. At $\rho\ll R_s$, the dipole cross section grows with
energy with the hard Pomeron intercept $\lambda=0.288$, while at large
separations, $\rho\gg R_s$, (\ref{eq:wuestsigma}) becomes independent of
energy. For more discussion on (\ref{eq:wuestsigma}), refer to the original
work \cite{Wuesthoff1}. 

We now turn our attention to the $q\bar q$ nucleus cross section
$\sigma_{q\bar q}^A(\rho,x)$, Eq.\ (\ref{eq:sigmanuc}), which is shown by the
two lower curves in fig.\ \ref{fig:sigmaa} (left). 
In all calculations, we employ realistic parameterizations of nuclear 
densities from \cite{Jager}.
In addition to the
expected suppression due to nuclear shadowing, one also sees that the
saturation value of $\sigma_{q\bar q}^A$, which is almost at its
geometrical limit $2\pi R_A^2$, is energy dependent. Moreover, $\sigma_{q\bar
q}^A(\rho\to\infty,x)$ is a decreasing function of energy.  This is a
consequence of the gluon shadowing in (\ref{eq:sigmanuc}). At very small $x$,
gluon shadowing becomes strong, (see fig.\ \ref{fig:gshad}) and the $q\bar q$
nucleus cross section lies below its geometrical limit.  The stronger the
gluon shadowing, the smaller the saturation value of $\sigma_{q\bar q}^A$. 
However, $R_G\to1$ at $\rho\to0$ 
since $\widetilde Q^2\sim1/\rho^2\to\infty$ (see
fig.~\ref{fig:gshad}). 

The nuclear suppression of the dipole cross section is plotted on the 
right-hand side 
of fig.\ \ref{fig:sigmaa}.  Small sizes are less affected by the nucleus than
large sizes. This illustrates the effect of color filtering \cite{bbgg},
which is the mechanism behind nuclear broadening of transverse momenta (see
section \ref{sec:broad}).  While small $q\bar q$-pairs, which have large
intrinsic transverse momenta according to the uncertainty principle,
propagate through the nucleus almost undisturbed, large pairs (small
transverse momenta) are absorbed, i.e.\ the coherence of the fluctuation is
disturbed and the $\gamma^*$ is freed. This effect leads to an increase
of the mean transverse momentum.

\section{Gluon shadowing in nuclei}\label{sec:glueshadow}

The nuclear shadowing for gluons needed as input for the $q\bar
q$-nucleus
cross section in (\ref{eq:sigmanuc}) is calculated in the LC Green function
approach developed in \cite{kst2}, where gluon shadowing is calculated from
shadowing of the $|q\bar qG\ra$ Fock component of a longitudinally polarized
photon.  In this section, we briefly review the approach of \cite{kst2} and
present the results of our calculation for gluon shadowing as function of
$x$, $Q^2$ and the length $L$ of the path in the nuclear medium. 

Longitudinal photons can serve to measure the gluon density because they
effectively couple to color-octet-octet dipoles. This can be understood in
the following way: the light-cone wave function for the transition
$\gamma^*_L\to q\bar q$ does not allow for large, aligned jet configurations. 
Thus, unlike the transverse case, all $q\bar q$ dipoles from longitudinal
photons have size $1/Q^2$ and the double-scattering term vanishes like
$\propto 1/Q^4$. The leading-twist contribution for the shadowing of 
longitudinal
photons arises from the $|q\bar q G\ra$ Fock state of the photon. Here again,
the distance between the $q$ and the $\bar q$ is of order $1/Q^2$, but the
gluon can propagate relatively far from the $q\bar q$-pair. In addition,
after radiation of the gluon, the pair is in an octet state. Therefore, the
entire $|q\bar qG\ra$-system appears as a $GG$-dipole, and the shadowing
correction to the longitudinal cross section is just the gluon shadowing we
want to calculate.

\begin{figure}[t]
\centerline{
  \scalebox{0.8}{\includegraphics{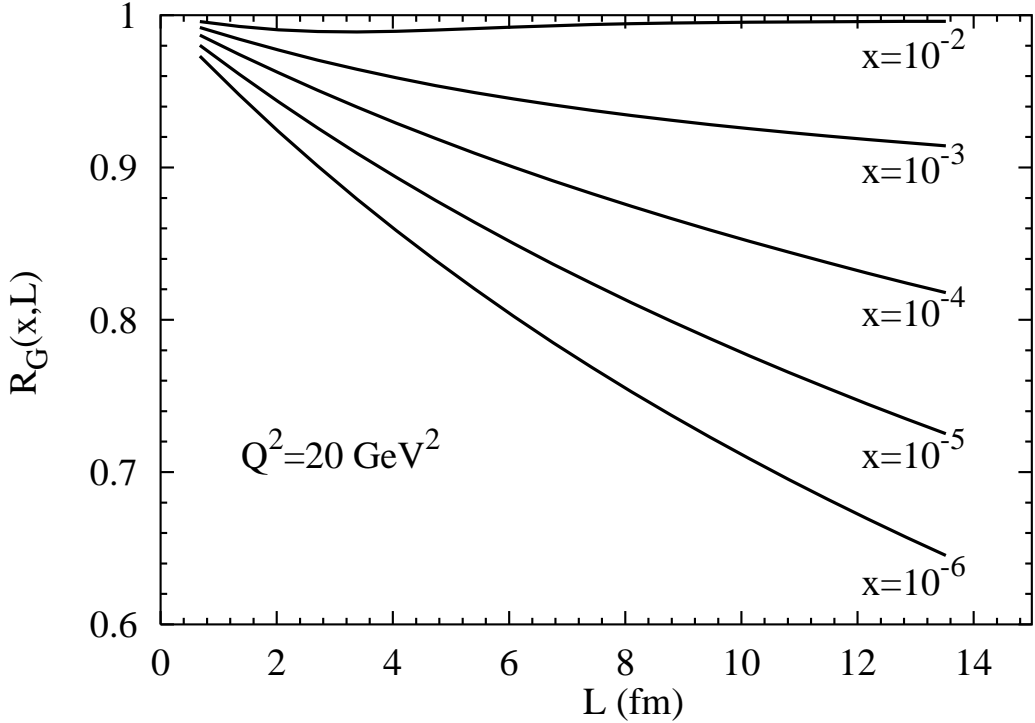}}
 }
\center{\parbox[b]{13cm}{
\caption{\label{fig:gshadvsl}\em
 Gluon shadowing vs.\ the length of the nuclear medium $L=2\sqrt{R_A^2-b^2}$,
where $b$ is the impact parameter and $R_A$ the nuclear radius. 
All curves are for $Q^2=20$ GeV$^2$ but for different values of~$x$. }}}
\end{figure}

A critical issue for determining the magnitude of gluon shadowing is the 
distance the gluon can propagate from the $q\bar q$-pair in impact parameter 
space, i.e. knowing how large the $GG$ dipole can become.  
This value could be extracted from single diffraction data in hadronic 
collisions in \cite{kst2} because these data allow the 
diffractive gluon radiation (the triple-Pomeron contribution in Regge
phenomenology) to be unambiguously singled out.
The diffraction cross section ($\propto \rho^4$) is even
more sensitive to the dipole size than the total cross section ($\propto
\rho^2$) and is therefore a sensitive probe of the mean transverse
separation. It was found in \cite{kst2} that the mean dipole size
must be of the order of $r_0=0.3\,\fm$, considerably smaller than a light 
hadron. A rather small gluon cloud of this size surrounding the valence 
quarks is the only way that is known to resolve the long-standing problem of 
the small small size of the triple-Pomeron coupling.  
The smallness of the $GG$ dipole is incorporated into the LC approach by 
a nonperturbative interaction between the gluons. 

Note that the small value of $r_0$ dictated by data for diffraction is 
consistent with
the results of other approaches. Indeed, the same small size characterizing
gluonic fluctuations was found in the instanton liquid model \cite{shuryak},
in the QCD sum rule analysis of the gluonic formfactor of the 
proton \cite{braun}, and it also follows from lattice calculations \cite{pisa}.
Note that the value of $r_0$ also limits the $Q^2$-range where the 
approximation $q\bar qG\approx GG$ is valid. One has to ensure that 
$Q^2\gg 1/r_0^2$, otherwise the $q\bar q$ pair is not pointlike compared to 
the size of the entire Fock state.

Our results for gluon shadowing as a function of the length of 
the nuclear medium at impact parameter $b$ are shown in 
fig.~\ref{fig:gshadvsl}. 
The calculations are performed for lead with a uniform nuclear density of
$\rho_A=0.16\,\fm^{-3}$.  Details are presented in {\ref{app:gshad}. The 
small size of the $GG$ dipole leads to a rather weak gluon shadowing
(except for specific reactions where the $q\bar q$ pair is 
colorless \cite{kth}).  For most
values of $x$, gluon shadowing increases as a function of $L$ as one would
expect.  At the largest value of $x=0.01$, however, gluon shadowing becomes
smaller as $L$ increases, and $R_G$ approaches $1$. Although this behavior
seems to be counterintuitive, it can be easily understood by noting that at
$x=0.01$ the coherence length of the $|q\bar qG\ra$-Fock state becomes very
small and the formfactor of the nucleus suppresses shadowing \cite{krt2}. 
The curves shown in fig.\ \ref{fig:gshadvsl} are the ones
which actually enter our calculation for DY via the $q\bar q$-nucleus cross
section (\ref{eq:sigmanuc}). The values of $x$ entering our calculation are
$x\approx 10^{-3}$ for RHIC and $x\approx 10^{-6}$ for LHC. 

We also calculate gluon shadowing as function of $x$ at fixed $Q^2$ and as
a function of $Q^2$ at fixed $x$, integrated over the impact parameter $b$. The
results are shown in fig.\ \ref{fig:gshad}.
 \begin{figure}[t]
\centerline{
  \scalebox{0.43}{\includegraphics{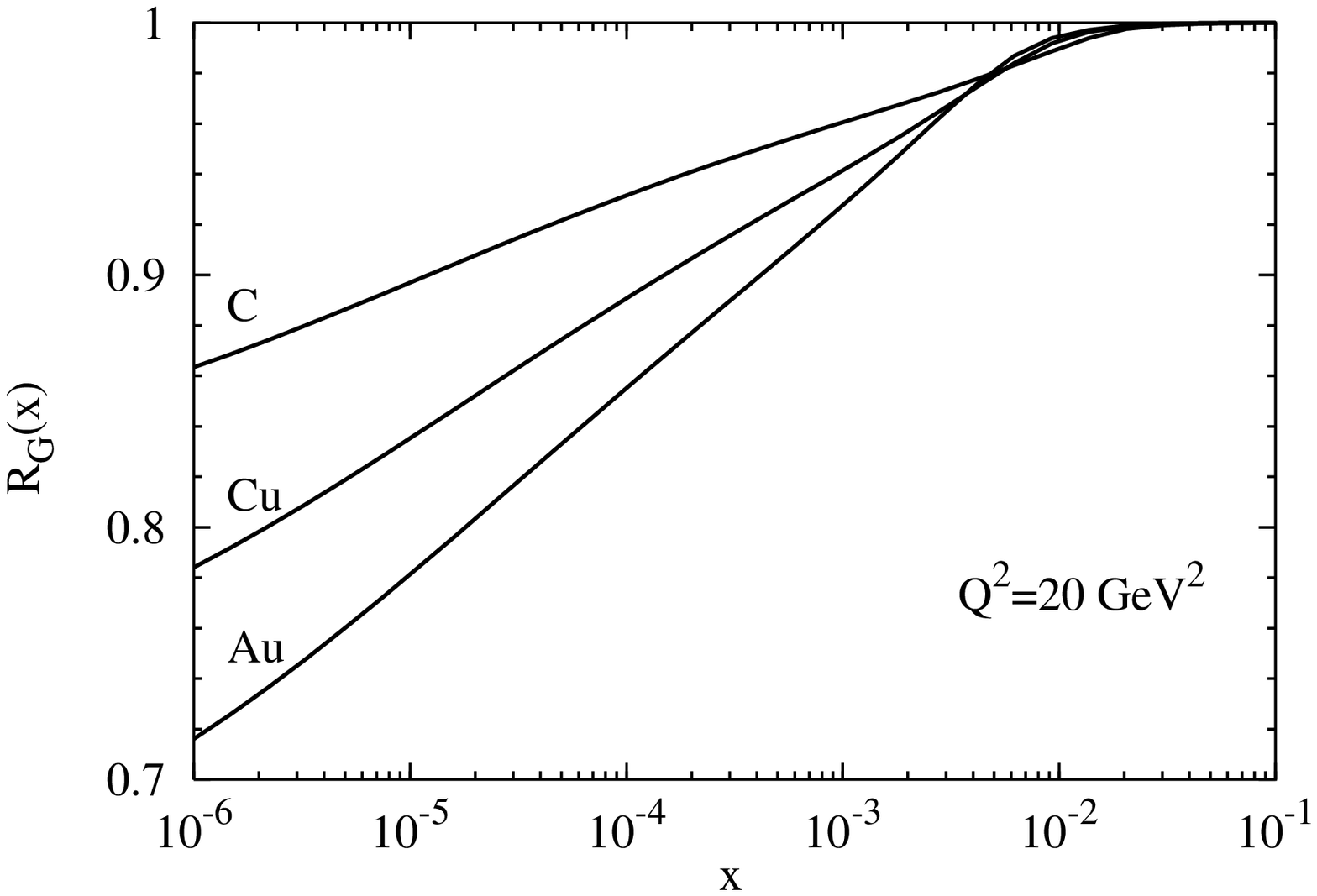}}
  \scalebox{0.43}{\includegraphics{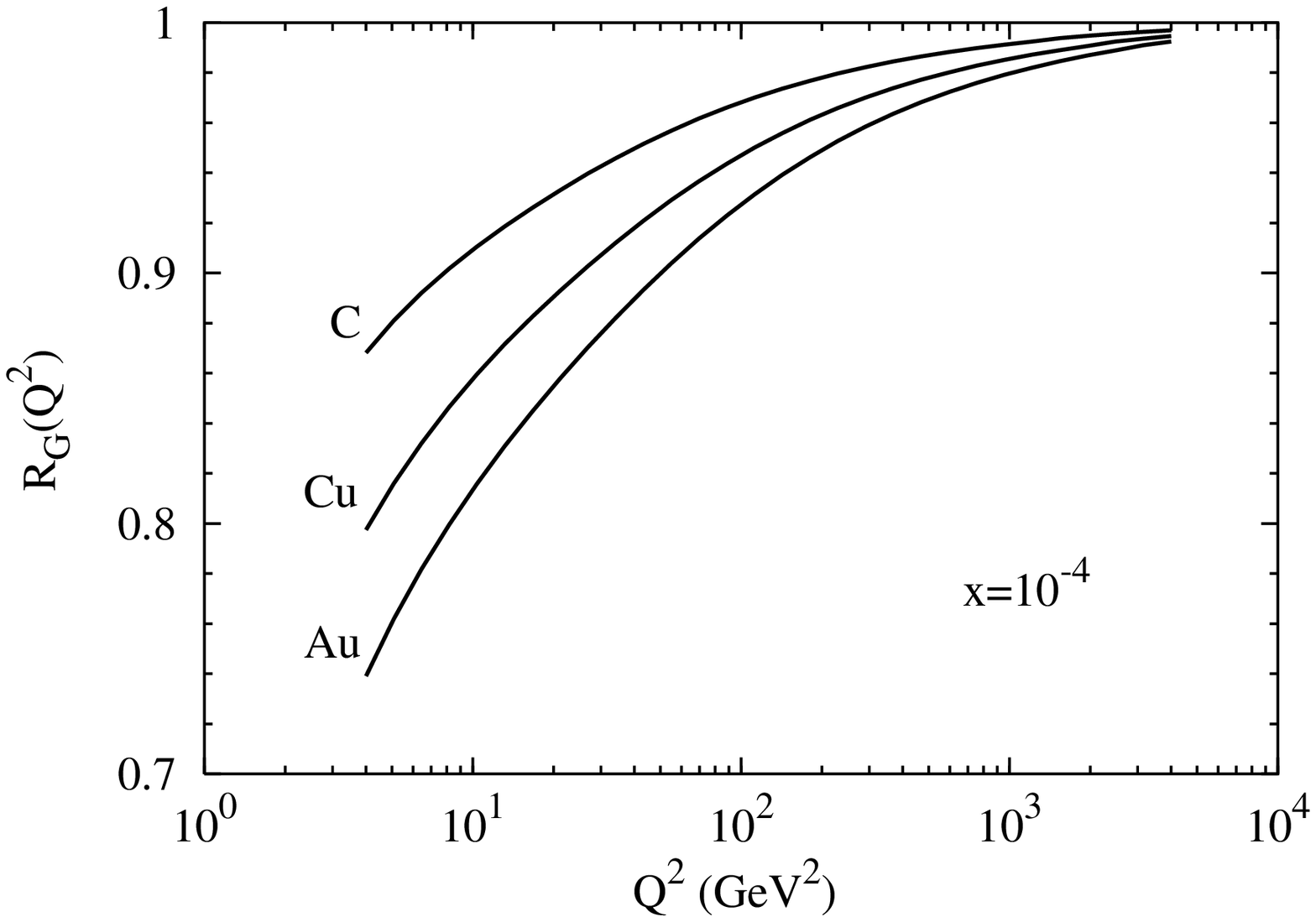}}
 }
\center{\parbox[b]{13cm}{
\caption{\label{fig:gshad}\em
The $x$- and $Q^2$-dependence of gluon shadowing for carbon, copper and
gold. The $x$-dependence is shown for
$Q^2=20$ GeV$^2$, while the figure on the right is calculated for
$x=10^{-4}$.}}}
 \end{figure}
In the left-hand 
plot, one observes that gluon shadowing vanishes for $x>0.01$. This
happens because the lifetime of the $|q\bar qG\ra$-fluctuation becomes smaller
than the mean internucleon distance of $\sim 2$~fm as $x$ exceeds $0.01$.
Indeed, in \cite{krt2} an average coherence length of slightly less than 
$2$~fm was found for the $|q\bar qG\ra$-state at $x=0.01$ and large $Q^2\gg
1/r_0^2$. Note that gluon shadowing sets in at a smaller value of $x$ than
quark shadowing because the mass of a $|q\bar q G\ra$-state is larger than
the mass of a $|q\bar q\ra$-state. This delayed onset of gluon shadowing was
already found in \cite{kst2}.  We also point out that gluon shadowing
is even weaker than quark shadowing in the $x$-range plotted, because the
small size of the $GG$-dipole overcompensates the Casimir factor in the
$GG$-proton cross section, $\sigma^N_{GG}=(9/4)\sigma^N_{q\bar q}$.  The plot
on the right-hand side 
of fig.\ \ref{fig:gshad} shows the $Q^2$-dependence of gluon
shadowing and clearly demonstrates that gluon shadowing is a leading-twist
effect, with $R_G$ only very slowly (logarithmically) approaching unity as
$Q^2\to\infty$. 
 
\section{Nuclear shadowing for DY pair production in
\boldmath$pA$- and \boldmath$DA$-collisions} \label{sec:shadow}

Nuclear shadowing for the DY process was first observed in proton-nucleus
($pA$) collisions at large $x_F$ \cite{e772} in the E772 experiment. 
The shadowing effect will also be present in the energy
range of RHIC and LHC. Since RHIC will probably first 
measure the DY cross section from deuterium-nucleus ($DA$) rather than 
$pA$ collisions, we
perform calculations for both, $pA$ and $DA$ collisions. 

The dipole formulation provides the following explanation of shadowing in
the DY process. When the coherence time is long, one of the projectile quarks 
develops a $\gamma^*q$-fluctuation long before it reaches the target. If the
transverse momentum transfer from the target is large enough to resolve the
fluctuation, the virtual photon is freed and eventually is observed as a
lepton pair in the detector.  In the case of a nuclear target, the set of struck
nucleons compete to free the virtual photon. If the $|\gamma^*q\ra$-state
has a very small transverse size, it can propagate through the entire
nucleus because none of the bound nucleons can provide a kick strong enough
to resolve the $|\gamma^*q\ra$ structure in the incident quark. These small
fluctuations have the same small probability to interact with any nucleon, so 
they will not be shadowed. On the other hand, if the fluctuation is 
sufficiently large in 
size, only a small momentum transfer is necessary to resolve the photon.  Thus, 
the coherence of a large fluctuation will be destroyed with high probability
already in the first collision on the surface of the nucleus.  Nucleons deeper
in the nucleus do not add much to the probability of freeing the $\gamma^*$. 
Thus, this probability nearly saturates for these extremely large fluctuations,
and the DY cross section will scale like $A^{2/3}$.  From these
considerations, we can find two necessary conditions for shadowing
\cite{krt2}, \begin{itemize} \item{The $\gamma^*q$ fluctuation must have a
lifetime long enough to allow for at least two scatterings during the 
coherence time $t_c$.} \item{The $\gamma^*q$ fluctuation must have a large
freeing cross section\footnote{One should distinguish between the freeing and 
the total cross sections of a fluctuation.  The latter is always large for a 
colored
quark and all its fluctuations, while the former is driven by the difference
between scattering amplitudes of different fluctuations ($|q\ra$ and
$|q\gamma^*\ra$ in our case).  Since it is only the quark that interacts in 
each of these Fock states, the freeing cross section is controlled by the 
relative
transverse displacement of the quarks within different fluctuations. This is
how the dipole cross section comes about.}, i.e.\ its transverse size must
be sufficiently large.} \end{itemize} The first condition is assumed to be  
fulfilled throughout 
this paper, where we consider only the case of infinite $t_c$.  The
dependence of shadowing on the fluctuation size is encoded in the $q\bar
q$-nucleus cross section (\ref{eq:sigmanuc}).

\begin{figure}[t]
\centerline{
  \scalebox{0.43}{\includegraphics{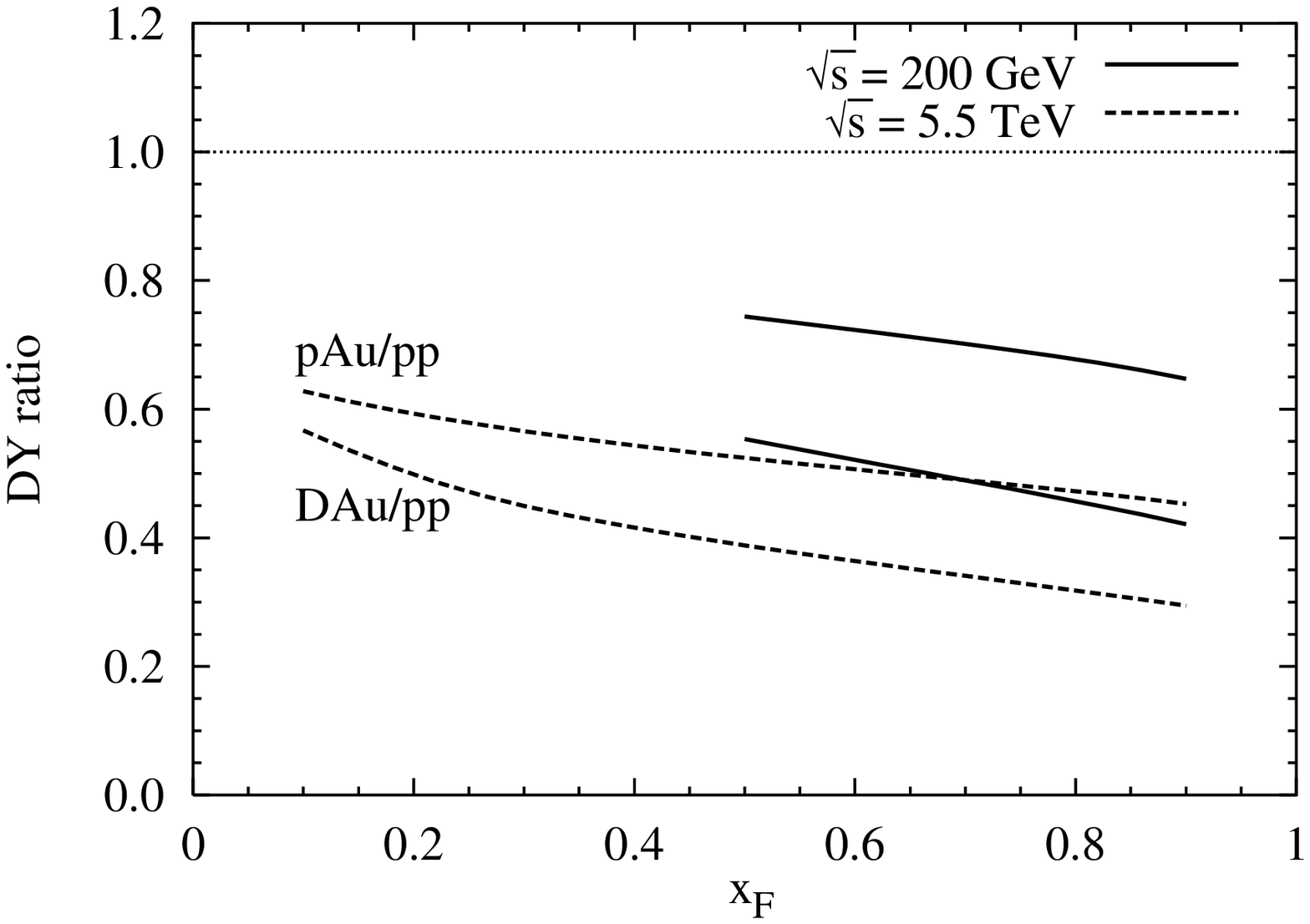}}
  \scalebox{0.43}{\includegraphics{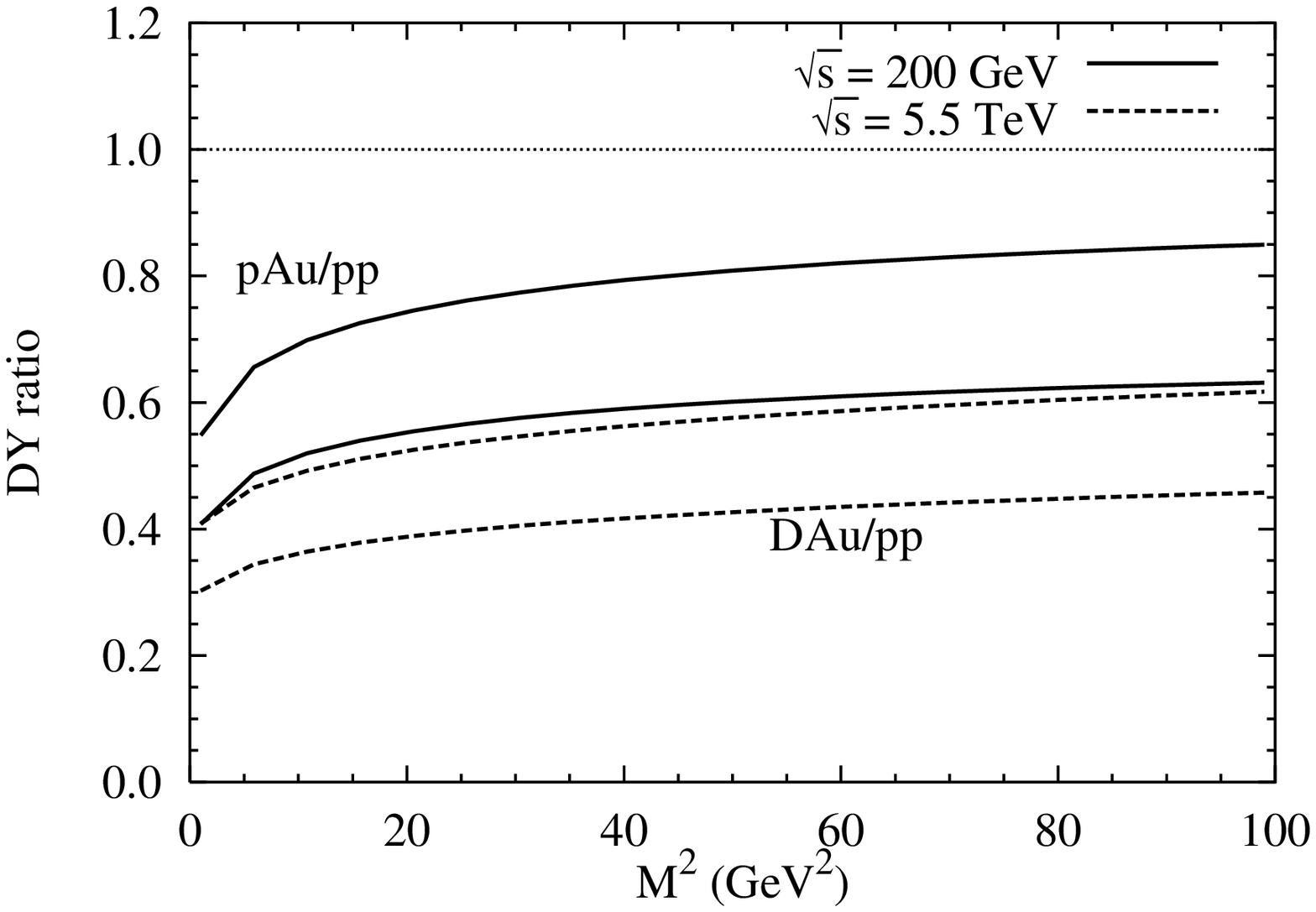}}
 }
\center{\parbox[b]{13cm}{
\caption{\label{fig:total}\em
  Shadowing for the total DY cross section 
in proton -- gold (upper curves) and deuterium -- gold 
(lower curves) collisions
at the energies of RHIC and LHC 
as function of Feynman $x_F$ and dilepton mass $M^2$, respectively.
The left-hand figure is calculated 
for $M=4.5$ GeV. The figure on the right for 
$x_F=0.5$. 
}}}
\end{figure}

Note that since the $\gamma^*q$-fluctuation is formed long before it 
reaches the the target,
the dilepton is unaffected by quark energy loss. Thus, the entire 
suppression of the DY cross section at very low $x_2$, say $x_2<0.001$, is due
to shadowing, and we do not need to worry about energy loss. This is different
at the lower
fixed-target energies \cite{e772}, where the observed depletion of the DY 
cross section originates from a combination of shadowing and energy loss
\cite{eloss1,eloss2}. The complimentary behavior of shadowing and energy loss
is discussed in more detail in \cite{eloss2}: long $t_c$ means that only 
shadowing occurs, while for short $t_c$, one observes only energy loss.

To obtain the cross section for an incident hadron, the partonic cross 
section (\ref{eq:dylctotal}) has to be weighted with the quark (and antiquark)
distributions, $q(x)$, of the projectile hadron, and one has to include 
in addition the factors necessary to account for 
the decay of the virtual photon into the dilepton.  For an incident proton,
this cross section becomes
 \beqn
\label{eq:dylchadr}
\frac{d^4\sigma(pA\to l\bar lX)}
{dM^2\,dx_F}&=&\frac{\alpha_{em}}{3\pi M^2}
\frac{x_1}{x_1+x_2}\int_{x_1}^1\frac{d\alpha}{\alpha^2}
\sum_fZ_f^2\left\{q_f
\left(\frac{x_1}{\alpha},M^2\right)+
q_{\bar f}\left(\frac{x_1}{\alpha},M^2\right)\right\}\\
\nonumber
& & \qquad\qquad\qquad\qquad\qquad\qquad\qquad\qquad\times
\frac{d\sigma(qA\to \gamma^*X)}{d\ln\alpha},
 \eeqn
where $Z_f$ is the charge of a quark of flavor $f$.  We assume that the 
the same expression (\ref{eq:dylchadr}) applies for both proton and deuteron 
projectiles, so that the only difference between these cases is that the 
flavor sum ranges over the quarks of the proton {\it and} neutron in the case 
of an incident deuteron.   
Nuclear effects in the deuterium structure function and finite-size 
effects are neglected, and isospin symmetry is assumed. 

For a calculation that can actually be compared to data, we employ the CTEQ5L 
parameterization \cite{cteq} (taken from CERNLIB \cite{cernlib})
for $q_{f,\bar f}$. Note that the projectile quark distributions enter 
at large $x=x_1/\alpha > x_1$, where they are well known. Thus, 
the uncertainty 
arising from the choice of parton distributions is minimal. 
However, these parton distributions are different for the proton and deuterium,
so that the $pA$ and $DA$ DY cross section are not trivially related.
Shadowing can now be obtained by evaluating (\ref{eq:dylchadr}) and 
(\ref{eq:dylctotal}), with $\sigma_{q\bar q}^A(\rho,x_2)$ taken from 
(\ref{eq:sigmanuc}), and dividing by $A$ times the analogous calculation with 
the $q\bar q$-proton cross section (\ref{eq:wuestsigma}). 
In the case of the deuterium projectile, we divide
by $2A$.  The nuclear density parameterizations are taken from \cite{Jager}.

The result as function of $x_F$ and dilepton mass $M$ 
at different energies is shown in fig.~\ref{fig:total}.
For each energy (RHIC and LHC) we calculated $pAu$ and $DAu$ collisions and
normalized both to $pp$ collisions. Note that the $DAu$ curve is always below 
the $pAu$ curve (for a given energy).  This is because of the different flavor 
structure of deuterium and the fact that $d$-quarks are weighted 
with a factor $Z_d^2=1/9$ in (\ref{eq:dylchadr}), compared to the
factor $Z_u^2=4/9$ for $u$-quarks.  For the RHIC energy of 
$\sqrt{s}=200\,\GeV$, we calculate only for $x_F>0.5$
to make sure that the fluctuation lifetime significantly exceeds the nuclear
radius. At the very high LHC energy of $\sqrt{s}=5.5\,\TeV$, the coherence
time is much larger than the nuclear radius for any value of $x_F$ (except at
the very endpoints).  Thus, the entire $x_F$-range is shadowed. Shadowing is
especially strong at LHC energies at large $x_F$, where $x_2$ can become as
small as $x_2\approx 10^{-6}$. At such low $x_2$, the effects of gluon
shadowing leads to a sizable additional suppression of the DY cross
section.  Without the gluon shadowing contribution in (\ref{eq:sigmanuc}), 
shadowing of the DY reaction at LHC would be strongly underestimated.
The mass dependence of shadowing in DY is shown in the plot on the right-hand
side of
fig.\ \ref{fig:total}. The weak dependence on $M$ reflects the fact that 
shadowing for DY is a leading-twist effect, just as for DIS. Indeed, 
configurations with $\alpha\to 1$ in (\ref{eq:dylctotal}) and 
(\ref{eq:dylchadr}) are the analog of Bjorken's aligned 
jet configurations in DIS \cite{bhq}, which make shadowing persist as
$M\to\infty$. 

\begin{figure}[t]
\centerline{
  \scalebox{0.43}{\includegraphics{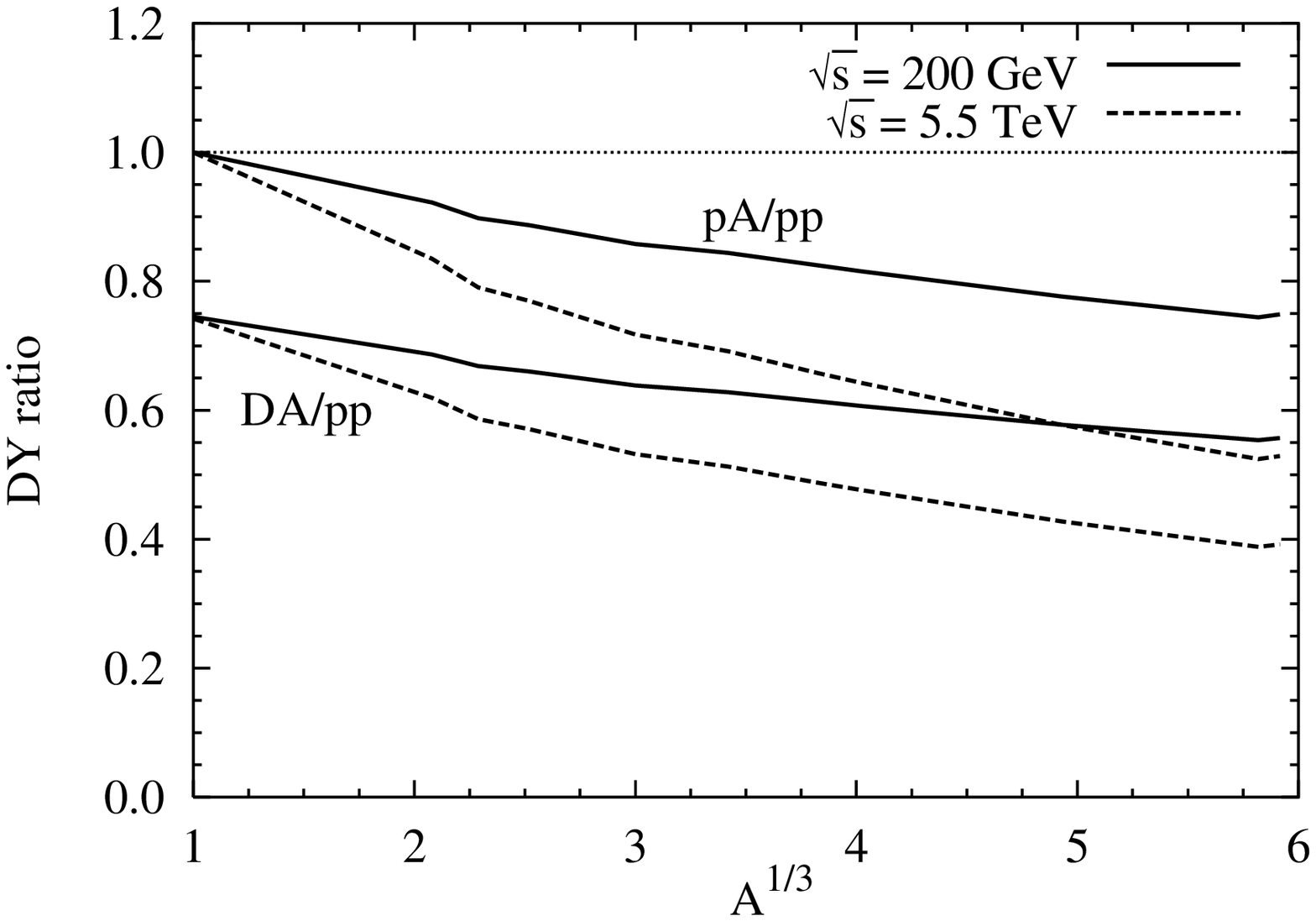}}
  \scalebox{0.43}{\includegraphics{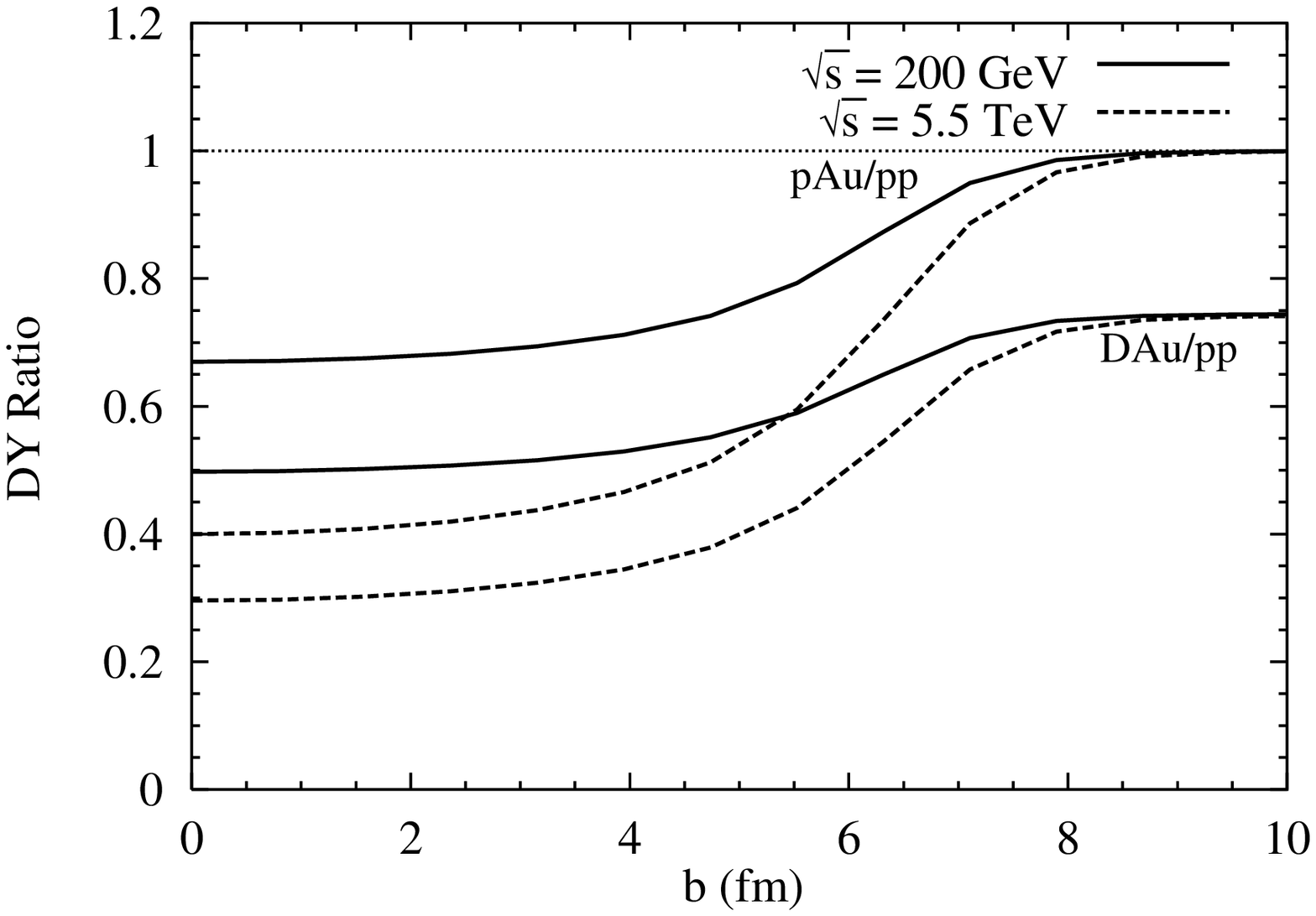}}
 }
\center{\parbox[b]{13cm}{
\caption{\label{fig:total2}\em
  Shadowing for the total DY cross section as a function of $A^{1/3}$ and 
impact parameter $b$.
Both figures are calculated for $M=4.5$ GeV and $x_F=0.5$. 
In each plot, the lower pair of curves is for deuterium -- gold scattering
($DAu$) and the two upper curves are for proton -- gold collisions ($pAu$).
}}}
\end{figure}

We also investigate the $A$- and the impact parameter dependence of shadowing,
with the results shown in fig.\ \ref{fig:total2}.  The amount of shadowing, 
{\it i.e.} the difference from unity in fig.\ \ref{fig:total2}, is to a good 
approximation proportional to $A^{1/3}$. The deuterium curves in 
fig.\ \ref{fig:total2} do not of course go to unity at $A=1$ or $b\to\infty$;
the flavor suppression remains in these limiting cases.
We point out that it is a special advantage of the dipole approach
that it naturally predicts the impact parameter dependence of nuclear effects.
To obtain the $b$ dependence, one simply eliminates the $b$ integral in
(\ref{eq:sigmanuc}) and divides by the nuclear thickness $T(b)$ instead of $A$.

\section{Nuclear modification of the DY transverse momentum distribution in
\boldmath$pA$- and \boldmath$DA$-collisions} \label{sec:broad}

The differential DY cross section as a function of the dilepton
transverse momentum $q_T$ can be calculated in the dipole formulation as well. 
At the energies relevant for RHIC and LHC, the
transverse momentum distribution of DY pairs from $p(D)+A$ collisions 
can be written in 
frozen approximation \cite{kst1,krt3},
 \beq
\frac{d^4\sigma(pA\to l\bar lX)}
{dM^2\,dx_F\,d^2q_{T}}=\frac{\alpha_{em}}{3\pi M^2}
\frac{x_1}{x_1+x_2}\int_{x_1}^1\frac{d\alpha}{\alpha^2}
\sum_fZ_f^2\left\{q_f
\left(\frac{x_1}{\alpha}\right)+
q_{\bar f}\left(\frac{x_1}{\alpha}\right)\right\}
\frac{d\sigma(qA\to \gamma^*X)}
{d\ln\alpha\,d^2q_{T}}\,,
\label{eq:dylcnucl}
 \eeq
in analogy to Eq.~(\ref{eq:dylchadr}). The 
differential cross section for heavy-photon radiation in a quark-nucleus
collision was derived in \cite{kst1},
 \beqn\nonumber\label{eq:dylcdiff}
\frac{d\sigma(qA\to \gamma^*X)}{d\ln\alpha d^2q_{T}}
&=&\frac{1}{(2\pi)^2}
\int d^2\rho_1d^2\rho_2\, 
\exp[{\rm i}\vec q_{T}\cdot(\vec\rho_1-\vec\rho_2)]
\Psi^*_{\gamma^* q}(\alpha,\vec\rho_1)
\Psi_{\gamma^* q}(\alpha,\vec\rho_2)\\
&\times&
\frac{1}{2}
\Bigl[\sigma^A_{q\bar q}(\alpha\rho_1,x_2)
+\sigma^A_{q\bar q}(\alpha\rho_2,x_2)
-\sigma^A_{q\bar q}(\alpha|\vec\rho_1-\vec\rho_2|,x_2)\Bigr]\ .
 \eeqn
In this expression, the $z$-axis is parallel to the momentum of the
projectile quark.
Integrating Eq.~(\ref{eq:dylcdiff}) over $q_{T}$ we arrive at the cross section
Eq.~(\ref{eq:10}).
Three of the four integrations in
(\ref{eq:dylcdiff}) can be performed
analytically for an arbitrary form of the dipole cross
section $\sigma^N_{q\bar q}$ \cite{thesis}. 
The details of calculations are moved to \ref{appdypp}.
Since the remaining integration still has to be performed over an oscillating
function, the $q_{T}$-range in which numerical calculations are feasible
is limited. We calculate up to $q_{T}=10$ GeV, which covers the 
experimentally interesting region.

\begin{figure}[t]
\centerline{
  \scalebox{0.8}{\includegraphics{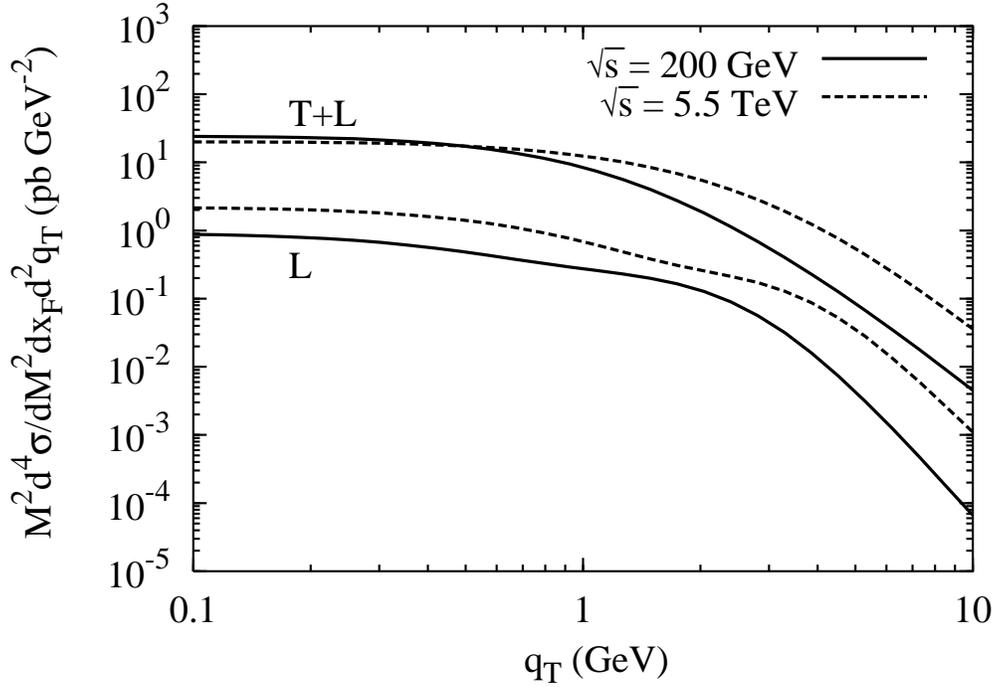}}
 }
\center{\parbox[b]{13cm}{
\caption{\label{fig:abs}\em
 Absolute value of the differential DY cross section for $pAu$ scattering,
divided by $A(=197)$. 
The curves are predictions for RHIC ($\sqrt{s}=200$ GeV)  and LHC
($\sqrt{s}=5.5$ TeV) for the dilepton mass of $M=4.5$ GeV and $x_F=0.5$.  
The upper two curves show the sum of transverse and longitudinal cross section
($T+L$) and the lower two curves ($L$) separately show the 
longitudinal cross section.
}}}
 \end{figure}

As in the preceding section, we perform calculations for $pA$ 
and for $DA$ scattering.
Our results for the differential DY cross section for 
transverse and longitudinal pairs are shown in fig.\ \ref{fig:abs}. 
We show curves only for $pA$ in this figure, because the difference between
$pA$ and $DA$ is hardly visible on the logarithmic scale. 
Even though $x_2$ increases as function of $q_T$, at $q_T=10$ GeV the 
coherence time $t_c$ is still larger than the nuclear radius.
As already observed in \cite{krt3}, the differential cross section does not 
diverge at zero transverse momentum because of the flattening of the dipole 
cross section at large separations. On the partonic level, we reproduce
the same asymptotic behavior that is expected in the standard parton model,
\beq\label{eq:asymp}
\frac{d\sigma_T(qp\to\gamma^*X)}{d^2q_{T}}
\propto\frac{1}{q_{T}^4}\quad{\rm for}\;\;q_{T}\to\infty
\eeq
and
\beq
\frac{d\sigma_L(qp\to\gamma^*X)}{d^2q_{T}}
\propto\frac{1}{q_{T}^6}\quad{\rm for}\;\;q_{T}\to\infty.
\eeq
Embedding the partonic cross section (\ref{eq:dylcdiff}) into the hadronic
environment as in (\ref{eq:dylcnucl}) will lead to a somewhat steeper
decay at large $q_{T}$, reflecting the decrease of the structure function 
$F_2$ of the projectile with $x_1$ and the increase of $x_1$ with $q_{T}$.
While this effect is quite strong at $x_F=0$ and fixed target energies,
it becomes negligible at large $x_F$ and the high {\em cm.} energies
at RHIC and LHC. Indeed, even at 
$q_{T}=10$ GeV, the asymptotic limit is not yet fully reached.
The $q_{T}$ dependence is still slightly less steep than in 
(\ref{eq:asymp}).

To see the effect of nuclear shadowing and antishadowing we divide the
nuclear differential cross section by $A$ times that of the nucleon
($2A$ for $DA$ scattering). Then, 
nuclear effects manifest themselves as a deviation from unity.  The 
results of
calculations (see \ref{appdypp}) for gold at the energies of RHIC and LHC
are presented in fig.~\ref{fig:rel} for the unpolarized (top) and
longitudinally polarized (bottom) DY cross section ratios. Also, the difference
between $pA$ (left) and $DA$ (right) now becomes clearly visible. As already 
explained in Sect.~\ref{sec:shadow}, this difference is due to the larger 
abundance of $d$-quarks in deuterium. Note that we neglected nuclear effects in
deuterium and assumed isospin symmetry for the $DA$ curves. 

\begin{figure}[htb]
  \scalebox{0.7}{\includegraphics{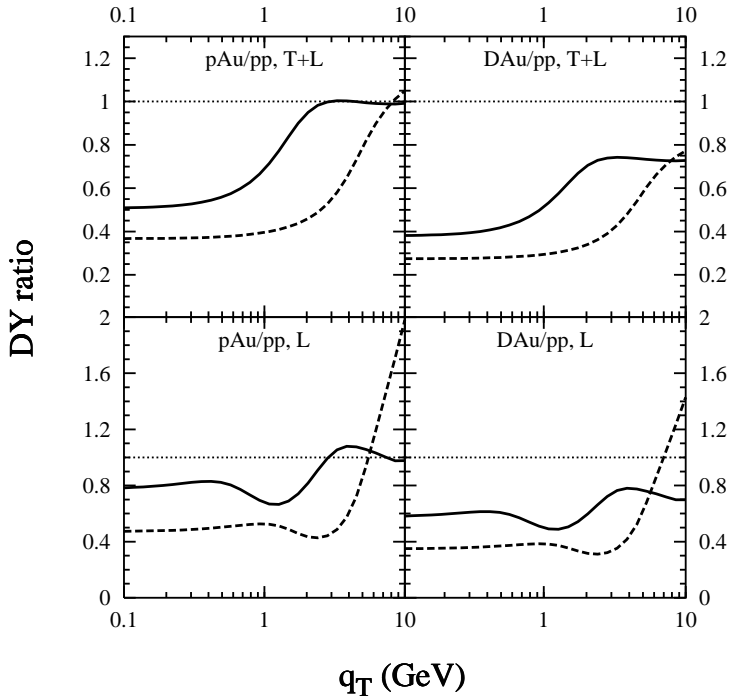}}\hfill
  \raise2.2cm\hbox{\parbox[b]{2.0in}{
   \caption{\label{fig:rel}\em
 Nuclear effects on the DY transverse momentum distribution.
  Curves show the DY cross sections for $pAu$ (left) and deuterium -- 
gold (right) collisions per nucleon divided
  by  the DY cross section from $pp$ scattering.
Solid curves are predictions for RHIC ($\sqrt{s}=200$ GeV)  and 
dashed for LHC
($\sqrt{s}=5.5$ TeV)
  Calculations are for the same kinematics as fig.\ \ref{fig:abs}.
  	} 
  }
}
 \end{figure}

For low transverse momenta, we expect DY dilepton production to be
shadowed. Note that shadowing for longitudinal $q\bar q$ pairs is smaller
than for transverse pairs because the longitudinal cross section is
dominated by small distances in the dipole cross section. However, gluon
shadowing, whose onset we observe in the RHIC predictions and which 
should become the dominant
effect at LHC, is about the same for longitudinal and transverse pairs.
Indeed, we predict rather different
shadowing effects for longitudinal and transverse dileptons at RHIC, but
about the same at LHC.

It is interesting that the effect of antishadowing, the so-called Cronin 
enhancement predicted in \cite{kst1}, disappears at the energy of RHIC after 
inclusion of
gluon shadowing, which was disregarded in \cite{kst1}. 
This reminds one of the missing Cronin enhancement in charged particle 
multiplicities that was measured at RHIC \cite{Drees}. However,
the RHIC data cannot be explained by gluon shadowing, because the 
$x$ of these data is too large.  Some antishadowing is still
possible at large $q_{T} \sim 10\,\GeV$ at the energy of LHC as a result of 
the substantial rise of the dipole cross section with energy and the 
corresponding relative enhancement of the multiple interactions responsible 
for the Cronin effect. This expectation is confirmed by fig.~\ref{fig:rg}, 
which shows the results of calculations with (solid curves) and without 
(dashed curves) gluon shadowing.

 \begin{figure}[htb]
  \scalebox{0.7}{\includegraphics{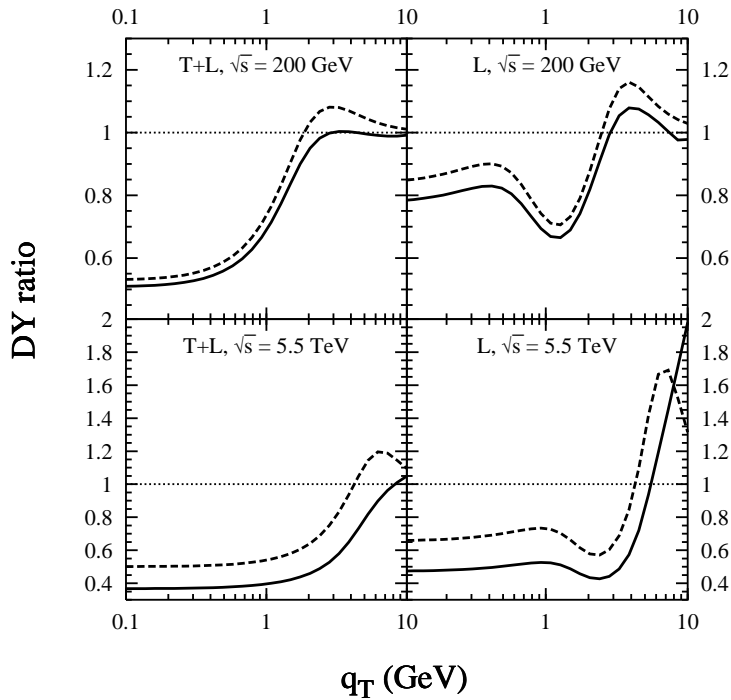}}\hfill
  \raise1.2cm\hbox{\parbox[b]{2.0in}{
   \caption{\label{fig:rg}\em
  The influence of gluon shadowing on the DY cross section. Dashed curves
are calculated without gluon shadowing, {\it i.e.}\ $R_G=1$ in 
(\ref{eq:sigmanuc}), while solid curves include gluon shadowing.
The influence on the longitudinal DY cross section is shown separately
in the two left-hand 
plots $(L)$. The two plots on the right show the DY ratio for 
the sum of the transverse and longitudinal cross sections $(T+L)$. 
Calculations are for the same kinematics as fig.\ \ref{fig:abs}.
  	} 
  }
}
 \end{figure}
The DY process with the production of a longitudinally polarized photon
manifests stronger effects of antishadowing (fig.~\ref{fig:rel} bottom and
fig.~\ref{fig:rg} right), as was earlier observed in \cite{kst1}. 
However, this enhancement of the longitudinal cross section will be 
difficult to see 
in experiments because the transverse cross section is so much larger
than the longitudinal one. All nuclear effects are expected to vanish at 
very large $q_T$. 

One can also study the moments of the transverse momentum distribution.  A
frequently measured characteristic of nuclear effects is the broadening of the
mean value of the DY transverse momentum squared, which is the difference
between the values of mean transverse momentum squared measured in $pA$ and
$pp$ collisions,
 \beq
\delta\la q_{T}^2\ra =
\la q_{T}^2\ra_A -
\la q_{T}^2\ra_N\ ,
\label{del-q}
 \eeq
 where
 \beq
\la q_{T}^2\ra_{A(N)}  = 
\frac{\int_0^{q_{T}^{max}} d^2q_{T}\,q_{T}^2\,
d\sigma^{pA}_{DY}d^2q_{T}}
{\int_0^{q_{T}^{max}} d^2q_{T}\,
d\sigma^{pN}_{DY}/d^2q_{T}}\ .
\label{mean-q}
 \eeq
 $d\sigma^{pA}_{DY}/d^2q_{T}$ is the proton-nucleus
DY cross section given by Eq.~\ref{eq:dylcnucl}.

It is easy to understand from fig.\ \ref{fig:rel} that a nuclear target leads 
to a 
larger mean transverse momentum of DY dileptons than a proton target: low
$q_{T}$ pairs, corresponding to large arguments of the dipole cross section
($\alpha\rho$), are shadowed, while high transverse momentum pairs are not 
(color filtering \cite{bbgg}). 
However, the actual numerical value
of broadening, i.e.\ the increase of the square mean transverse momentum,
depends on the maximum $q_{T}$ included in the analysis. This is not an
artifact of our approach, as this is also the case in experiment. 

According to (\ref{eq:asymp}) the numerator in (\ref{mean-q}) diverges at
large $q_{T}$ for the transverse cross section. 
Even after averaging over the projectile parton distribution, the integral
is very slowly converging, and one has to introduce
an upper cutoff $q_{T}^{max}$ since there is a maximal transverse momentum
accessible in experiment. On the other hand, the large-$q_T$ tail
of the differential cross section
should be the same for nuclear and nucleon targets since no 
nuclear effects are
expected at large $q_{T}$. For this reason one may think that
the divergence  would cancel in the difference in
Eq.~(\ref{del-q}) and render the result cutoff independent. 
This might be true if no nuclear effects occurred in the
integrated DY cross section, i.e. if $\sigma^{pA}_{DY} =
A\,\sigma^{pN}_{DY}$. However this is never the case. For long
coherence time, $t_c\gg R_A$, shadowing diminishes the DY nuclear cross
section, i.e. the denominator in the first term in Eq.~(\ref{del-q}). As a
result, the high-$q_{T}$ tail of the nuclear $q_{T}$ distribution is
renormalized and undercompensated by the second term in Eq.~(\ref{del-q}).
This is why there is sensitivity to the upper cutoff $q_{T}^{max}$ in our
results, and this sensitivity is even more pronounced at the higher energies 
where shadowing increases. On the other hand,
for short $t_c\to0$, where  shadowing vanishes, energy loss has a similar 
effect of 
suppressing the DY cross section on nuclei \cite{kn,eloss1,eloss2}.

Note that in at least some experiments \cite{report,Jen-Chieh,Alde}, the
transverse momentum broadening is extracted 
from the data by fitting the points with the functional form \cite{Kaplan}
\beq\label{eq:form}
\frac{d\sigma_{DY}^{pA(N)}}{d^2q_T}=\frac{\sigma^{pA(N)}_0}
{\left(1+\frac{q_T^2}{(q_0^{pA(N)})^2}\right)^n},
\eeq
where typically $n=6$ for both a proton and a nuclear target. 
The fast decay of the differential cross section in Eq.~(\ref{eq:form})
is due to the increase of $x_1$ as function of $q_T$. As mentioned above,
this effect vanishes at large $x_F$ and high $s$. The $q_T$-distributions
at RHIC and LHC (at large $x_F$) will be much flatter than (\ref{eq:form}).

\begin{figure}[htb]
\centerline{
  \scalebox{0.43}{\includegraphics{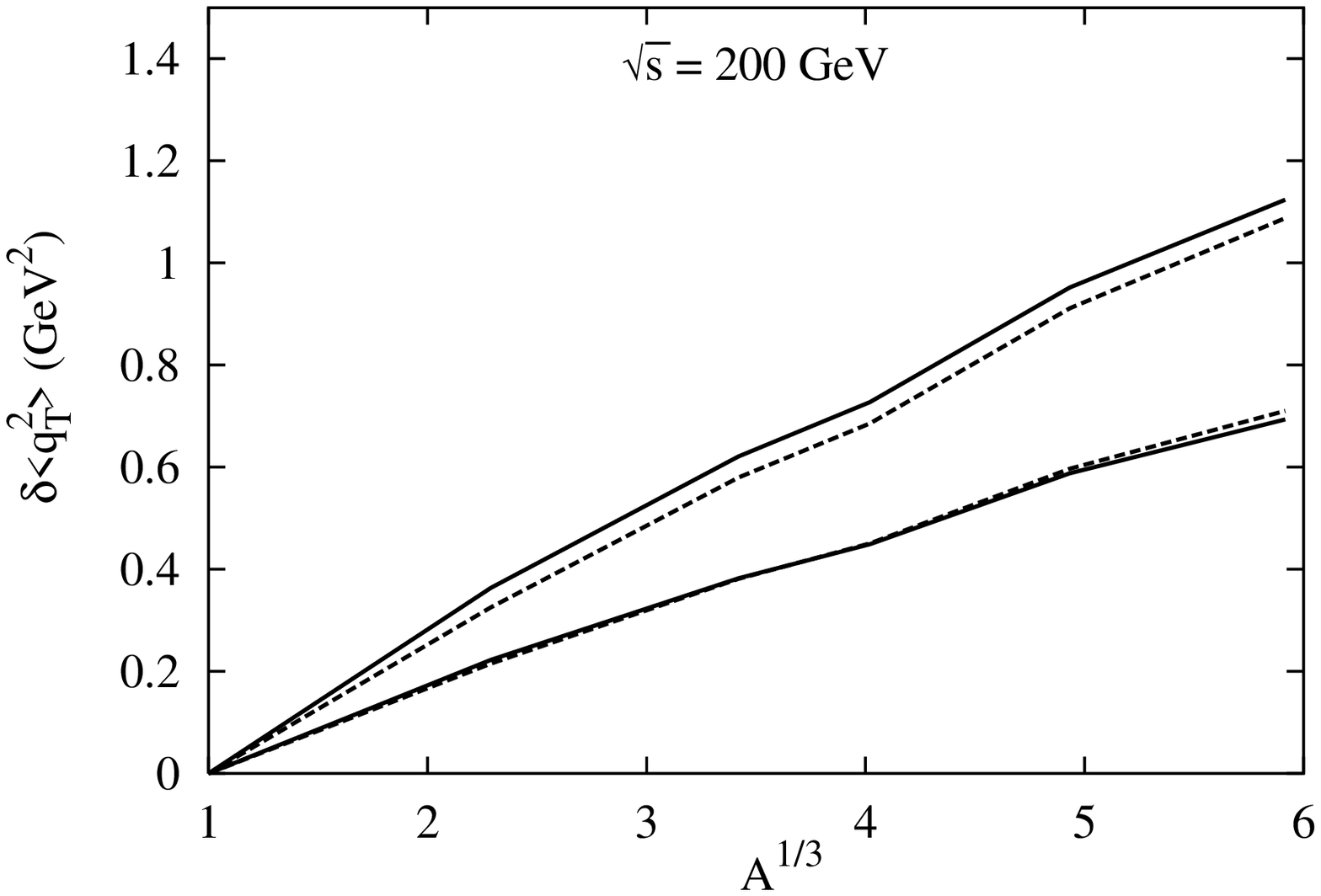}}
  \scalebox{0.43}{\includegraphics{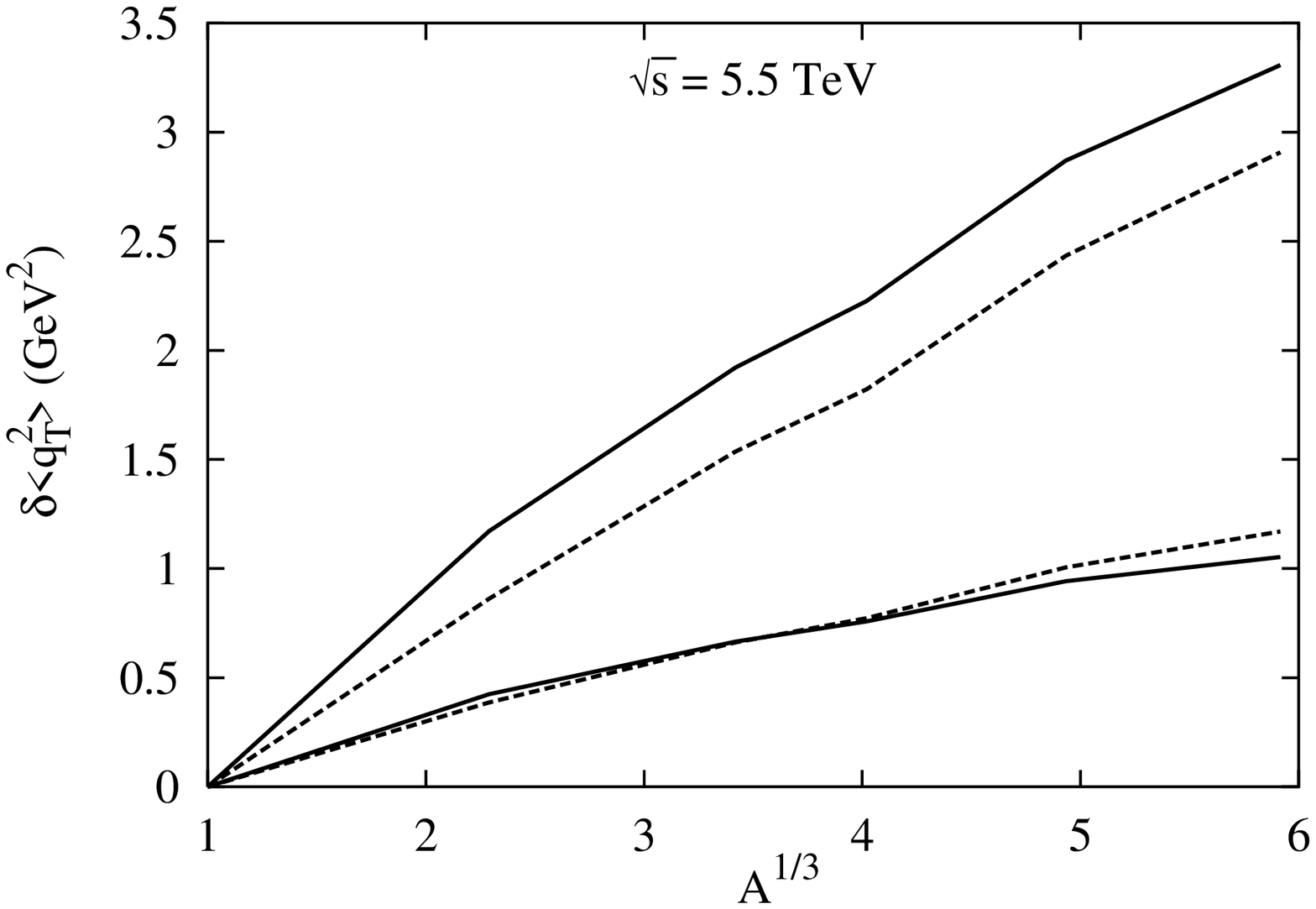}}
 }
\center{\parbox[b]{13cm}{
\caption{\label{fig:broadsvsa}\em
  Nuclear broadening for DY dileptons from $pAu$ 
at RHIC and LHC. Broadening depends on 
the transverse momentum cutoff $q_{T}^{max}$. In each plot, the lower pair of
curves is calculated for $q_{T}^{max}=5$ GeV, while the upper pair is for
$q_{T}^{max}=10$ GeV. The solid curves include gluon shadowing, the dashed 
ones do not.
  Calculations are for the same kinematics as fig.\ \ref{fig:abs}.
}}}
 \end{figure}

In fig.~\ref{fig:broadsvsa} we compare $A$-dependences of the broadening
$\delta\la q_{T}^2\ra$ calculated with different cutoffs, $q_{T}^{max} =
5$ (bottom curves) and $10\,\GeV$ (upper curves). 
The main observations are first that broadening is roughly proportional to the 
path length in the nuclear medium. This is true for any value of $q_{T}^{max}$. 
Furthermore, $\delta\la q_{T}^2\ra$ can become quite large for heavy nuclei,
around $1$ GeV at RHIC and around $3$ GeV at LHC. While the influence of gluon 
shadowing on broadening is rather weak, the $q_{T}^{max}$-dependence is 
quite strong. This is studied in more detail in fig.\ \ref{fig:k2max}. 
Increasing the transverse momentum cutoff from $5$ GeV to $10$ GeV
at RHIC energy leads to
an increase of nuclear broadening of slightly more than $50$\%.
At LHC energies,  however, where one still has nuclear effects in the transverse
momentum distribution at rather large values of $q_{T}$, broadening 
increases by a factor of $3$. Therefore,
the DY process turns out to be a less than ideal tool to measure the 
broadening of the transverse momentum distribution for a quark propagating 
through nuclear matter. 

\begin{figure}[htb]
\centerline{
  \scalebox{0.43}{\includegraphics{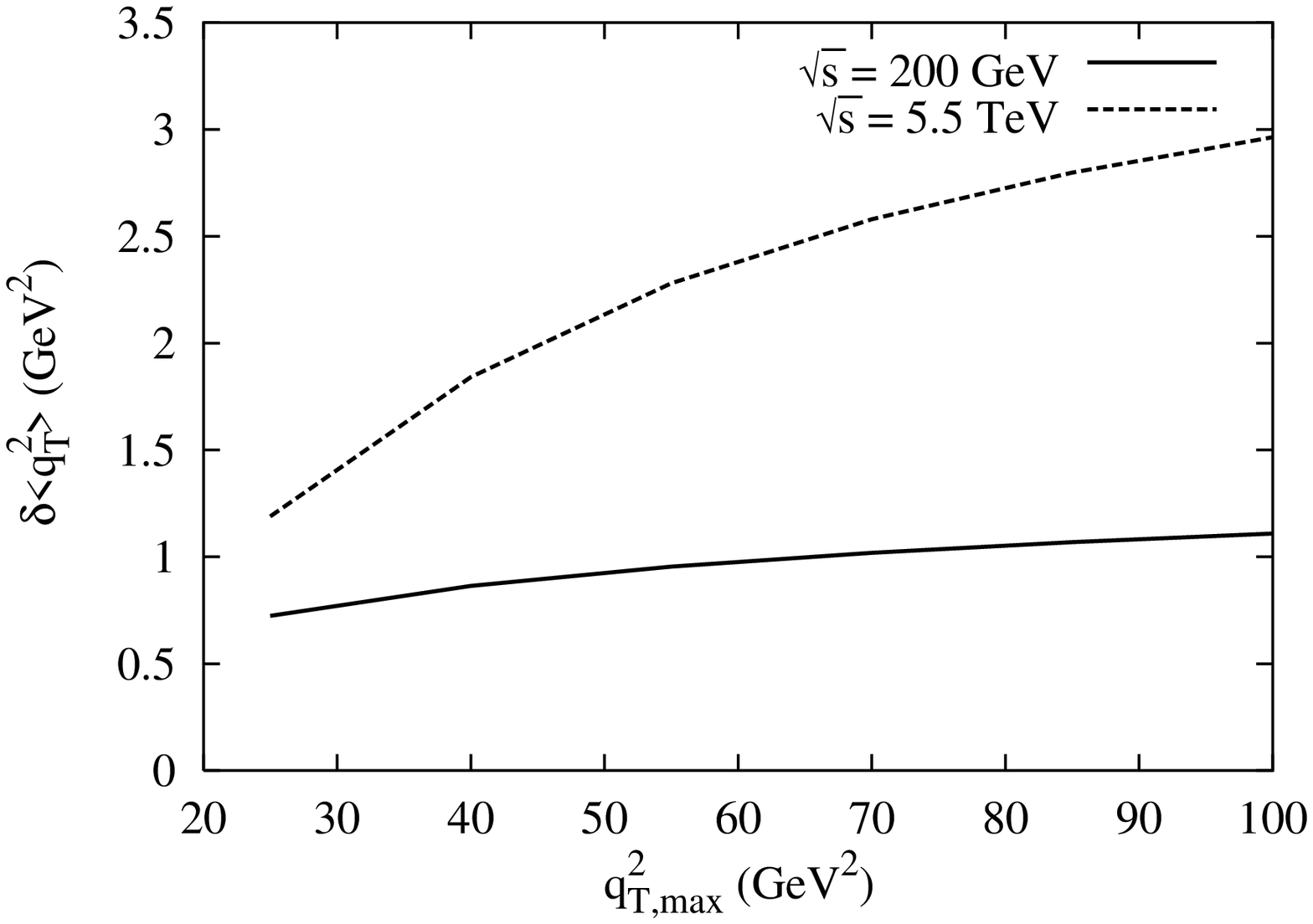}}
  \scalebox{0.43}{\includegraphics{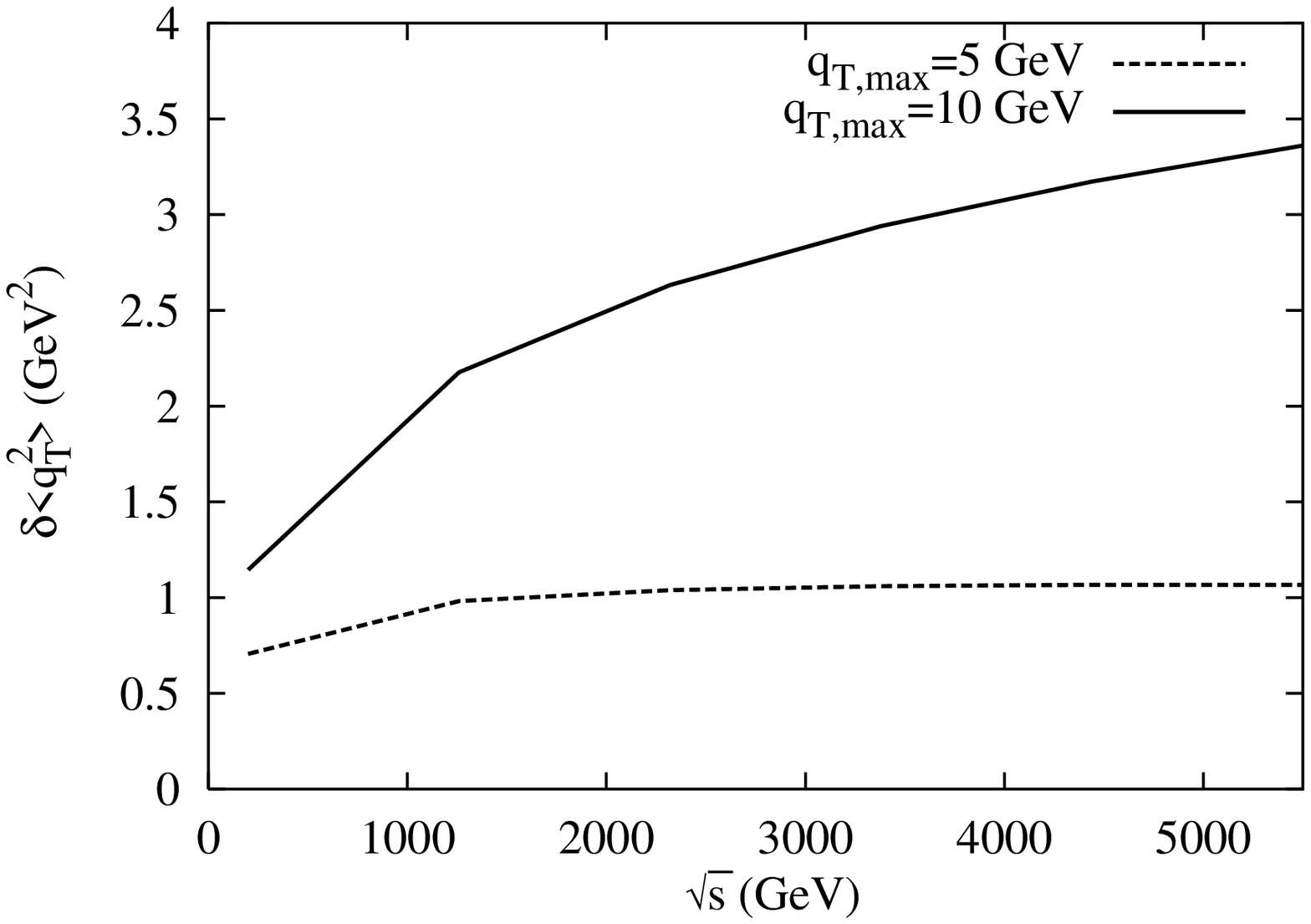}}
 }
\center{\parbox[b]{13cm}{
\caption{\label{fig:broadvss}\label{fig:k2max}\em
  The cutoff (left) and energy (right) 
dependence of transverse momentum broadening for $pAu$ scattering.
The calculations includes gluon shadowing. 
and are for the same kinematics as fig.\ \ref{fig:abs}.
}}}
 \end{figure}

Finally, we calculate the energy dependence of nuclear broadening, shown in 
the plot on the right in fig.\ \ref{fig:broadvss}. Again, calculations are 
performed for two different values
of the transverse momentum cutoff. Note that the shape of the curve depends
strongly on $q_{T}^{max}$. For a transverse momentum cutoff of $5$ GeV, there
is almost no energy dependence of $\delta\la q_{T}^2\ra$ above RHIC energy. 
The situation looks different if the transverse momentum cutoff is $10$ GeV. 
In this case, broadening does increase as function of energy. It will therefore
be difficult to draw conclusions from possible future data on the energy 
dependence of $\delta\la q_{T}^2\ra$, since presumably one will only 
see the cutoff dependence. 

Valuable insight into the relationship between shadowing and broadening is 
gained
if one performs the integration in the numerator of (\ref{mean-q})
analytically for $q_{T}^{max}\to\infty$. Without the projectile parton
distribution, the result reads  
 \beq
\la q_{T}^2\ra_{A(N)}  = \eta^2 +
\frac{C^{A(N)}(x_2)}{\sigma^{qA(N)}_{DY}(\alpha)}
\int d^2\rho\,
\Bigl|\Psi^{T,L}_{\gamma^*q}(\rho,\alpha)\Bigr|^2\ ,
\label{mean-q-rho}
 \eeq
where the wave function squared has the form given in Eqs.~(\ref{eq:dylctT}) or
(\ref{eq:dylctL}); $\eta^2=(1-\alpha)M^2+\alpha^2m_q^2$;
$\sigma^{qA}_{DY}(\alpha)=\int d^2q_{T}\,
\sigma^{qA}_{DY}(q_T,\alpha)$; and
 \beq
C^{A(N)}(x_2)=
\left.\frac{\partial\sigma^{A(N)}_{q\bar q}(\rho,x_2)}
{\partial\rho^2}\right|_{\rho\to0}\ .
\label{c}
 \eeq
Note that without gluon shadowing, all the difference between
a nucleus and a nucleon occurs in the denominator of the second term,
$\sigma^{qA(N)}_{DY}(\alpha)$. 
Including gluon shadowing, one has
\beq
C^A(x_2)=C^N(x_2)\int d^2b\; R_G(x_2,b)T(b)
=C^N(x_2)\left[A-\int d^2b\; \Delta R_G(x_2,b)T(b)\right].
\eeq
At the same time, the problem of the divergence
at large $q_{T}$ is moved to the integral of the LC wave function squared
in the second term, which has logarithmic singularity at $\rho\to0$.
Thus, the broadening Eq.~(\ref{del-q}) takes the form,
 \beq
\delta\la q_{T}^2\ra = 
\left(\Delta R_{DY}-
\frac{1}{A}\int d^2b\; \Delta R_G(x_2,b)T(b)
\right)
\frac{AC^N(x_2)}{\sigma_{DY}^{qA}(\alpha)}\,
\int d^2\rho\,
\Bigl|\Psi^{T,L}_{\gamma^*q}(\rho,\alpha)\Bigr|^2\ ,
\label{del-q-rho}
 \eeq
where $\Delta R_{DY}$ and $\Delta R_G$ are the nuclear suppressions for DY and 
for gluons, i.e.\ the difference from unity, 
$\Delta R_{DY}\equiv 1-\sigma^{qA}_{DY}/(A\sigma^{qN}_{DY})$ 
(see Eq.~\ref{eq:deltarg} for $\Delta R_G$).

We reach the interesting conclusion that if there is no shadowing,
the broadening of the transverse momentum
vanishes as well. This is a manifestation of a close relationship between
broadening and shadowing in the regime of $t_c\gg R_A$. Indeed, broadening
is interpreted in the LC dipole approach as color filtering.  Namely, the
mean size of a $q\bar q$ dipole that can be propagated through nuclear matter 
decreases with the size of the configuration;
correspondingly, the intrinsic transverse momentum of the surviving 
configurations rises. Shadowing occurs due to the same phenomenon.  Once one 
says that shadowing is negligibly small, this also means that the dipole
is too small to undergo multiple interactions. However, in this case no
color filtering occurs either.  This rather intuitive result looks very
nontrivial in the framework of parton model, where one can get broadening
even without shadowing \cite{mv}.

Note that gluon shadowing seems to reduce the amount of broadening in Eq.\ 
(\ref{del-q-rho}). This should be expected, since gluon shadowing reduces
the nuclear thickness, cf.\ Eq.\ (\ref{eq:sigmanuc}), and a more dilute medium
leads to less broadening. However, shadowing for DY, $\Delta R_{DY}$, 
increases with gluon shadowing. Numerically, we find that the
influence of gluon shadowing cancels to a large extent and that broadening is 
almost independent of $R_G$, see Fig.~\ref{fig:broadsvsa}.

In Eq.~(\ref{mean-q}) we avoided the (logarithmic) divergence in 
$\la q_T^2\ra$, related to the singular behavior of 
${\rm K}_1(\eta\rho)$ at small $\eta\rho$
for transverse photons, by introducing above an upper cutoff $q_{T}^{max}$ 
on the integrals over $q_{T}$.  These numerical results constitute 
quantitative predictions of the LC target rest frame formulation that may
be compared to experiment noting that $q_{T}^{max}$ is a physical parameter 
related to the acceptance of the spectrometer in the measurement.  
Motivated by the desire to understand these same numerical results analytically,
we next examine the theory employing certain simplifications and 
approximations.  

Our main approximation is to make the replacement of the fluctuation 
distribution by a Gaussian,
 \beq
|\Psi(\alpha,\rho)|^2\rightarrow{\alpha_{em}\over 2\pi^2}((1+(1-\alpha)^2)
\eta^2n^2{\rm e}^{-k_0^2\rho^2},
\label{approx2}
 \eeq
where $k_0^2=2\beta\eta^2$, with $n^2$ and $\beta$ chosen to give an acceptable 
match to the actual fluctuation distribution, which we simplify to be 
\beqn
|\Psi(\alpha,\rho)|^2={\alpha_{em}\over 2\pi^2}((1+(1-\alpha)^2)
\eta^2{\rm K}_1^2(\eta\rho),
\label{approx1}
\eeqn
noting that $m_q$ is small and the longitudinal contribution 
is about a 10\% correction to the momentum distribution.  Because the
distribution of fluctuations is integrable with the cutoff, the average
size of the fluctuation is meaningful, and with the Gaussian wave 
function, the mean-square transverse spread of the fluctuation is 
$\la \rho^2\ra=1/k_0^2$.

With such approximations, the integral in Eq.~(\ref{eq:dylcdiff}) can be 
evaluated analytically.  The result of carrying out the integrals over 
$\rho_1$, $\rho_2$, and $q_{T}$ gives
\beqn
&&\int d^2 q_{T}\, q_{T}^2\, \frac{d^3\sigma^{qN}_{DY}}{
 d(\ln \alpha)\,d^2q_{T}} \sim\pi 
\int d^2 b\, \left\{\frac{2\,C(\alpha )\,\alpha^2\,
T_A(b)}{k_0^2} + \frac{2\,\left[C(\alpha )\,\alpha^2\,T_A(b)\right]^2}
{\left[2\,k_0^2 + C(\alpha )\,\alpha^2\,T_A(b)\right]^2}\right\}\nonumber\\
&\approx& \pi
\int d^2 b\, \left\{\frac{2\,C(\alpha )\,\alpha^2\,
T_A(b)}{k_0^2} + \frac{\,\left[C(\alpha )\,\alpha^2\,T_A(b)\right]^2}
{2\,k_0^4}+~...\right\} ,
\label{mj6}
\eeqn
where we have omitted the prefactors in Eq.~(\ref{approx2}), and the 
expansion is useful for examining the limit of weak shadowing.
Likewise,
\beqn
&& \int d^2 q_{T}\, \frac{d^3\sigma^{qN}_{DY}}{
 d(\ln \alpha)\,d^2q_{T}} \sim \frac{2\,\pi}{k_0^2}
\int d^2 b\, \frac{C(\alpha )\,\alpha^2\,
T_A(b)}{2\,k_0^2 + C(\alpha )\,\alpha^2\,T_A(b)}\nonumber\\
&\approx& \pi
\int d^2 b\, \left\{\frac{\,C(\alpha )\,\alpha^2\,
T_A(b)}{k_0^4} - \frac{\,\left[C(\alpha )\,\alpha^2\,T_A(b)\right]^2}
{2\,k_0^6}+~...\right\} .
\label{mj7}
\eeqn
In these expressions we have introduced an effective $C=C(\alpha)$, defined 
independent of $\beta$ so that the Drell-Yan cross section for the GW color 
dipole cross section is reproduced in the $\rho^2$ approximation to it on a 
nucleon,
\beq
C(\alpha)={\int {\rm K}_1^2(\eta\rho)\sigma(\alpha\rho)d^2\rho\over n^2\int
{\rm e}^{-k_0^2\rho^2}\alpha^2 \rho^2 d^2\rho} .
\label{ceff}
\eeq

We could optimally match approximate theory to the exact one 
by choosing $n^2$ and $\beta$ numerically for a given $q_{T}^{max}$, but 
since our interest is insight rather than numerical precision at this 
point our conditions are simply the following: (1) Preserve the 
integral $\int \rho^2{\rm K}_1^2(\eta\rho)d^2\rho$, 
 \beq
\int^{\infty}_0\rho^2{\rm K}_1^2(\eta \rho) d^2\rho= n^2\int^{\infty}_0
\rho^2{\rm e}^{-k_0^2\rho^2} d^2\rho,
\label{rsq)}
 \eeq
determining  $n^2=16\beta^2/3$. (2) Adjust $\beta$ to preserve 
$\la \rho^2\ra$ using the {\it asymptotic} form for ${\rm K}_1(x)$.  
This gives $\beta=1$.  We determine the effective momentum cutoff 
corresponding to this Gaussian by comparing the exact numerical 
value for $\la q_{T}^2\ra_N$ for the nucleon to the approximate
value.  The latter is obtained from the small-$A$ limit of Eqs.~(\ref{mj6}) 
and (\ref{mj7}), 
\beq
\la q_{T}^2\ra_N=2\la\alpha^2(1+(1-\alpha )^2)\ra/\la
\alpha^2(1+(1-\alpha )^2)/k_0^2\ra,
\label{mj7a}
\eeq
where the brackets indicate the convolution (\ref{eq:dylcnucl}) with the 
quark distribution function.  We find that for $q_{T}^{max}=10$ GeV,
$\la q_{T}^2\ra_N$ agrees with the RHIC values to about 10\% and LHC values 
to about 30\%.  

For a nucleus, the integrals over impact parameter in Eq.~(\ref{mj6}) and 
Eq.~(\ref{mj7})
may be carried out analytically for a sharp-surface density model in which
the density is constant at $\rho_0$ out to radius $R_{1/2}$, and zero beyond.  
Then, from Eq.~(\ref{mj6}), we find
\beqn
&&\int d^2 q_{T}\, q_{T}^2\, \frac{d^3\sigma^{qN}_{DY}}{
 d(\ln \alpha)\,d^2q_{T}} 
\sim \frac{4\pi^2 R_{1/2}^2}{y^2}\,
\left[3{\ln}\left(1+y\right) +\frac{2}{3}y^3+
\frac{1}{2}y^2-2y-\frac{y}{1+y}\right]\ .
\label{mj10}
\eeqn
and from Eq.~(\ref{mj7})
\beqn
&& \int d^2 q_{T}\, \frac{d^3\sigma^{qN}_{DY}}{
 d(\ln \alpha)\,d^2q_{T}}\sim \frac{4\pi^2 R_{1/2}^2}{y^2k_0^2}\,
\left[{\ln}\left(1+y\right) +\frac{y^2}{2}-y\right]\ .
\label{mj11}
\eeqn
where $y=C(\alpha )\alpha^2\rho_0 R_{1/2}/k_0^2$.  To obtain $\la q_{T}^2\ra$
for the nucleus, it is necessary to perform a convolution of these expressions
with the quark distribution function of the nucleon projectile as in
Eq.~(\ref{eq:dylcnucl}).

It is interesting to examine the expansion of Eq.~(\ref{mj10}) and 
Eq.~(\ref{mj11}) in powers of $y$. Recalling that $k_0^2=1/\la\rho^2\ra$, and 
further noting that $L=2R_{1/2}$ is the diameter of the 
nucleus, this is essentially an expansion in $y=L/2\lambda$ where 
$\lambda=1/\la\sigma_{\bar qq}\ra\rho_0$ is the mean-free path of the 
fluctuation.  For most values of $\alpha$, the number of interaction 
mean-free paths in crossing the nucleus is tiny due to the large value of 
$M^2$ in $\eta^2 = (1-\alpha)M^2 + \alpha^2m_q^2$.  
However, for $\alpha\approx 1$, the mean transverse 
separation of the fluctuation may become relatively large (we find 
$\la\rho^2\ra^{1/2}=0.7$ fm for $m_q=0.2$ GeV).  In fact, for a large nucleus 
($A$=208), we see that for RHIC energies 
$C(\alpha\approx1)=2.5$ and $y\simeq 1.5$.  
Clearly, these larger fluctuations have a relatively small mean-free path 
$\lambda$ and are subject to appreciable color filtering in traversing the 
nucleus.  For LHC, $C(\alpha\approx1)=5.7$ and $y\simeq 3.4$.  At he same time, 
the amount by which $\la q_{T}^2\ra$ differs its 
value for a nucleon, Eq.~(\ref{mj7a}), grows and hence nuclear broadening also 
grows.  In this fashion, the physics of nuclear broadening is again seen to be 
directly related to color filtering for $\alpha \simeq 1$.  Since the 
expansion of Eqs.~(\ref{mj10}) and 
(\ref{mj11}) converges slowly in the region where the largest contributions to 
the nuclear broadening occur, it is necessary to evaluate the integrals over 
$b$ and $\alpha$ without making an expansion to calculate $\delta \la
q_{T}^2\ra$ with sufficient accuracy.

With these approximations (we have also omitted gluon shadowing, which
has been shown to have a weak effect on
$\delta\la q_{T}^2\ra$), we find for heavy nuclei $t_c\gg R_A$,
$\delta \la q_{T}^2\ra\propto\la T_A^2\ra$, {\it i.e.} 
$\delta \la q_{T}^2\ra$ is very nearly linear with $A^{1/3}$, just as we 
found in our more exact numerical studies.  The constants of proportionality
are about $2.2C(x_2)$ at RHIC and $1.1C(x_2)$ at LHC, which overestimate the 
slope of the exact results by about 20\% and 44\%, respectively.  Although 
linearity in $A^{1/3}$ would follows in the weak scattering 
limit, we again remark that our analytical calculations indicate 
substantial effects from higher-order multiple scattering.  

In the opposite limit of short $t_c\to0$, broadening of a quark is known 
to rise
linearly with the length of the path of the quark in nuclear matter before
the DY reaction occurs \cite{hk,jkt},
 \beq
\delta\la q_{T}^2\ra_q =
C(x_2)\,\la T_A\ra\ ,
\label{del-q-l}
 \eeq
where $\la T_A\ra = \int d^2b\,T^2_A(b)/A$ is the mean nuclear thickness.
At the same time, broadening for the DY pair, which carries only fraction
$\alpha$ of the quark momentum, is reduced,
 \beq
\delta\la q_{T}^2\ra_{DY} =
\la\alpha^2\ra\, \delta\la q_{T}^2\ra_q\ ,
\label{del-dy}
 \eeq
 where
 \beq
\la\alpha^2\ra = \frac{\int_0^1 
d\alpha\,\alpha^2 \int d^2\rho\,
\Bigl|\Psi_{q\gamma^*}(\rho,\alpha,M^2)\Bigr|^2\,
\sigma^N_{q\bar q}(\rho,x)}
{\int_0^1 d\alpha \int d^2\rho\,
\Bigl|\Psi_{q\gamma^*}(\rho,\alpha,M^2)\Bigr|^2\,
\sigma^N_{q\bar q}(\rho,x)}\ .
\label{mean-alpha}
 \eeq
This value depends on $M$ logarithmically, and for $M\sim 5\,\GeV$,
$\la\alpha^2\ra \approx 0.9$.

It is interesting to compare the nuclear dependences of broadening in the 
two limiting regimes $t_c\gg R_A$, Eq.~(\ref{del-q-rho}), and
$t_c\to0$, Eq.~(\ref{del-q-l}).
Expanding the nuclear cross section Eq.~(\ref{eq:signuc}) in 
$\sigma^N_{q\bar q}\,T_A(b)$, we can then perform the integration for 
longitudinal DY photons (for transverse photons it diverges).  Then we arrive 
at the
same expression Eq.~(\ref{del-q-l}), except tat it acquires an extra factor
 \beq
K=\frac{\la[\sigma^N_{q\bar q}]^2\ra}
{4\,\Bigl|\la\sigma^N_{q\bar q}\ra\Bigr|^2}\ ,
\label{k}
 \eeq
 where
 \beq
\la\dots\ra =
\int d^2\rho\,(\dots)\,
\Bigl|\Psi^{L}_{\gamma^*q}(\rho,\alpha)\Bigr|^2\ .
\label{mean}
 \eeq
 Applying the $\rho^2$ approximation Eq.~(\ref{eq:fs}) for the dipole
cross we get $K=4/5$. Thus, the broadening for longitudinal
DY pairs in the asymptotic regime
$t_c\gg R_A$ matches rather well the low energy regime.

The dependence of the broadening $\delta\la q_{T}^2\ra$ on the cut-off 
is an unpleasant property that brings uncertainty to the comparison of theory
with data. As we mentioned, it is related to the large $q_{T}$ behavior
Eq.~(\ref{eq:asymp}) of the
DY cross section, leading to a logarithmic divergence in the integral over
$q_{T}$ weighted with $q_{T}^2$,
 \beq
\la\sigma^{pA}_{DY}\,q_{T}^2\ra \equiv
\int\limits_0^{q_{T}^{max}} d^2q_{T}\,q^2_{T}\,
\sigma^{pN}_{DY}(q_{T},\alpha)\ ,
\label{sig-q}
 \eeq
 which is the numerator in (\ref{mean-q}). Since this integral has exactly the
same divergence at large $q_{T}^{max}$ for nuclear and nucleon targets,
it must cancel in the difference,
 \beq
\delta\la\sigma_{DY}\,q_{T}^2\ra =
\la\sigma^{pA}_{DY}\,q_{T}^2\ra -
A\,\la\sigma^{pN}_{DY}\,q_{T}^2\ra\ ,
\label{del-sig-q}
 \eeq
 and the result should be independent of the upper cut-off
$q_{T}^{max}$ when it is sufficiently large. One can also normalize this
difference dividing both terms by $A\,\la\sigma_{DY}\ra$.
Unfortunately, the result is not an exact measure of the broadening of the
transverse momentum of a quark propagating through a nucleus.
However, these quantities are independent of the experimental acceptance
($q_{T}^{max}$), and this fact makes it a better observable than the 
broadening, Eq.~(\ref{del-q}), to compare with theory.
 
\section{Polarization of DY pairs}\label{sec:polarization}

In the preceding section, we separately calculated the DY cross section for
transverse and longitudinal photons. In experiment, the different polarizations
can be distinguished by investigating the angular distribution of DY pairs.
The most general form of the DY angular distribution reads \cite{LT1},
 \beq\label{eq:angular}
\frac{d^4\sigma}{dx_FdM^2d\cos\theta d\phi}\propto 1+\lambda\cos^2\theta
+\mu\sin(2\theta)\cos(\phi)+\frac{\nu}{2}\sin^2(\theta)\cos(2\phi),
 \eeq
where $\theta$ is the angle between the muon and the $z$-axis in the rest
frame of the virtual photon, and $\phi$ is the azimuthal angle.  Of course,
$\lambda$ and $\phi$ depend on the choice of $z$-axis in the dilepton 
center-of-mass frame.  Since the dipole approach is formulated in the target rest
frame, it is convenient to put the $z$-axis in the direction of the radiated
photon \cite{bhq}.  The target rest frame and the dilepton center-of-mass
frame are then related by a boost in $z$-direction, so that the transverse
polarizations are the same in the target rest frame and in the photon rest
frame. Note that in the dilepton center-of-mass frame, the $z$-axis is
antiparallel to the target momentum. This frame is called the $u$-channel
frame and the curves we present are valid in this frame. 

The $\phi$-dependence of 
the cross section is difficult to measure. At RHIC, only the value of $\lambda$ 
can be measured \cite{Jen-Chieh}. 
Since $\lambda=+1$ for transverse and $\lambda=-1$ for longitudinal,
one obtains after integration over the azimuthal 
angle,
\beq\label{eq:lambda}
\lambda=\frac{\sigma_T-\sigma_L}{\sigma_T+\sigma_L}.
\eeq

\begin{figure}[t]
\centerline{
  \scalebox{0.43}{\includegraphics{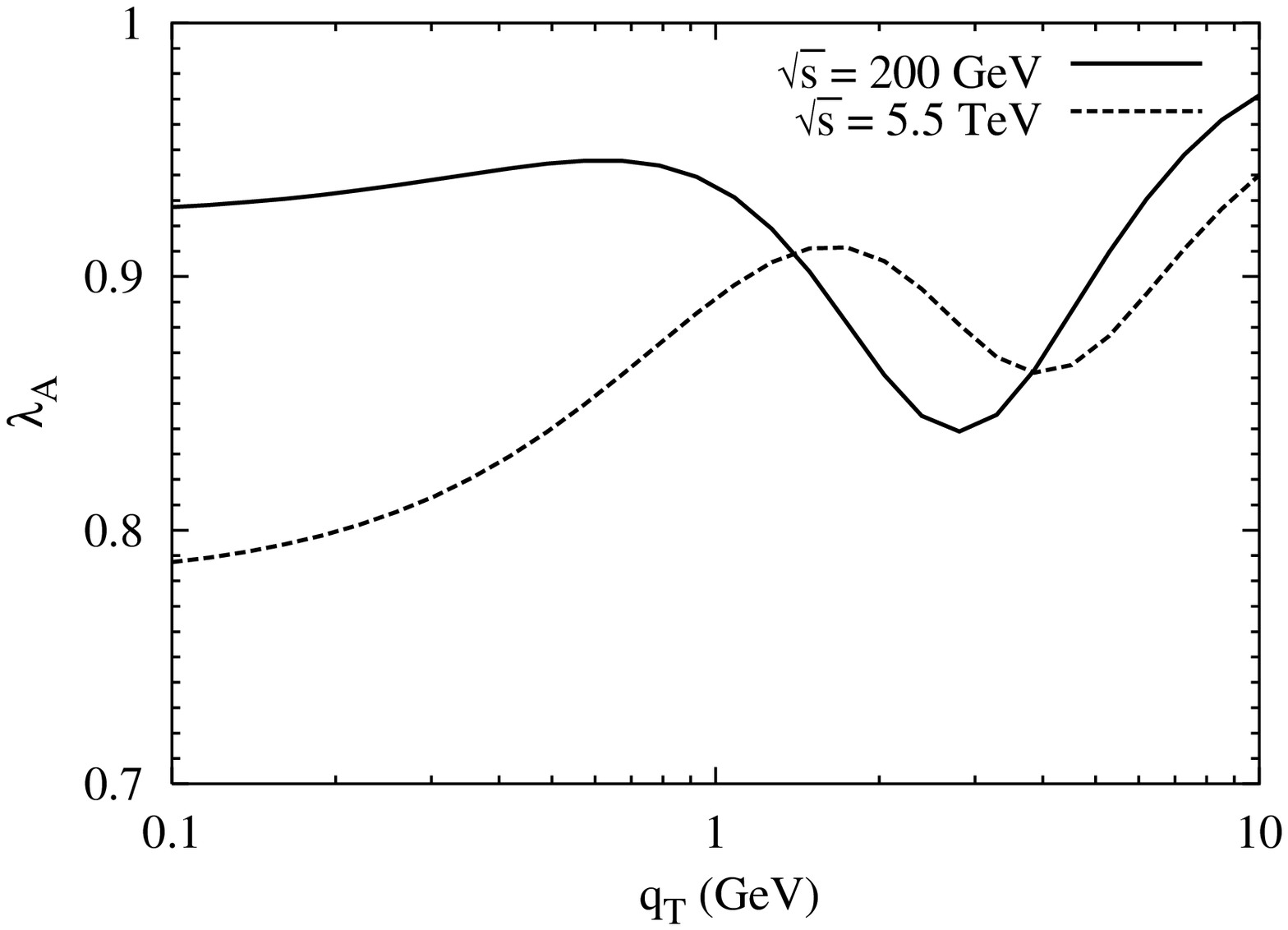}}
  \scalebox{0.43}{\includegraphics{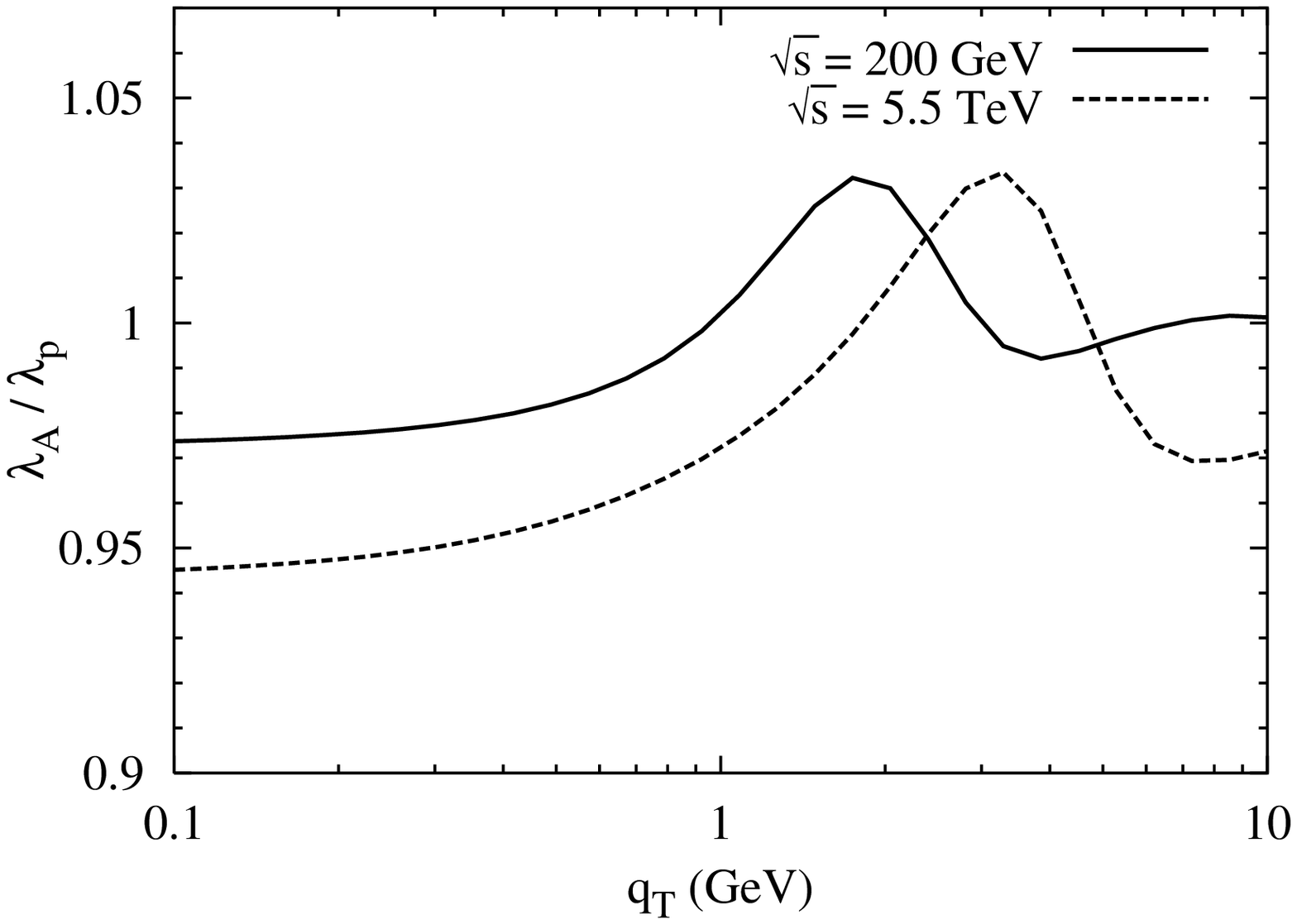}}
 }
\center{\parbox[b]{13cm}{
\caption{\label{fig:lambda}\em
  The parameter $\lambda$ (\ref{eq:lambda})
in $DAu$ collisions at RHIC and LHC (left). The figure on the right 
illustrates nuclear effects  on $\lambda$, by showing the ratio of $\lambda$
for $DAu$ scattering and for $pp$ scattering. Calculations are for the same 
kinematics as in fig.~\ref{fig:abs} and include gluon shadowing.
}}}
\end{figure}

Our results for $\lambda$ as function of $q_T$ at RHIC and LHC energies 
are shown in fig.~\ref{fig:lambda}. Note that $q_T^2=x_1x_2S-M^2$ is 
a Lorentz invariant variable and therefore has the same value for all choices
of the $z$-axis. At low $q_{T}$, $\lambda$ 
shows a clear deviation from unity. This deviation increases with energy,
because the longitudinal cross section is more strongly dominated by small
distances than the transverse. Since the dipole cross section grows faster
with energy at small separations, the relative amount of longitudinal
photons at small $q_{T}$ increases with energy, see fig.~\ref{fig:abs}.
At very large transverse momentum, $\lambda$ will eventually approach unity
because the longitudinal cross section falls off steeper than the transverse
one (\ref{eq:asymp}). It is also interesting to investigate nuclear effects on
$\lambda$. For $pp$ collisions, $\lambda$ has been calculated previously in
\cite{krt3}. One sees from fig.~\ref{fig:lambda}, that the nucleus leads to
a stronger suppression at small $q_{T}$ than a proton target. This is because 
longitudinal photons are less shadowed at $q_{T}\to 0$ than transverse, 
fig.~\ref{fig:rel}. Nuclear effects on $\lambda$ vanish at very large 
transverse momentum, as expected.

An interesting feature of our result is that the parton model relation
 \beq\label{eq:ltrelation}
1-\lambda-2\nu=0,
 \eeq
which is known as Lam-Tung relation \cite{LT1}, is not fulfilled in the
dipole approach. The violation of the Lam-Tung relation becomes apparent from
the behavior of $\lambda$ as $q_{T}\to 0$. Kinematics require that the
double-spin-flip amplitude $\nu$ vanishes at $q_{T}=0$, thus $\lambda$
should approach unity.  Remarkably, this relation holds to order $\alpha_s$ in
the QCD improved parton model. Note that (\ref{eq:ltrelation})  even holds
separately for the annihilation and for the Compton corrections in the parton
model. Even at order $\alpha_s^2$, the corrections to (\ref{eq:ltrelation})
are very small \cite{Mirkes}, by order of magnitude $1-\lambda-2\nu\sim
10^{-2}$.  Experimentally, the Lam-Tung relation is clearly violated
\cite{LTexp}, $1-\lambda-2\nu\sim -1$. However, all experimental data lie at
large $x_2$ and are therefore not relevant for our calculation. At present, the
reason for the violation of (\ref{eq:ltrelation}) is not known. Possibly, higher
twist effects, included by default in our dipole approach, can provide an
explanation \cite{Rainer}.  However, it is not yet possible to apply the
resummation technique \cite{resumcol} for logs in $(q_{T}/M)$ to higher
twists, which would be necessary to compare to future data on DY angular
distributions from PHENIX. 

The dipole formulation provides a much easier way to calculate the DY transverse
momentum distribution even at low $q_{T}$. It is not surprising that the
result from the ${\cal O}(\alpha_s)$ parton model calculation is not
reproduced in our approach, because the dipole picture is not an expansion in
powers of the strong coupling constant. Instead, all contributions from higher
order graphs that are enhanced by a large $\log(1/x_2)$ are contained in the
dipole cross section.  Moreover, if we use a phenomenological parameterization
of $\sigma^N_{q\bar q}$, some higher twists and nonperturbative effects are
contained as well. The Lam-Tung relation is violated in our calculation
because of the flattening of the dipole cross section. Indeed, for a
quadratically rising $\sigma^N_{q\bar q}(\rho,x)=C(x)\rho^2$, $\lambda$ would
vanish at $q_{T}=0$, as can be seen from (\ref{eq:dylcdiff}). Note that the
remaining coefficients for the DY angular distribution (\ref{eq:angular}) in
the dipole formulation can be obtained from Eqs. (22)-(25) of \cite{bhq}. It
is easy to check that (\ref{eq:ltrelation}) is fulfilled only with $\sigma^N_{q\bar
q}(\rho,x)=C(x)\rho^2$. Thus, the Lam-Tung relation is also violated in
$pp$-collisions \cite{krt3}.
Finally, we would like to stress that the behavior of
$\lambda$ at low $q_{T}$ depends heavily on the large $\rho$ behavior of the
dipole cross section, which is not well constrained by DIS and diffractive
data. However, even if (\ref{eq:wuestsigma}) is unrealistic at large
separations, the qualitative behavior of $\lambda$ will remain the same for
any flattening parameterization. We therefore believe that it is worthwhile to
measure $\lambda$ at RHIC, because such data could give us information about
the dynamics beyond the conventional parton model. 

\section{DY process in heavy ion collisions}\label{sec:aa}

As we pointed out in the introduction it is impossible to predict nuclear
shadowing for the DY process in nuclear collisions from the parton model using 
data for DIS and the DY reaction on proton and nuclear targets. Indeed, the
nuclear effects to be predicted are presented in the form,
 \beq
R^{AB}_{DY}(x_1,x_2)\equiv
\frac{\sigma^{AB}_{DY}(x_1,x_2)}
{AB\,\sigma^{NN}_{DY}(x_1,x_2)}\ ,
\label{shad-ab}
 \eeq
 where
 \beqn
\sigma^{AB}_{DY}(x_1,x_2) &=&
N\sum\limits_f Z_f^2
\int d^2b\int d^2s\,T_A(\vec s)T_B(\vec b-\vec s)\Bigl[
q_f^v(x_1)R^A_v(x_1,\vec s)\,\bar q_f^s(x_2)R^B_s(x_2,\vec b-\vec s) 
\nonumber\\ &+& 
\bar q_f^s(x_1)R^A_s(x_1,\vec s)\,q_f^v(x_2)R^B_v(x_2,\vec b-\vec s) +
q_f^s(x_1)R^A_s(x_1,\vec s)\,\bar q_f^s(x_2)R^B_s(x_2,\vec b-\vec s)
\nonumber\\ &+& 
\bar q_f^s(x_1)R^A_s(x_1,\vec s)\,q_f^s(x_2)R^B_s(x_2,\vec b-\vec s)
\Bigr]\ ,
\label{ab-x-sect}\\
\sigma^{NN}_{DY}(x_1,x_2) &=&
N\,\sum\limits_f Z_f^2\Bigl[
q_f^v(x_1)\,\bar q_f^s(x_2) + 
\bar q_f^s(x_1)\,q_f^v(x_2) 
\nonumber\\ &+&
 q_f^s(x_1)\,\bar q_f^s(x_2)+
\bar q_f^s(x_1)\,q_f^s(x_2)\Bigr]\ ,
\label{nn-x-section}
 \eeqn
and $\vec b$ is the impact parameter of the beam ($A$) and target ($B$)
nuclei.   The first two terms in Eq.~(\ref{ab-x-sect}) correspond to (i) 
annihilation of a valence quark of the beam with a sea antiquark of the
target; (ii) a sea antiquark of the beam with a valence quark of the
target. The third and fourth terms correspond to (iii) both the quark and
antiquark originating from either the sea of the beam or target. 
Provided that the Drell-Yan $K$-factor is independent of $A$ and $B$,
the overall normalization factor $N$ in
these expressions 
is irrelevant for shadowing Eq.~(\ref{shad-ab}) since it
cancels in the ratio. This assumption is supported by
experimental data on the $K$ factor for various heavy ion collisions 
\cite{Kfactor}. 

Note that a generalization of 
Eqs.~(\ref{eq:dylcnucl}) --
(\ref{eq:dylcdiff}) to the case of $AB$ scattering
including higher twist effects, would be much
more complicated than Eqs.~(\ref{ab-x-sect}) --
(\ref{nn-x-section}). Such
higher twist effects, however, will be important only for nuclear effects in
the DY transverse momentum distribution in $AB$, which are beyond the scope 
of this paper.

All parton distributions in Eqs.~(\ref{ab-x-sect}) --
(\ref{nn-x-section}) are taken at the same virtuality $Q^2=M^2$. For the
sake of simplicity, we assume that the shadowing, $R_s^A$, is the same for sea
quarks and antiquarks, and we neglect the isospin noninvariance of the sea
distribution at moderately small $x\sim 0.1$ \cite{nusea}, which can easily be
taken into account. 

The nuclear shadowing $R^A_{q_1}(x,\vec s)$ is given in Sect.~\ref{subsec:az}
below.  Note that $R_q^A(x,b)$ at small $x$ is a
function of nuclear thickness. It vanishes at large impact parameters
on the nuclear periphery, but it reaches its maximum at $b=0$. Data 
for DIS or the DY
reaction in $pA$ collisions provide information only about the $b$-integrated
shadowing effect.
Knowledge of only such integrated shadowing is not sufficient for the
calculation of shadowing in an $AB$ collision, Eq.~(\ref{ab-x-sect}).
Nevertheless, it was assumed in
\cite{esk-01} that shadowing is independent of impact parameter,
$R_q^A(x,Q^2,b)=R_q^A(x,Q^2)$. Clearly, such an {\it ad hoc} assumption
is not justified and leads to basic consequences which cannot be
accepted. For instance, it precludes any dependence of shadowing effects
on centrality in heavy ion collision. 
					
As we demonstrated above, the LC dipole approach provides direct access
to the impact parameter dependence of shadowing effects.  In the following 
subsection, we calculate shadowing for valence and sea quarks and compare 
the $b$-integrated result to the EKS98 parameterization \cite{ekr}. 
Of course, the impact parameter dependence of shadowing is taken into 
account in Sect.~\ref{subsec:aa}, where we 
predict shadowing for the DY reaction in a nuclear collision in the integrated
form Eq.~(\ref{ab-x-sect}), as well as a function of centrality. In order to
calculate $R^{AB}_{DY}(x_1,x_2,b)$ for a collision with impact parameter
$b$, one should just eliminate the integration over $\vec b$ in the
numerator of Eq.~(\ref{ab-x-sect}), and replace
 \beq
A\,B \Rightarrow T_{AB}(b)=\int d^2s\,T_A(s)\,T_B(\vec b-\vec s)
\label{t_ab}
 \eeq
in the denominator of Eq.~(\ref{shad-ab}). Comparison with the
minimal-bias events for DY dilepton production in heavy ion collisions
would serve as a rigorous test of the theory. 

\subsection{Nuclear shadowing for sea and valence quarks}\label{subsec:az}

Since we rely on the factorization relations Eqs.~(\ref{shad-ab}) --
(\ref{nn-x-section}), we can calculate shadowing $R_s^A(x,Q^2)$
and $R_v^A(x,Q^2)$ in DIS, which looks somewhat simpler as it does not
include a convolution with the initial quark distribution.

Shadowing for sea quarks is calculated with
Eq.~(\ref{eq:sigmanuc}) and is given by
 \beq
R_s(x,Q^2,b) = \frac{
2\int_0^1 d\alpha\int d^2\rho\,
\Bigl|\Psi_{q\bar q}(\rho,\alpha,Q^2)\Bigr|^2\,
\left[1-\left(1-\frac{1}{2A}\,\sigma^N_{q\bar q}(\rho,x)
R_G(x,\lambda/\rho^2,b)
T_A(b)\right)^A\right]}
{T_A(b)\,\int_0^1 d\alpha\int d^2\rho\,
\Bigl|\Psi_{q\bar q}(\rho,\alpha,Q^2)\Bigr|^2\,
\sigma^N_{q\bar q}(\rho,x)}\ .
\label{sea}
 \eeq
The light-cone wavefunctions $\Psi_{q\bar q}$ for the transition
$\gamma^*\to q\bar q$ can be found in the literature, see 
{\em e.g.~}\cite{Wuesthoff1}.
Let us recall that this expression is valid only for the so-called
``frozen'' approximation, i.e. in the asymptotic regime $t_c\gg R_A$, which
takes place at very small $x$. In the transition region $t_c \lsim R_A$ one
should employ the LC Green function technique.  This was done in
\cite{krt1,krt2}, although gluon shadowing was neglected in those
calculations. At very small $x$ gluon shadowing is essential, as
demonstrated above (see comparison with data from the NMC experiment in
\cite{eloss2}). 

Nuclear shadowing for valence quarks has never been calculated. This shadowing
is usually believed to be small \cite{ekr}, if it occurs at all.  We shall
demonstrate, however, that shadowing for valence quarks is quite sizable, 
even stronger than the shadowing of sea quarks.  This is another new result of 
the present paper.  We note in this regard that the nuclear structure function 
$F_2(x)$ is different from the quark distribution function in an essential
way; namely, the former contains shadowing effects and therefore the baryon 
number sum rule is not applicable to it \cite{Brodsky}, 
a difference that may explain the 
discrepancy of our results compared to Ref.~\cite{ekr}.

Note that we call the ratio Eq.~(\ref{sea}) shadowing for sea quarks
because the dipole cross section $\sigma^N_{q\bar q}(\rho,x)$ includes only
the part that rises with energy, corresponding to gluonic exchanges in the
cross-channel. Therefore, this is the part of the sea generated via gluons
(there are also other sources of the sea, for instance the meson cloud of
the nucleon, but they vanish linearly with $x$ or faster).  The fact that
the color-dipole cross section includes only the part generated by gluons 
is the reason why it
should be used only at very small $x<0.01$, where the sea dominates.  This part
of the dipole cross section can be called the Pomeron in the language of Regge
phenomenology. In the same framework, one can relate the valence quark
distribution in the proton to the Reggeon part of the dipole cross section,
which has been neglected so far. 

Thus, to include valence quarks in the dipole formulation of DIS, one
should replace
 \beq
\sigma^N_{q\bar q}(\rho,x) \Rightarrow
\sigma^{\pom}_{q\bar q}(\rho,x) + 
\sigma^{\reg}_{q\bar q}(\rho,x)\ ,
\label{reggeons}
 \eeq
where the first (Pomeron) term corresponds to the gluonic part of the cross
section, which we have used so far. It is responsible for the sea quark part of
the nucleon structure function.
The second (Reggeon) term must reproduce the distribution of valence quarks
in the nucleon; this condition constraints its behavior at small $x$.
One can guess that it has the following form,
 \beq
\sigma^{\reg}_{q\bar q}(\rho,x) = 
\tilde N\,\rho^2\,\sqrt{x}\ ,
\label{sig-r}
 \eeq
where $\sqrt{x}$ should reproduce the known $x$ dependence of valence quark
distribution (as, in fact, motivated by Regge phenomenology), and the factor
$\rho^2$ is needed to respect the Bjorken scaling. The factor $\tilde N$ will
cancel in what follows.

\begin{figure}[t]
  \scalebox{0.65}{\includegraphics*{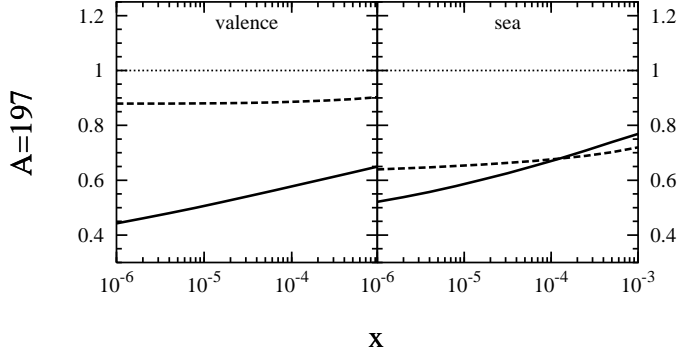}}\hfill
  \raise2cm\hbox{\parbox[b]{2.3in}{
    \caption{
      \label{fig:sv_shadow}\em
     Shadowing for sea and valence $u$-quarks in DIS off gold at $Q=4.5$ GeV. 
Solid lines are calculated
from Eqs.\ (\ref{v-shad-2}) -- (\ref{sea-2}), 
while dashed curves show the EKS98 parameterization 
\cite{ekr}.
    }
  }
}
\end{figure}

We are now in a position to calculate shadowing for valence quarks by
inserting the cross section Eq.~(\ref{reggeons}) into the eikonal
expression Eq.~(\ref{sea}).  Clearly, only the gluonic part of the dipole
cross section is subject to gluon shadowing, i.e. only the first (Pomeron)
term of Eq.~(\ref{reggeons}) should be multiplied by $R_G$.  Furthermore, if one
expands the numerator in powers of $\sigma^{\reg}_{q\bar q}$ and picks
out the linear term\footnote{The small size of $\sigma^{\reg}_{q\bar q}(\rho,x)$
at small $x$ motivates such an expansion; however, one should note
that it would not be proper to include the higher powers 
of the Reggeon cross section.  Indeed, the Reggeons correspond to
planar graphs.  These cannot be eikonalized since they lead to the so-called AFS
(Amati-Fubini-Stangelini)  planar graphs, which vanish at high energies
\cite{gribov}.}, then one arrives at the following expression for nuclear
shadowing of the valence quarks,
 \beq
R_v(x,Q^2,b) = \frac{
\int_0^1 d\alpha\int d^2\rho
\Bigl|\Psi_{q\bar q}(\rho,\alpha,Q^2)\Bigr|^2
\sigma^{\reg}_{q\bar q}(\rho,x)\,
\Bigl[1-\frac{1}{2A}\sigma^{\pom}_{q\bar q}(\rho,x)
R_G(x,\lambda/\rho^2,b)
T_A(b)\Bigr]^A}
{\int_0^1 d\alpha\int d^2\rho\,
\Bigl|\Psi_{q\bar q}(\rho,\alpha,Q^2)\Bigr|^2\,
\sigma^{\reg}_{q\bar q}(\rho,x)}\ .
\label{v-shad}
 \eeq
This shadowing is even stronger than the shadowing for sea quarks,
Eq.~(\ref{sea}). 
Indeed, for weak shadowing we can also expand Eqs.~(\ref{sea}) and
(\ref{v-shad}) in powers of $\sigma^{\pom}_{q\bar q}$. Then 
one obtains a
shadowing correction $1-R_v = {1\over2}\,\sigma_{eff}\,T_A$ in
(\ref{v-shad})
which is twice as large as that for sea quarks $1-R_s =
{1\over4}\,\sigma_{eff}\,T_A$
in (\ref{sea}). Here $\sigma_{eff}=\la\sigma^2_{q\bar
q}\ra/\la\sigma_{q\bar q}\ra$ \cite{eloss2}. 

These estimates rely, however, on the $\rho^2$-approximation 
for the dipole cross section and on the assumption that shadowing is weak.
The result of our calculation, including gluon shadowing and a realistic 
parameterization (\ref{eq:wuestsigma}) of the dipole cross section, is shown
in fig.~\ref{fig:sv_shadow}. 
We show only the $b$-integrated shadowing, which is given by
\beqn
\lefteqn{R_v(x,Q^2) =}\\ 
\nonumber &&
\frac{\int d^2b\, T_A(b)
\int_0^1 d\alpha\int d^2\rho
\Bigl|\Psi_{q\bar q}(\rho,\alpha,Q^2)\Bigr|^2
\sigma^{\reg}_{q\bar q}(\rho,x)\,
\Bigl[1-\frac{1}{2A}\sigma^{\pom}_{q\bar q}(\rho,x)
R_G(x,\lambda/\rho^2,b)
T_A(b)\Bigr]^A}
{A\int_0^1 d\alpha\int d^2\rho\,
\Bigl|\Psi_{q\bar q}(\rho,\alpha,Q^2)\Bigr|^2\,
\sigma^{\reg}_{q\bar q}(\rho,x)}\ 
\label{v-shad-2}
 \eeqn
for valence quarks and by
\beqn
\lefteqn{R_s(x,Q^2) =}\\
\nonumber&& \frac{
2\int d^2b\int_0^1 d\alpha\int d^2\rho\,
\Bigl|\Psi_{q\bar q}(\rho,\alpha,Q^2)\Bigr|^2\,
\left[1-\left(1-\frac{1}{2A}\,\sigma^N_{q\bar q}(\rho,x)
R_G(x,\lambda/\rho^2,b)
T_A(b)\right)^A\right]}
{A\,\int_0^1 d\alpha\int d^2\rho\,
\Bigl|\Psi_{q\bar q}(\rho,\alpha,Q^2)\Bigr|^2\,
\sigma^N_{q\bar q}(\rho,x)}\ .
\label{sea-2}
 \eeqn
for sea quarks. Shadowing for valence quarks is still 
stronger than it is for sea quarks, but not by a factor of $2$. However, 
valence quark shadowing calculated in the LC approach is much stronger 
than in the parameterization of \cite{ekr}. 
Note that the authors of \cite{ekr} force their fit to the nuclear valence quark 
distribution to fulfill the baryon number sum rule, which is however violated
by shadowing effects. Therefore valence quark shadowing is underestimated in
\cite{ekr}.

Unfortunately, it will be 
impossible to extract the low-$x$ valence quark distribution of a nucleus
from DY experiments, because the nuclear structure function is 
dominated by sea quarks. Maybe neutrino-nucleus scattering experiments 
could provide data that help to differentiate shadowing for sea and 
valence quarks.

\subsection{Modification of the DY cross section 
in heavy-ion collisions}\label{subsec:aa}

We can now make use of Eqs.~(\ref{shad-ab}) -- (\ref{nn-x-section}) and
predict nuclear effects for the cross section of DY lepton pair production
in heavy ion collisions.
We perform calculations at large $x_F$, where the structure function of
the target nucleus, say nucleus $B$ in (\ref{ab-x-sect}), enters at 
small $x_2\ll 0.1$ and is therefore subject to shadowing. We calculate 
the shadowing ratios $R_{s,v}$ in (\ref{ab-x-sect}) as function of
impact parameter from Eqs.~(\ref{sea}) and (\ref{v-shad}). 
However, shadowing considered in the previous
section is not the only nuclear effect affecting the ratios
$R_{s,v}(x_1,x_2)$.  Indeed, the parton distributions of the
projectile nucleus, $A$, are sampled at $x_1\gsim 0.1$,
i.e.\ in the region where the EMC \cite{michele} effect must be taken
into account. This leads to a $10-20\%$ suppression at medium-large $x_1$ and
a strong enhancement at $x_1\approx x_F\to1$ due to Fermi motion. 
Also, a small $2-3\,\%$ enhancement is known to exist at $x_1\sim 0.1$. 
All these are medium effects 
caused by the difference between the properties of bound and free nucleons. 
Since the nuclear density, apart from its surface, is approximately the same
for all nuclei except for the lightest ones, one can assume that these medium
effects are about the same for all bound nucleons in all nuclei. This
assumption is supported by data, which displays no strong
$A$-dependence of the EMC effect from medium through heavy nuclei
\cite{michele}. Of course, there is no contradiction here with our previous
statement about a strong impact parameter dependence of nuclear shadowing. 

\begin{figure}[t]
\centerline{
  \scalebox{0.43}{\includegraphics{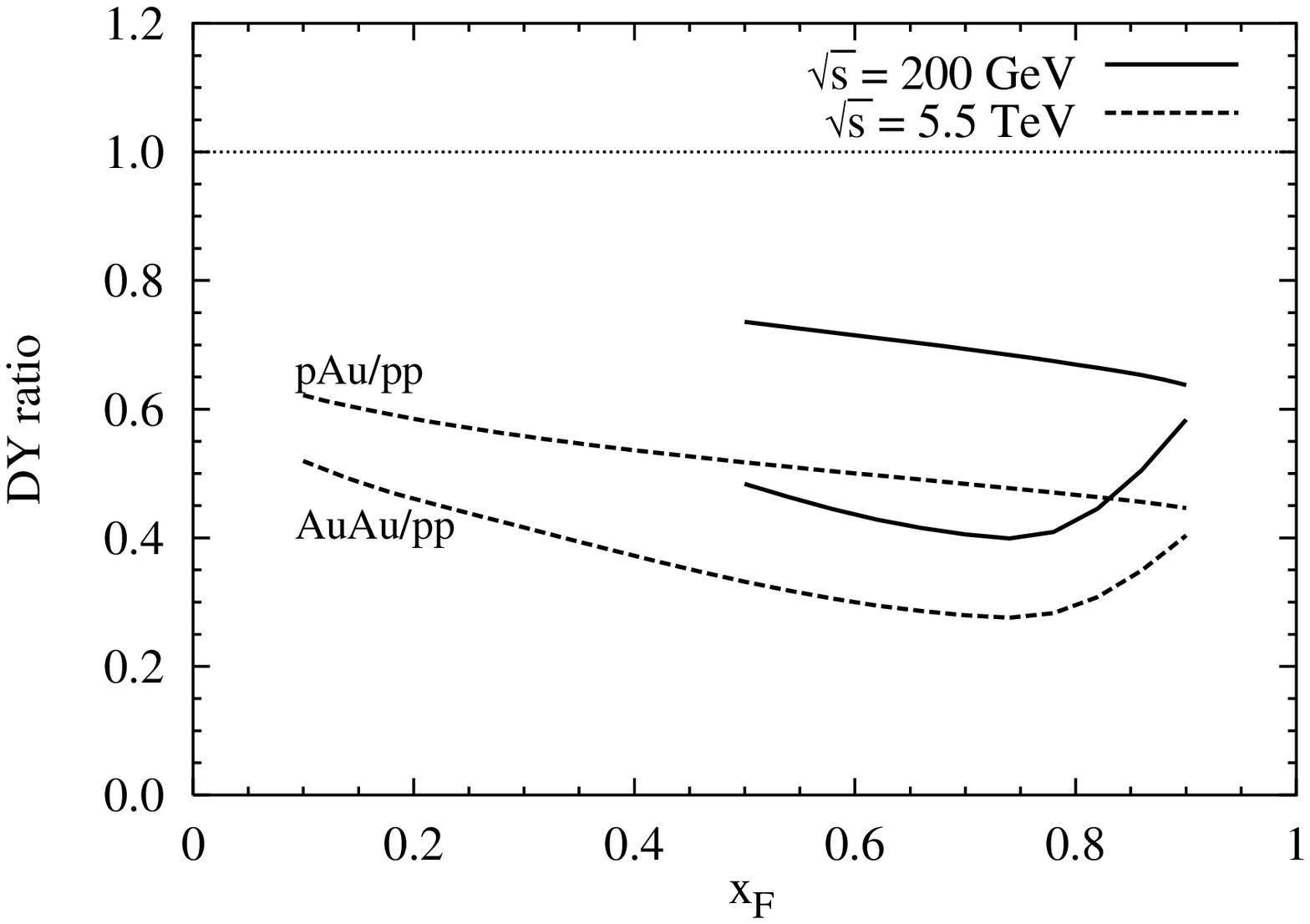}}
  \scalebox{0.43}{\includegraphics{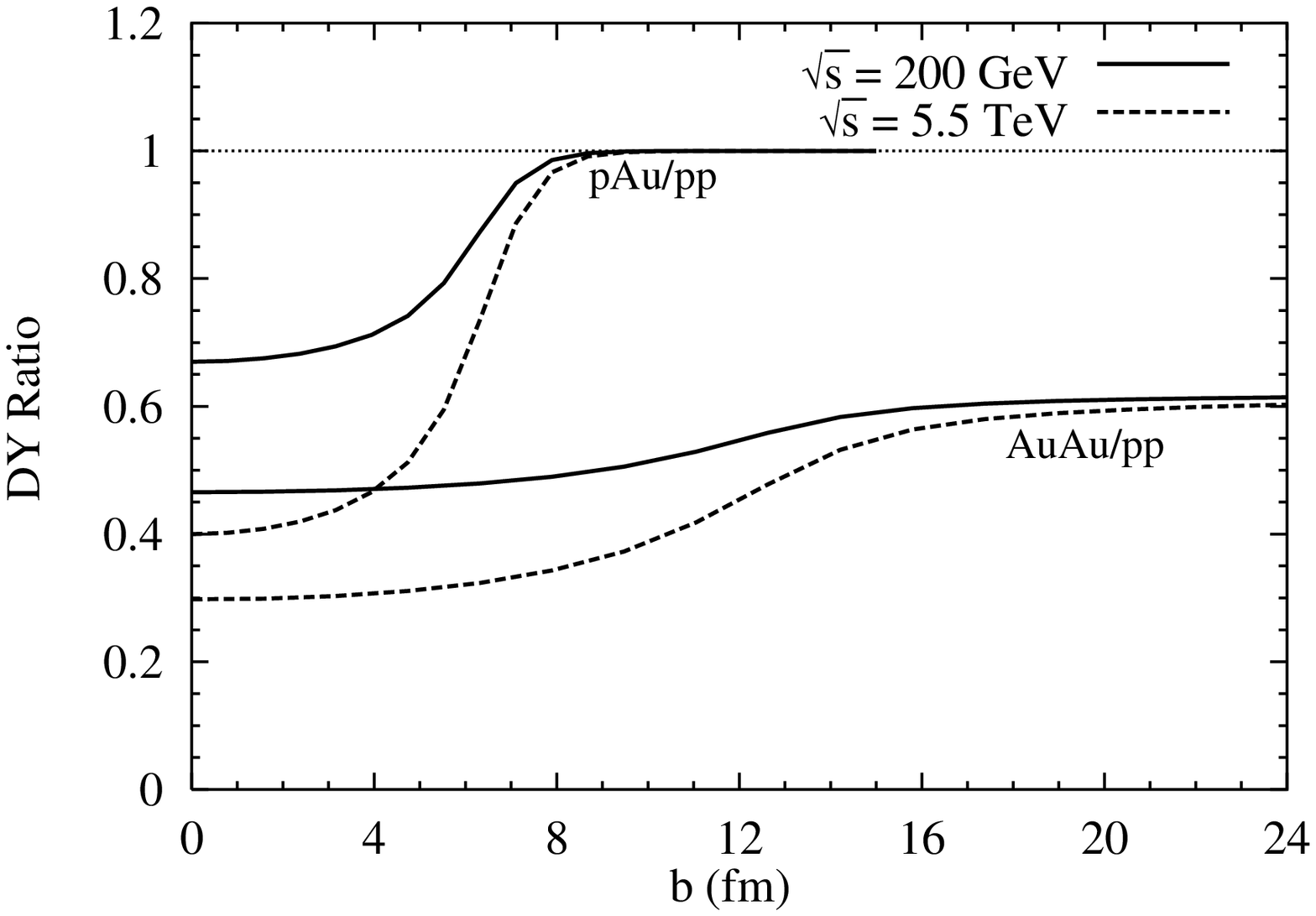}}
 }
\center{\parbox[b]{13cm}{
\caption{\label{fig:aatotal}\em
Nuclear effects on the DY process in heavy ion collisions at $M=4.5$ GeV.
For comparison, the corresponding curves for $pAu$ scattering 
(fig.\ \ref{fig:total}) are also displayed. The plot on the right, 
which shows the impact parameter dependence of nuclear effects, is calculated
at $x_F=0.5$. The curves for $AuAu$ collisions do not approach $1$ at large $b$
because of the different flavor composition of a nucleus. Note that all curves
are divided by the values for $pp$ scattering.
}}}
\end{figure}

For our actual calculations, we employ the EKS98 \cite{ekr} parameterization
of the nuclear parton distributions for nuclear effects at large $x$ 
(EMC, Fermi motion). For the parton densities of a proton, 
which are
needed as a baseline for EKS98 and for calculation of the denominator 
(\ref{nn-x-section}), we use the CTEQ5L parameterization \cite{cteq}. 
Note that, 
unlike in the $pA$ case, parton densities of the proton are not only needed 
at large $x$, but also at very small $x\ll 0.1$. The CTEQ parameterization is 
applicable
down to $x=10^{-5}$, which is sufficient for RHIC ($x_2\approx 0.001$) but not 
for LHC,
where values as low as $x\approx 10^{-6}$ are reached. For the LHC calculation, 
we therefore switch to GRV98LO
\cite{grv}, which is applicable down to $x=10^{-9}$.
All evolution codes are taken from CERNLIB PDFLIB 8.04 \cite{cernlib}.

Our predictions for the DY modification in
$AA$ collisions at RHIC and LHC are shown in fig.~\ref{fig:aatotal}.
For comparison, we also display the analogous curves for $pA$ scattering from 
Sect.~{\ref{sec:shadow}. The plot on the left shows that DY dilepton 
production in $AuAu$ collisions is suppressed much more strongly than in $pAu$
collisions at the same kinematics, except for large values of $x_F$, where
Fermi motion makes the nuclear suppression vanish. However, the strong 
suppression in $AuAu$ collisions is not just a combination of shadowing and 
the EMC effect. As one can see from the plot on the right, DY from $AuAu$ is 
still suppressed by about 40\% compared to $pp$
at very large $b$, where such nuclear effects are
absent. Indeed, a nonnegligible part of the suppression in nucleus-nucleus
collisions is due to the different flavor composition of nuclei. A heavy nucleus
like gold ($A=197$, $Z=79$) has more neutrons than protons, thus the average 
nucleon in this nucleus has an excess of $d$-quarks over $u$-quarks, compared 
to a proton. Since $d$-quarks enter the lowest-order expression for the DY 
cross section (\ref{nn-x-section}) with a weight factor of 
$Z^2_d=1/9$ (compared
to $Z^2_u=4/9$) one observes an additional suppression. 
Finally, we mention that 
the suppression at large $b$ could be even stronger than 
in fig.\ \ref{fig:aatotal},
because neutrons are predominantly located at large impact parameter. 
This effect is not taken into account in our calculation, where all 
nucleons are
assumed to be uncorrelated.

\section{Summary and outlook}\label{sec:outlook}

We have presented an analysis of nuclear effects in DY dilepton production in
$pA$ and $AB$ collisions. All calculations are performed within the light
cone dipole formalism for the DY process, because this approach suggests a
very simple and intuitive treatment of nuclear effects.  It essentially
simplifies at high energies, where the coherence length for the DY process
substantially exceeds the nuclear size. Then one can employ the eikonal
formalism to describe multiple interactions, since the different eigenstates of
interaction do not mix.  This regime, relevant to the energies of RHIC and
LHC, is considered throughout the paper. Formulas for the DY cross
section are sufficiently simple to incorporate realistic nuclear densities and
a realistic parameterization of the dipole cross section. The predictions for
RHIC and LHC presented here can therefore be compared to future data and
serve as a test of the theory. 

Since we assume the coherence length to be long, gluon shadowing becomes
important. The coherence length for higher Fock states is shorter than that
for the lowest Fock state, and it is usually the order of,
or shorter than, the nuclear radius. In this case, simple eikonalization 
cannot be applied to
gluon shadowing and we employ the LC Green function formalism, which takes
account of the variations in size of the projectile fluctuations as they 
propagate through the nucleus. The gluon shadowing incorporated in our
calculations is especially important at very high energies, or small
$x_2\ll1$, and it makes the $q\bar q$-nucleus cross section saturate at a
value significantly smaller than the geometrical limit of $2\pi R_A^2$. 

Not surprisingly, gluon shadowing leads to a stronger suppression of the total 
DY cross section than one would obtain from quark shadowing alone.
Note that while fixed target experiments at medium-high energies find 
suppression of the DY cross section only at large Feynman-$x_F$, partially
caused by onset of shadowing and energy loss, we expect the entire
$x_F>0$-range shadowed at RHIC and LHC. For the DY transverse momentum
distribution, gluon shadowing leads to nontrivial modifications.  At low
lepton-pair transverse momentum, gluon shadowing enhances the
suppression already expected from quark shadowing, but at intermediate
transverse momentum, it strongly reduces the enhancement from the Cronin
effect.  This observation resembles the missing Cronin enhancement in 
charged
particle multiplicities \cite{Drees}.  However, the latter effect cannot be
due to gluon shadowing, because the $x$ of the data is too large. 
The observed nuclear suppression is probably
caused by final state interactions with the produced matter and is related
to the induced energy loss and absorption. 
This will be studied in more detail in a forthcoming publication.

Furthermore, we calculate the nuclear broadening of the mean transverse
momentum squared for DY dileptons produced in $pA$ collisions. This
quantity turns out to be divergent for radiation of transversely polarized
DY photons. We demonstrate that this problem is caused by nuclear shadowing
in the total DY cross section and that the two phenomena, broadening and
shadowing, are closely related.  As a result, the predicted broadening of
the transverse momentum squared depends strongly on the upper cutoff on
$q_{T}$.  This is not purely a problem of the theoretical approach, as this 
cutoff
dependence is present also in the experimental analysis. We found that
$\la\delta q_{T}^2\ra$ can vary up to a factor of three, depending on the
cutoff. We suggest a different observable in which the divergent tails of 
the $q_{T}$ distribution cancel and therefore render the result
independent of the cutoff.

We separately analyze the differential DY cross sections for transversely
and longitudinally polarized pairs. In both cases, the differential cross
section $d\sigma/d^2q_{T}$ does not diverge as $q_{T}\to 0$. This result follows
naturally in the dipole approach as a consequence of the saturating dipole
cross section. In the parton model, a more complicated resummation of logs
in $q_{T}/M$ is necessary in order to render the DY cross section finite as
$q_{T}\to 0$. On the partonic level and for large transverse momenta, we 
reproduce the behavior
expected from perturbative QCD, namely $d\sigma_T/d^2q_{T}\propto q_{T}^4$
for transverse pairs and $d\sigma_L/d^2q_{T}\propto q_{T}^6$ for
longitudinal. 

Experimentally, the transverse and the longitudinal DY cross
sections can be distinguished by investigating the angular distribution of
DY pairs. We calculate the parameter $\lambda$, Eq.\ (\ref{eq:lambda}),
which characterizes the relative contribution of transverse and
longitudinal pairs to the DY cross section as a function of the dilepton
transverse momentum and, in addition, investigate nuclear effects on
$\lambda$. Although these nuclear effects do not turn out to exceed $6\%$, the
different $q_{T}$-dependence of the transverse and longitudinal cross sections
leads to a nonmonotonic behavior of $\lambda$ that can be observed in
future experiments.

Finally, we present estimates for nuclear effects in heavy ion
collisions at the energies of RHIC and LHC. We make use of QCD
factorization, but calculate nuclear shadowing for sea and valence quarks
separately within the LC dipole approach. Contrary to usual expectations, we
found considerable shadowing for valence quarks, stronger than
for sea quarks.
We calculate nuclear suppression of DY dilepton production for
$AuAu$ collisions as function of $x_F$ and impact parameter $b$ and find
considerably stronger suppression in $AuAu$ collisions than in $pAu$ 
collisions. Even at large impact parameter, the nucleus-nucleus DY 
cross section is reduced compared to $pp$ as a result of flavor effects.

We leave for further study the following problems:  (i) Since the 
approximation of a long coherence length employed for the DY process at RHIC is
satisfied only at $x_F\gsim 0.5$, one must use the LC Green function technique 
in order to cover the entire range of $x_F$. It can also provide a proper
interpretation for the Fermilab data \cite{e772}. (ii) Development of the
LC dipole approach for the DY process in heavy ion collisions is still a
challenge; (iii) Nuclear modification of the transverse momentum
distribution and polarization effects should be also calculated for
nucleus-nucleus collisions.

\bigskip
{\bf Acknowledgments:}
 We are grateful to Jen-Chieh Peng for informing us about details of
experiments and for useful discussions. B.Z.K.\ thanks the High Energy and
Nuclear Theory groups at Brookhaven National Laboratory for hospitality
during his visit when the present paper was completed. J.R.\ is grateful to
Rainer Fries and Arthur Hebecker for discussion on the Lam Tung relation.
This work was partially supported by the Gesellschaft f\"ur
Schwer\-ionen\-for\-schung Darmstadt (GSI), grant GSI-OR-SCH, and by the
U.S.~Department of Energy at Los Alamos National Laboratory under Contract
No.~W-7405-ENG-38.

\def\appendix{\par
 \setcounter{section}{0}
\setcounter{subsection}{0}
 \def\thesection{Appendix \Alph{section}}
\def\thesubsection{\Alph{section}.\arabic{subsection}}
\def\theequation{\Alph{section}.\arabic{equation}}
\setcounter{equation}{0}}

\appendix

\section{Calculation of gluon shadowing}\label{app:gshad}

Gluon shadowing is given by the shadowing for longitudinal photons as
\beq
R_G(x,Q^2)=\frac{G_A(x,Q^2)}{AG_N(x,Q^2)}
\equiv 1-\frac{\Delta\sigma_L^{\gamma^*A}(x,Q^2)}{A\sigma_L^{\gamma^*p}(x,Q^2)}.
\eeq
and is calculated according to the
formulas derived in \cite{kst2}. In this appendix, we give some details of
our
calculation. Our starting point is Eq.\ (90) of \cite{kst2}:
\beq\label{start}
\Delta\sigma_L^{\gamma^*A}(x,Q^2)=\int d^2b \int_{-\infty}^\infty dz_1
\int_{-\infty}^\infty dz_2 \Theta(z_2-z_1)\rho_A(b,z_1)\rho_A(b,z_2)
\Gamma(x,Q^2,z_2-z_1)
\eeq
where
\beqn\label{Gamma}\nonumber
\Gamma(x,Q^2,\Delta z)&=&
\Re \int d\ln(\alpha_G)\frac{16\alpha_{em}\left(\sum_F Z_q^2\right)\alpha_s(Q^2)
C^2_{eff}}{3\pi^2Q^2\widetilde b^2}\\
\nonumber
&\times&
[(1-2\zeta-\zeta^2){\rm e}^{-\zeta}+\zeta^2(3+\zeta){\rm E}_1(\zeta)]\\
&\times 
&\left[\frac{t}{w}+\frac{\sinh(\Omega\Delta z)}{t}\ln\!\left(1-\frac{t^2}{u^2}
\right)+\frac{2t^3}{uw^2}+\frac{t\sinh(\Omega\Delta z)}{w^2}+\frac{4t^3}{w^3}
\right]
\eeqn
is given by Eq.\ (112) of \cite{kst2} and
\beqn
\Delta z&=&z_2-z_1,\\
\Omega&=&\frac{{\rm i}B}{\alpha_G(1-\alpha_G)\nu},\\
B&=&\sqrt{\widetilde b^4-{\rm i}\alpha_G(1-\alpha_G)\nu C_{eff}\rho_A},\\
\nu&=&\frac{Q^2}{2m_Nx},\\
\zeta&=&{\rm i} xm_N\Delta z,\\
t&=&\frac{B}{\widetilde b^2},\\
u&=&t\cosh(\Omega\Delta z)+\sinh(\Omega\Delta z),\\
w&=&(1+t^2)\sinh(\Omega\Delta z)+2t\cosh(\Omega\Delta z),\\
\widetilde b^2&=&(0.65\,\,{\rm GeV})^2+\alpha_GQ^2.
\eeqn
The limits for the $\alpha_G$-integration are $x\le\alpha_G\le0.1$, with
$\alpha_G$ being the momentum fraction of the gluon relative to its parent
quark. 
We use a running coupling constant \cite{Dok}
\beq
\alpha_s(Q^2)=\frac{4\pi}{9\ln\!\left(\frac{Q^2+0.25\,{\rm GeV}^2}{(200\,{\rm
MeV})^2}\right)}
\eeq
with freezing at low scales.
In
(\ref{Gamma}), the gluon-gluon-nucleon cross section is parameterized in the
form
\beq
\sigma^N_{GG}(\rho,\widetilde x)=C_{eff}(\widetilde x)\rho^2.
\eeq
Note that $\sigma^N_{GG}$ is sampled at the energy $\widetilde x=x/\alpha_G$.
For $\alpha_G\to x$, $\widetilde x$ can become greater than 0.1 and the dipole
formulation is no longer valid. To overcome this problem, we employ the
prescription
\beq
\widetilde x={\rm min}(x/\alpha_G,0.1).
\eeq
\begin{figure}[t]
\centerline{
  \scalebox{0.43}{\includegraphics{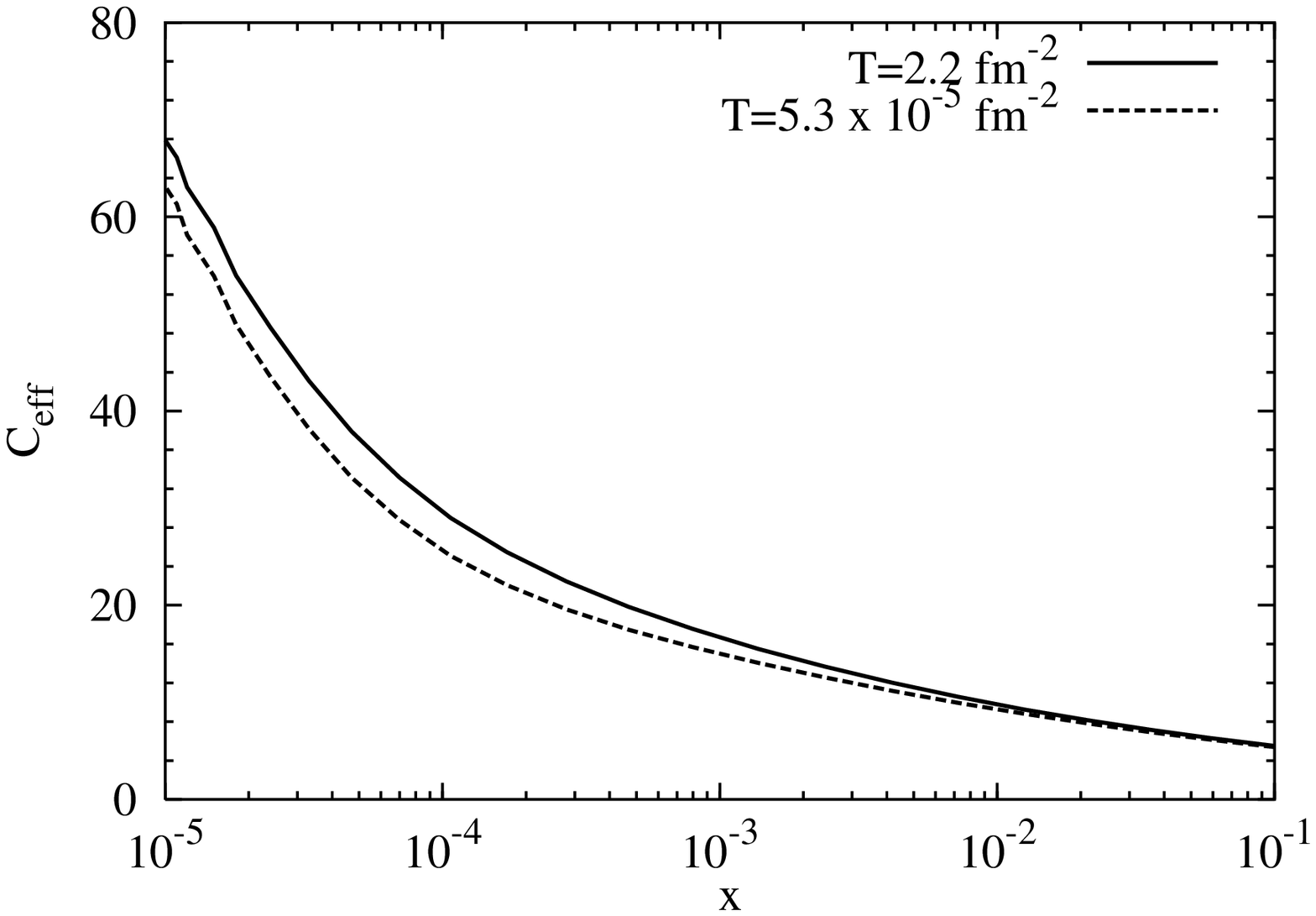}}
  \scalebox{0.43}{\includegraphics{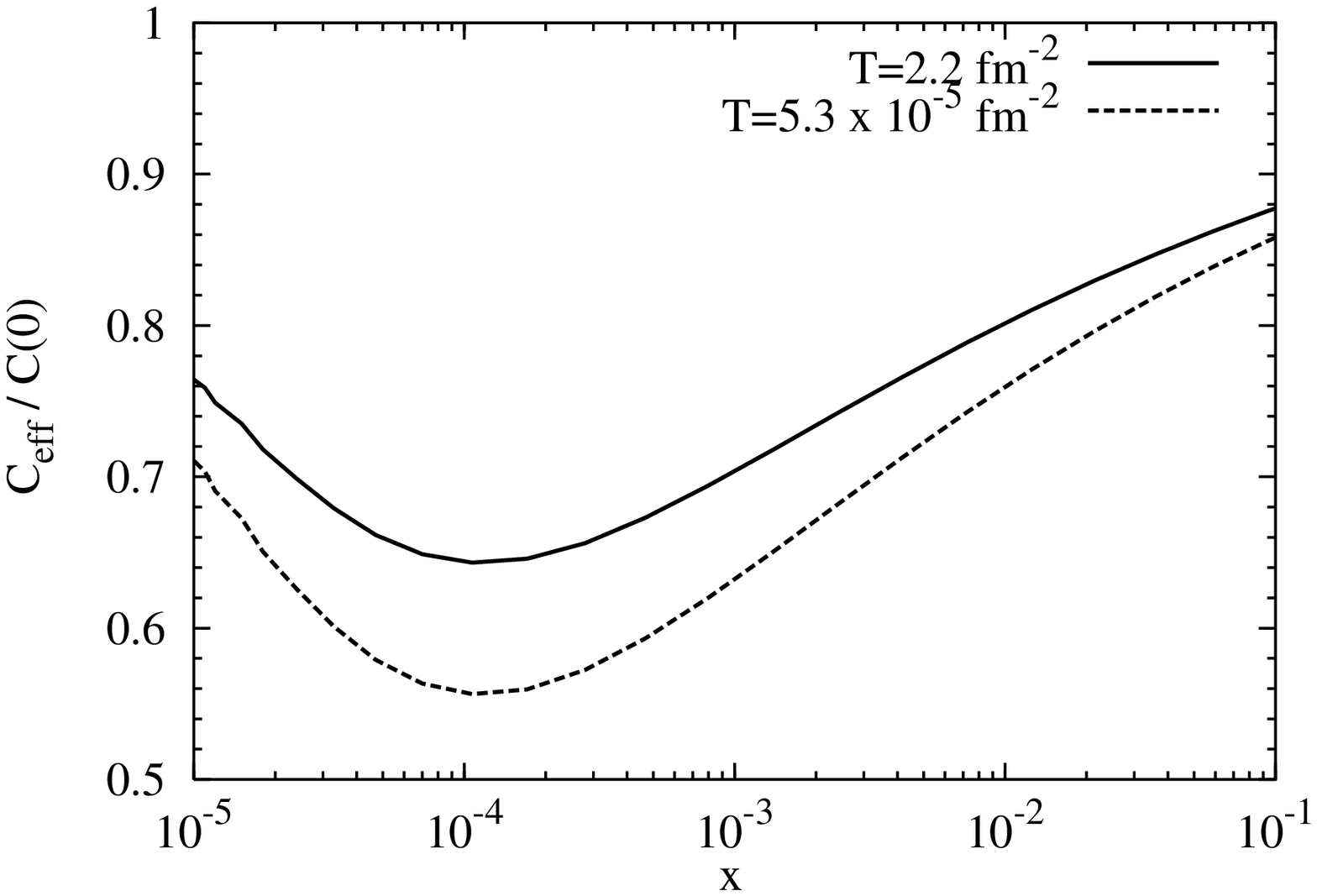}}
 }
\center{\parbox[b]{13cm}{
\caption{\label{fig:ceff}\em
  The $x$-dependence of $C_{eff}$ for gold at $Q^2=20$ GeV$^2$. It is
  also shown how much $C_{eff}$ is suppressed compared to $C(\rho=0)$.
 The two curves in each plot correspond to impact parameter $b=0$ (solid) 
and very large impact parameter (dashed), respectively.
}}}
\end{figure}

The parameter $C_{eff}$ is then determined from the asymptotic condition
\beqn\lefteqn{\label{asymp}\nonumber
\frac{\int d^2b d^2\rho\left|\Psi_{qG}(\rho)\right|^2
\left(1-\exp\!\left(-\frac{1}{2}
C_{eff}(\widetilde{x})\rho^2\,T_A(b)\right)\right)}
{\int d^2\rho\left|\Psi_{qG}(\rho)\right|^2C_{eff}(\widetilde x)\rho^2}}
& & \hphantom{XXXXXXXXXX}
\\
\nonumber
\\
&=&\frac{\int d^2b d^2\rho\left|\Psi_{qG}(\rho)\right|^2
\left(1-\exp\!\left(-\frac{9}{8}\sigma^N_{q\bar q}(\rho,\widetilde{x})
\,T_A(b)\right)\right)}
{\int d^2\rho\left|\Psi_{qG}(\rho)\right|^2
\frac{9}{4}\sigma^N_{q\bar q}(\rho,\widetilde x)},
\eeqn
where $\sigma^N_{q\bar q}(\rho,\widetilde x)$ is the dipole cross 
section in the
saturation model of \cite{Wuesthoff1}.
The LC wave function for radiation of a quark from a gluon, including the
nonperturbative interaction introduced in \cite{kst2}, reads
\beq
\left|\Psi_{qG}(\rho)\right|^2=\frac{4\alpha_s}{3\pi^2}
\frac{\exp\!\left(-\widetilde{b}^2\rho^2\right)}{\rho^2}.
\eeq
The choice of $C_{eff}$ differs from the one made in \cite{kst2}, where
$C_(\rho=0)=d\sigma^N_{GG}(\rho)/d\rho^2|_{\rho=0}$ was employed 
as the effective C. 
The prescription (\ref{asymp}) is more realistic, because the $C_{eff}$ is
determined by those values of $\rho$ which are most important for shadowing.
Since the dipole cross section levels off at large separations, $C_{eff}$ will
be lower than $C(\rho=0)$. This is illustrated in fig.\ \ref{fig:ceff}.
The $\sigma_L^{\gamma^*p}$ in the denominator of (\ref{start}) is calculated
with $C_{eff}(x)$ (instead of $\widetilde x$).

With a constant nuclear density $\rho_A$, one can integrate
(\ref{start}) twice by parts,
\beq
\Delta\sigma_L^{\gamma^*A}(x,Q^2)=\frac{\pi}{12}\rho_A^2\int_0^{2R_A} dL
\left(L^3-12R_A^2L+16R_A^3\right)
\Gamma(x,Q^2,L),
\eeq
with $L=2\sqrt{R_A^2-b^2}$.

The $q\bar q$-nucleus cross section in the long coherence time limit is 
calculated from the formula
\beq\label{signuc2}
\sigma_{q\bar q}^A(\rho,x)=
2\int d^2b\left\{1-\left(1-\frac{\sigma^N_{q\bar q}(\rho,x)
T_A(b)R_G(x,Q^2,b)}{2A}\right)^A\right\},
\eeq
where gluon shadowing as function of impact parameter $b$ is given by
\beq\label{A.20}
R_G(x,Q^2,b)
=1-\frac{\Delta\sigma_L^{\gamma^*A}(x,Q^2,b)}{T_A(b)\sigma_L^{\gamma^*p}(x,Q^2)},
\eeq
and
\beq\label{A.21}
\Delta\sigma_L^{\gamma^*A}(x,Q^2,b)=\rho_A^2\int_0^{L} dz
\left(L-z\right)
\Gamma(x,Q^2,z),
\eeq
Furthermore, in (\ref{signuc2}) $R_G$ is evaluated at the scale
\beq
Q^2=\frac{1}{\rho^2}+4\,{\rm GeV}^2.
\eeq

\begin{figure}[t]
  \scalebox{0.5}{\includegraphics{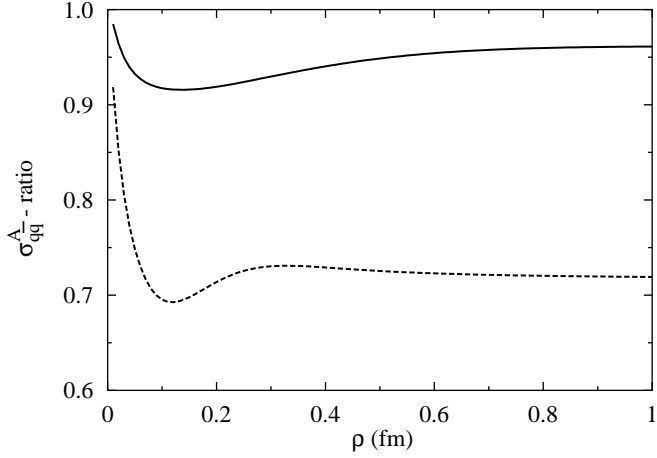}}\hfill
  \raise0.5cm\hbox{\parbox[b]{2.44in}{
   \caption{\label{sig_kern2}\em
   This figure illustrates the influence of gluon shadowing on the
  $q\bar q$-nucleus cross section. It
  shows the $q\bar q$-nucleus cross section for gold
  calculated with gluon
  shadowing, Eq.\ (\ref{signuc2}),
  divided by the corresponding quantity without gluon shadowing, i.e.
  $R_G\to 1$. The solid curve is for $x=10^{-3}$ (RHIC), while the dashed curve
  is for $x=10^{-6}$ (LHC).
  	} 
  }
}

\end{figure}

\section{Calculation of the DY transverse momentum
distribution}\label{appdypp}

\setcounter{equation}{0}

The differential DY cross section is expressed as a four-fold
Fourier integral (\ref{eq:dylcdiff})
\beqn\nonumber
\frac{d\sigma(qp\to \gamma^*X)}{d\ln\alpha d^2{q_{T}}}
&=&\frac{1}{(2\pi)^2}
\int d^2\rho_1d^2\rho_2\, \exp[{\rm i}\vec {q_{T}}\cdot(\vec\rho_1-\vec\rho_2)]
\Psi^*_{\gamma^* q}(\alpha,\vec\rho_1)\Psi_{\gamma^* q}(\alpha,\vec\rho_2)\\
&\times&
\frac{1}{2}
\left\{\sigma^N_{q\bar q}(\alpha\rho_1)
+\sigma^N_{q\bar q}(\alpha\rho_2)
-\sigma^N_{q\bar q}(\alpha(\vec\rho_1-\vec\rho_2))\right\},
\eeqn
where
\beqn
\Psi^{*T}_{\gamma^* q}(\alpha,\vec\rho_1)\Psi^T_{\gamma^* q}(\alpha,\vec\rho_2)
&=& \frac{\alpha_{em}}{2\pi^2}\Bigg\{
     m_f^2 \alpha^4 {\rm K}_0\left(\eta\rho_1\right)
     {\rm K}_0\left(\eta\rho_2\right)\nonumber\\
   &+& \left[1+\left(1-\alpha\right)^2\right]\eta^2
   \frac{\vec\rho_1\cdot\vec\rho_2}{\rho_1\rho_2}
     {\rm K}_1\left(\eta\rho_1\right)
     {\rm K}_1\left(\eta\rho_2\right)\Bigg\}\ ,
\label{eq:dylctT}\\
 \Psi^{*L}_{\gamma^* q}(\alpha,\vec\rho_1)\Psi^L_{\gamma^* q}(\alpha,\vec\rho_2)
&=& \frac{\alpha_{em}}{\pi^2}M^2 \left(1-\alpha\right)^2
  {\rm K}_0\left(\eta\rho_1\right)
     {\rm K}_0\left(\eta\rho_2\right)\ . 
\label{eq:dylctL}
 \eeqn
 and $\eta^2 = (1-\alpha)M^2 + \alpha^2m_q^2$.
The Fourier-integral is inconvenient for numerical calculations, but one
can perform three of the integrations analytically for arbitrary
$\sigma^N_{q\bar q}(\alpha\rho)$. 

Consider the K$_0$-part first. With help of the relation
\beq\label{k0relation}
{\rm K}_0(\eta\rho)=\frac{1}{2\pi}\int d^2l\frac{{\rm e}^{{\rm i}\vec
l\cdot\vec\rho}}{l^2+\eta^2},
\eeq
one finds
\beqn\nonumber\label{k0part}
\frac{d\sigma(qp\to \gamma^*X)}{d\ln\alpha d^2{q_{T}}}
\Bigg|_{{\rm K}_0{\rm -part}}
&=&\frac{\alpha_{em}}{2\pi^2}
\left[m_f^2 \alpha^4+2M^2\left(1-\alpha\right)^2\right]
\frac{1}{(2\pi)^2}\int d^2\rho_1d^2\rho_2
\frac{d^2l_1}{2\pi}\frac{d^2l_2}{2\pi}\\
\nonumber&\times&
\frac{{\rm e}^{{\rm i}\vec {q_{T}}\cdot(\vec\rho_1-\vec\rho_2)}\,
{\rm e}^{-{\rm i}\vec l_1\cdot\vec\rho_1}\,
{\rm e}^{{\rm i}\vec l_2\cdot\vec\rho_2}}
{\left(l_1^2+\eta^2\right)\,\left(l_2^2+\eta^2\right)}\\
&\times&
\frac{1}{2}
\left\{\sigma^N_{q\bar q}(\alpha\rho_1)
+\sigma^N_{q\bar q}(\alpha\rho_2)
-\sigma^N_{q\bar q}(\alpha(\vec\rho_1-\vec\rho_2))\right\}.
\eeqn
Note that the term in the curly brackets consists of three contributions, which
depend either only on $\rho_1$ or on $\rho_2$ or on the difference 
$\vec\rho_1-\vec\rho_2$. Thus, the integral (\ref{k0part}) can be split into
three terms. In the integral that arises from the 
$\sigma^N_{q\bar q}(\alpha\rho_1)$ part, the $\rho_2$-integration is trivially
performed and leads to a two dimensional delta-function 
$\delta^{(2)}(\vec {q_{T}}-\vec l_2)$. This makes it possible to perform also 
the
$l_2$ integration. The integration over $l_1$ gives just the MacDonald function
K$_0$ (\ref{k0relation}).
Thus one is left with a two-fold integration over $\rho_1$. Provided the 
dipole cross section
depends only on the modulus of $\rho$, one can use the relation
\beq
{\rm J_0}=\frac{1}{2\pi}\int d\phi\,{\rm e}^{{\rm i}\vec
l\cdot\vec\rho}
\eeq
to perform one more integration. Here, J$_0$ is a Bessel function of first
kind. The contribution arising from $\sigma^N_{q\bar q}(\alpha\rho_2)$
is calculated in exactly the same way. For the 
$\sigma^N_{q\bar q}(\alpha(\vec\rho_1-\vec\rho_2))$-part one has 
to introduce the
auxiliary variable $\vec d=\vec\rho_1-\vec\rho_2$, before the procedure
described above can be applied.

The K$_1$ part is calculated in a similar way. Note that
\beq\label{k1relation}
{\rm K}_1(\eta\rho)=-\frac{1}{\eta}\frac{d}{d\rho}{\rm K}_0(\eta\rho).
\eeq
The K$_1$-part reads
\beqn\nonumber\label{k1part}
\frac{d\sigma(qp\to \gamma^*X)}{d\ln\alpha d^2{q_{T}}}\Bigg|_{{\rm K}_1{\rm -part}}
&=&\frac{\alpha_{em}}{2\pi^2}
\left[1+\left(1-\alpha\right)^2\right]
\frac{1}{(2\pi)^2}\int d^2\rho_1d^2\rho_2
\frac{d^2l_1}{2\pi}\frac{d^2l_2}{2\pi}\\
\nonumber&\times&
\frac{{\rm e}^{{\rm i}\vec {q_{T}}\cdot(\vec\rho_1-\vec\rho_2)}\,
{\rm e}^{-{\rm i}\vec l_1\cdot\vec\rho_1}\,
{\rm e}^{{\rm i}\vec l_2\cdot\vec\rho_2}}
{\left(l_1^2+\eta^2\right)\,\left(l_2^2+\eta^2\right)}
\vec l_1\cdot\vec l_2\\
&\times&
\frac{1}{2}
\left\{\sigma^N_{q\bar q}(\alpha\rho_1)
+\sigma^N_{q\bar q}(\alpha\rho_2)
-\sigma^N_{q\bar q}(\alpha(\vec\rho_1-\vec\rho_2))\right\}.
\eeqn
Like the K$_0$-part, the complete integral (\ref{k1part}) is split into three
pieces, corresponding to the three terms in the curly brackets. Again, one
integration over $\rho$ is immediately performed, leading to
$\delta$-functions, which allows one to do one integration over $l$. With the
second $l$-integration, one recovers the MacDonald function K$_1$ via
(\ref{k1relation}). For the K$_1$-part, one also needs the relation
\beq
{\rm J}_1(z)=-\frac{d}{dz}{\rm J}_0(z).
\eeq
Although the calculation is slightly more cumbersome for the
$\sigma^N_{q\bar q}(\alpha(\vec\rho_1-\vec\rho_2))$-part, 
all calculations are easily
performed.

Finally, one finds
\beqn\nonumber
\frac{d\sigma(qp\to \gamma^*X)}{d\ln\alpha d^2{q_{T}}}
&=&\frac{\alpha_{em}}{2\pi^2}\left\{
\left[m_f^2 \alpha^4+2M^2\left(1-\alpha\right)^2\right]
\left[\frac{1}{{q_{T}}^2+\eta^2}{\cal I}_1-\frac{1}{4\eta}{\cal I}_2\right]
\right.\\
&+&\left.\left[1+\left(1-\alpha\right)^2\right]
\left[\frac{\eta {q_{T}}}{{q_{T}}^2+\eta^2}{\cal I}_3-
\frac{{\cal I}_1}{2}+\frac{\eta}{4}{\cal
I}_2\right]\right\},
\eeqn
with
\beqn
{\cal I}_1&=&\int_0^\infty dr r {\rm J}_0({q_{T}}r){\rm K}_0(\eta r)
\sigma^N_{q\bar q}(\alpha r)\\
{\cal I}_2&=&\int_0^\infty dr r^2 {\rm J}_0({q_{T}}r){\rm K}_1(\eta r)
\sigma^N_{q\bar q}(\alpha r)\\
{\cal I}_3&=&\int_0^\infty dr r {\rm J}_1({q_{T}}r){\rm K}_1(\eta r)
\sigma^N_{q\bar q}(\alpha r).
\eeqn
The remaining integrals are evaluated numerically with the Numerical Recipes
\cite{numrec} routines.

\end{document}